\documentclass[a4paper,useAMS,usenatbib,usegraphicx]{mn2eadapt}

\usepackage[latin1]{inputenc}
\usepackage{amsmath}
\usepackage{amsfonts}
\usepackage{amssymb}
\usepackage[english]{babel}
\usepackage{times}
\usepackage{color}
\usepackage[colorlinks=false]{hyperref}
\hypersetup{pdfborder=0 0 0}
\usepackage{rotating}

\newcommand{\Msun}{M_{\odot}}

\newcommand{\Lsun}{L_{\odot}}
\newcommand{\kms}{km~s$^{-1}$}

\newcommand{\OI}{O~{\sc i}}

\newcommand{\CI}{C~{\sc i}}
\newcommand{\CII}{C~{\sc ii}}
\newcommand{\NaI}{Na~{\sc i}}

\newcommand{\SII}{S~{\sc ii}}

\newcommand{\SiII}{Si~{\sc ii}}

\newcommand{\CaII}{Ca~{\sc ii}}
\newcommand{\TiII}{Ti~{\sc ii}}

\newcommand{\CrII}{Cr~{\sc ii}}

\newcommand{\FeII}{Fe~{\sc ii}}

\newcommand{\Fefs}{$^{56}$Fe}
\newcommand{\Cofs}{$^{56}$Co}
\newcommand{\Nifs}{$^{56}$Ni}

\newcommand{\aap}{A\&A}
\newcommand{\mnras}{MNRAS}
\newcommand{\apj}{ApJ}
\newcommand{\apjs}{ApJS}
\newcommand{\apjl}{ApJ}
\newcommand{\aj}{AJ}

\newcommand{\nature}{Nature}

\newcommand{\sci}{Sci}

\begin{document}

\title[Dim SN Ia 2005bl: abundances and density profile.]{Spectral analysis of the 91bg-like Type~Ia SN~2005bl:\\Low luminosity, low velocities, incomplete burning.}
\author[Hachinger et al.]{Stephan Hachinger$^{1}$, Paolo A. Mazzali$^{1,2,3}$, Stefan Taubenberger$^{1}$,
\protect\vspace{0.2cm}\\{\upshape\LARGE R\"udiger Pakmor$^{1}$, Wolfgang Hillebrandt$^{1}$}\\
$^1$Max-Planck-Institut f\"ur Astrophysik, Karl-Schwarzschild-Str.\ 1, 85748 Garching, Germany\\
$^2$Scuola Normale Superiore, Piazza dei Cavalieri 7, 56126 Pisa, Italy\\
$^3$Istituto Nazionale di Astrofisica-OAPd, vicolo dell'Osservatorio 5, 35122 Padova, Italy}

\date{arXiv v3, 2012-12-23. The published paper is available at \href{http://www.blackwell-synergy.com}{www.blackwell-synergy.com}.}

\pubyear{2009}
\volume{}
\pagerange{}

\maketitle

\begin{abstract}
The properties of underluminous type Ia supernovae (SNe Ia) of the 91bg subclass have yet to be theoretically understood. Here, we take a closer look at the structure of the dim SN Ia 2005bl. We infer the abundance and density profiles needed to reproduce the observed spectral evolution between $-$$6$d and $+$$12.9$d with respect to $B$ maximum. Initially, we assume the density structure of the standard explosion model W7; then we test whether better fits to the observed spectra can be obtained using modified density profiles with different total masses and kinetic energies. Compared to normal SNe Ia, we find a lack of burning products especially in the rapidly-expanding outer layers ($v$$\gtrsim$$15000$\kms). The zone between $\sim$$8500$ and $15000$\kms\ is dominated by oxygen and includes some amount of intermediate mass elements. At lower velocities, intermediate mass elements dominate. This holds down to the lowest zones investigated in this work. This fact, together with negligible-to-moderate abundances of Fe-group elements, indicates large-scale incomplete Si burning or explosive O burning, possibly in a detonation at low densities. Consistently with the reduced nucleosynthesis, we find hints of a kinetic energy lower than that of a canonical SN Ia: The spectra strongly favour reduced densities at $\gtrsim$$13000$\kms\ compared to W7, and are very well fitted using a rescaled W7 model with original mass ($1.38$$\Msun$), but a kinetic energy reduced by $\sim$$30\%$ (i.e. from $1.33$$\cdot$$10^{51}$erg to $0.93$$\cdot$$10^{51}$erg).
\end{abstract}


\section{Introduction}
\label{sec:introduction}

Type Ia supernovae (SNe Ia) play a key role in modern astrophysics. They are invaluable as distance indicators for cosmology (e.g. \citealt{per97,per99,rie98,ast06,woo07}) because of the high accuracy with which the absolute luminosity of most SNe Ia can be inferred. The luminosity varies among different objects, but the variations correlate with distance-independent light-curve parameters such as the decline in magnitudes in the $B$-band within 15 days after $B$ maximum \citep{phi93}. Luminosity calibration techniques exploiting this fact are mostly applied to ``normal'' SNe Ia \citep{bra93} not showing poorly-understood peculiarities. These SNe supposedly emerge from a homogeneous sample of progenitors, which are thought to be C-O white dwarfs (WDs) accreting matter from a non-degenerate companion star (single-degenerate scenario).

In the single-degenerate paradigm, the smooth variations among ``normal'' SNe Ia \citep{bra93,nug95} can be explained within a delayed-detonation scenario \citep{kho91}: an initially subsonic explosion (deflagration) undergoes a deflagration-detonation transition (DDT) and proceeds as a supersonic detonation afterwards. The efficiency and extent of burning in the initial deflagration may then vary from object to object, which affects the nucleosynthesis and causes the observed variability \citep{maz07}.

Extremely sub- or superluminous SNe Ia \citep[e.g.][]{fil92bg,phi92,lei93,how06}, on the other hand, are more difficult to explain. Here, progenitors deviating from the Chandrasekhar mass may play a role, or some explosions might result from a merger of two WDs (double-degenerate scenario). Progenitor systems producing peculiar SNe Ia might also produce some rather ``normal'' explosions, contaminating the sample of homogeneous explosions used for distance determination. Clarifying which explosion scenarios lead to SNe Ia at which rates is therefore important for supernova cosmology, but it will also be of value for other fields. Studies concerned with the binary progenitors and population synthesis \citep[e.g.][]{rui09}, observed supernova rates \citep{gre08} or the impact of supernovae on their surroundings \citep[e.g.][]{sat07} will profit from understanding the origin of peculiar supernovae.

Thus motivated, we analyse the 91bg-like SN 2005bl \citep{tau08}. SNe of the 91bg subclass are dim and decline rapidly \citep[e.g.][]{fil92bg,lei93,tur96,gar04}. They were used, with other SNe, to infer the slope of the relation between luminosity and decline rate of SNe Ia \citep{phi93}, but later it became clear that dim SNe decline even more rapidly than expected from a linear luminosity$-$decline-rate relation among normal SNe \citep{phi99,tau08}. Spectroscopically, 91bg-like SNe show characteristic peculiarities, such as low line velocities around $B$ maximum \citep[e.g.][]{fil92bg} and clear spectral signatures of \TiII, indicating lower ionisation \citep{maz97bg}. All these properties together are consistent with a low mass of newly-synthesised \Nifs. To date, no elaborate explosion models have convincingly reproduced 91bg-like SNe Ia. Pure deflagration models show even lower expansion velocities than observed in these objects, especially when little \Nifs\ is produced \citep[cf.][]{sah08}. Delayed-detonation models might explain 91bg-like objects within a unified scenario for SNe Ia \citep{maz07}. Yet, there are hints of qualitative differences. One example are the improved fits to spectra of SN~1991bg of \citet{maz97bg}, enabled by a reduction in ejecta mass and kinetic energy with respect to canonical values. Ultimately, only refined analyses of photometric and spectroscopic properties can constrain explosion models.

We use a spectral synthesis code to analyse the structure and abundance stratification of SN 2005bl, reproducing its observed spectral evolution. The ``abundance tomography`` method \citep{ste05}, which we use, exploits the fact that the optically thick region of the ejecta becomes smaller as time progresses. Thus, deeper and deeper layers contribute to spectrum formation. Modelling a time series of spectra, we infer the abundance profile from the outer envelope to as deep a layer as possible. We then test whether variations in mass or explosion energy are needed to explain the differences between spectra of normal and dim SNe Ia. This is done performing abundance tomography with various density profiles, and assessing the quality of the resulting spectral fits. The range in masses and energies sampled by the modified models starts at $0.5$$\cdot$$M_{\textrm{Ch}}$ / $\sim$5$\cdot$$10^{50}$erg  and extends to $1.45$$\cdot$$M_{\textrm{Ch}}$ / $\sim$2$\cdot$$10^{51}$erg ($M_{\textrm{Ch}}$: Chandrasekhar mass, $1.4\Msun$). This choice has been motivated by parameters inferred for observed extreme SNe Ia of all kinds \citep{maz97bg,how06}.

The paper is structured as follows: First, we give a short introduction to the methods employed (Sec. \ref{sec:method}). We then present the models for SN~2005bl (Sec. \ref{sec:models}), discuss and assess them (Sec. \ref{sec:discussion}), and finally draw conclusions (Sec. \ref{sec:conclusions}).

\section{Method}
\label{sec:method}

The radiative transfer code we use and the abundance tomography method have already been described \citep{ste05,maz08eo}. Thus, we focus on aspects necessary for an understanding of the present study.

\subsection{Radiative transfer}

We use a 1D Monte Carlo (MC) radiative transfer code (\citealt{abb85}, \citealt{maz93a}, \citealt{luc99}, \citealt{maz00} and \citealt{ste05}) to compute SN spectra from a given density and abundance profile. The aim is to infer the chemical structure adjusting the abundances within the envelope until an optimal fit to the observed spectra is obtained.

The code computes the radiative transfer through the SN ejecta above an assumed photosphere. The densities within the envelope are calculated from an initial density profile describing the state of the ejecta after homologous expansion has set in, which is a few seconds after the explosion (e.g. \citealt{roe05}). The ejecta expand radially with $r=v\cdot t$, where $r$ is the distance from the centre, $t$ the time from explosion (see beginning of Sec. \ref{sec:models}), and $v$ the velocity. Radius and velocity can be used interchangeably as coordinates.

From the photosphere, which is located at an adjustable $v_{\textrm{ph}}$, thermal radiation [$I_{\nu}^{+}=B_{\nu}(T_{\textrm{ph}})$] is assumed to be emitted into the atmosphere. This is of course quite a crude approximation to the pseudo-continuous radiation field deep in the ejecta \citep{sau06}. Notable deviations mainly appear in the red and infrared, where a departure of the flux level from that of the observed spectra sometimes cannot be avoided. 

The radiation is simulated as ''photon packets``, which undergo Thomson scattering as well as line excitation-deexcitation processes, treated in the Sobolev approximation. The process of photon branching is included, which implies that the transitions for excitation and deexcitation can be different. In a branching event, the photon packet is not split up. Instead, it is emitted as a whole with a new frequency corresponding to a possible downward transition. This ''indivisible packet`` approach \citep{luc99} enforces radiative equilibrium. The downward transition is randomly selected, taking into account effective emission probabilities. Thus, if a large number of packets are simulated, the distribution of decays reflects the actual one.

A modified nebular approximation, which mimics effects of non-local thermodynamic equilibrium (NLTE), is used to calculate the action of the radiation field onto the gas. For each of the $30$ zones into which the envelope is discretised here, a radiation temperature $T_R$ and an equivalent dilution factor $W$ are calculated. These quantities mostly determine the excitation and ionisation state \citep{abb85}. Only the variables describing the state of the gas are discretised; for the paths and redshifts of photons, and for the positions of interaction surfaces of lines, continuous values are allowed.

The code iterates the radiation field and the gas conditions. Furthermore, $T_\textrm{ph}$ is automatically modified so as to match a given output luminosity $L$, taking backscattering into account. After convergence, the emerging spectrum is obtained from a formal integral solution of the transfer equation \citep{luc99}. 

\begin{table}
\scriptsize
\caption{Density models used in this work, and their total kinetic energy $E'_\textrm{k}$ and mass $M'$. The models are named according to the scaling factors for kinetic energy and mass with respect to W7 $\left(\frac{E'_\textrm{k}}{E_{\textrm{k},W7}}\textrm{ and }\frac{M'}{M_{W7}}\right)$, and sorted according to their kinetic energy $E'_\textrm{k}$.}
\label{tab:scaledmodels}
\centering
\begin{tabular}{lccccc}
 Model $\!\!\!\!$ & $\!\!\!\!$ $E'_{\textrm{k}} / E_{\textrm{k},W7}$ $\!\!\!\!$ & $\!\!\!\!$ $M'\!/ M_{W7}$ $\!\!\!\!$ &  $\!\!\!\!$ $E'_{\textrm{k}}$ [$10^{51}$ erg] $\!\!\!\!$ & $\!\!\!\!$ $M'$ [$\Msun$] $\!\!\!\!$ & $\!\!\!\!$ $\frac{E'_\textrm{k}}{M'}$ \!\!\!\!\! \Bigg/ \!\!\!\!\! $\left(\frac{E_{\textrm{k}}}{M}\right)_{\!W7\!\!\!\!\!}$ \\
 \hline
 w7e0.35	&	0.35	&	1.00	&	0.47	&	1.38	&  0.35 \\
 w7e0.5m0.5	&	0.50	&	0.50	&	0.66	&	0.69	&  1.0 \\
 w7e0.5m0.7	&	0.50	&	0.70	&	0.66	&	0.97	&  0.7 \\
 w7e0.5		&	0.50	&	1.00	&	0.66	&	1.38	&  0.5 \\
 w7e0.5m1.25	&	0.50	&	1.25	&	0.66	&	1.73	&  0.4 \\
 w7e0.7m0.7	&	0.70	&	0.70	&	0.93	&	0.97	&  1.0 \\
 w7e0.7		&	0.70	&	1.00	&	0.93	&	1.38	&  0.7 \\
 w7e0.7m1.25	&	0.70	&	1.25	&	0.93	&	1.73	&  0.6 \\
 w7e0.7m1.45	&	0.70	&	1.45	&	0.93	&	2.00	&  0.5 \\
 w7m0.7		&	1.00	&	0.70	&	1.33	&	0.97	&  1.4 \\
 w7		&	1.00	&	1.00	&	1.33	&	1.38	&  1.0 \\
 w7m1.25	&	1.00	&	1.25	&	1.33	&	1.73	&  0.8 \\
 w7m1.45	&	1.00	&	1.45	&	1.33	&	2.00	&  0.7 \\
 w7e1.45m1.45	&	1.45	&	1.45	&	1.93	&	2.00	&  1.0 \\
\hline
\end{tabular}
\end{table}

\subsection{Density profiles}
\label{sec:densityprofile}

As a first step, we adopt the density structure of the standard explosion model W7 \citep{nom84} as a basis for our calculations. We then repeat the abundance tomography with modified density profiles, changing the total mass and kinetic energy of the explosion. To achieve this, the values for each grid point in the W7 velocity-density structure are scaled uniformly (i.e. all velocities by one scaling factor, and all densities by another one) according to:
\begin{eqnarray}
\rho'& = &\rho_{W7}\cdot\left(\frac{E'_\textrm{k}}{E_{\textrm{k},W7}}\right)^{-3/2}\cdot\left(\frac{M'}{M_{W7}}\right)^{5/2}\\
v' & = & v_{W7}\cdot\left(\frac{E'_\textrm{k}}{E_{\textrm{k},W7}}\right)^{1/2}\cdot\left(\frac{M'}{M_{W7}}\right)^{-1/2}.
\end{eqnarray}

Here, $\rho'$ and $v'$ are the density and velocity coordinates of each grid point after the scaling. $E'_\textrm{k}$ and $M'$ are the new total kinetic energy and mass. 

The scaled density models used in this work are listed in Table \ref{tab:scaledmodels}, which gives an overview of the respective $\frac{E'_\textrm{k}}{E_\textrm{k}}$ and $\frac{M'}{M}$ ratios. We have not implemented every possible energy-mass combination within the limits given in Section \ref{sec:introduction}. Instead, we first constrained ourselves to a few test cases. Then, we sampled the $E'_\textrm{k}$$-$$M'$ plane more densely in the region where models of acceptable quality emerged (see Sec. \ref{sec:assessment}).

W7 naturally shows some differences with respect to more recent and realistic hydrodynamical simulations, and the scaled density profiles can also be expected to do so. However, it is possible to obtain good fits to spectra of ``normal`` SNe Ia like SN~2002bo \citep{ste05} using the W7 density structure. The results for the scaled profiles should therefore bring out possible differences between dim SNe Ia and normal ones.

\subsection{Abundance tomography}

The abundance tomography method \citep{ste05} uses a series of photospheric spectra to establish the abundance distribution within a supernova. The idea is that the opaque core of the expanding ejecta shrinks with time. Thus, a time series of spectra carries information about the abundances in the ejecta at different depths. In the picture adopted in our code, involving an approximate photosphere, the photosphere recedes to lower velocities with time. Deeper and deeper layers become visible, leaving their imprint on the spectra.

The earliest spectrum available can be used to obtain the photospheric velocity at that time and the abundances in the outer envelope. To this aim, we optimise the code input parameters to match that spectrum, as in a one-zone spectral model (e.g. \citealt{maz97bg}). The subsequent spectrum will carry the imprint of the material in the outer envelope and additionally that of the layers inside which the photosphere has receded. Because the abundances in the outer zone are already known, the abundances of the layers which have become visible can now be inferred, together with the new velocity of the photosphere. This procedure is then continued with later spectra. 

The optimum parameters inferred from a spectral model are subject to some uncertainty (see also the discussion in \citealt{maz08eo}). One important reason for this can be degeneracy, which makes the spectra appear similar for different parameter sets. The composition adopted for an outer layer in an early-epoch model may therefore be in conflict with a later spectrum, if the later spectrum is still influenced by the outer layers. In such cases, we revised the parameters for the outer layers so as to optimise the earlier and later spectra at the same time.

\section{Models}
\label{sec:models}

\begin{figure*}
   \centering
   \includegraphics[width=14.5cm]{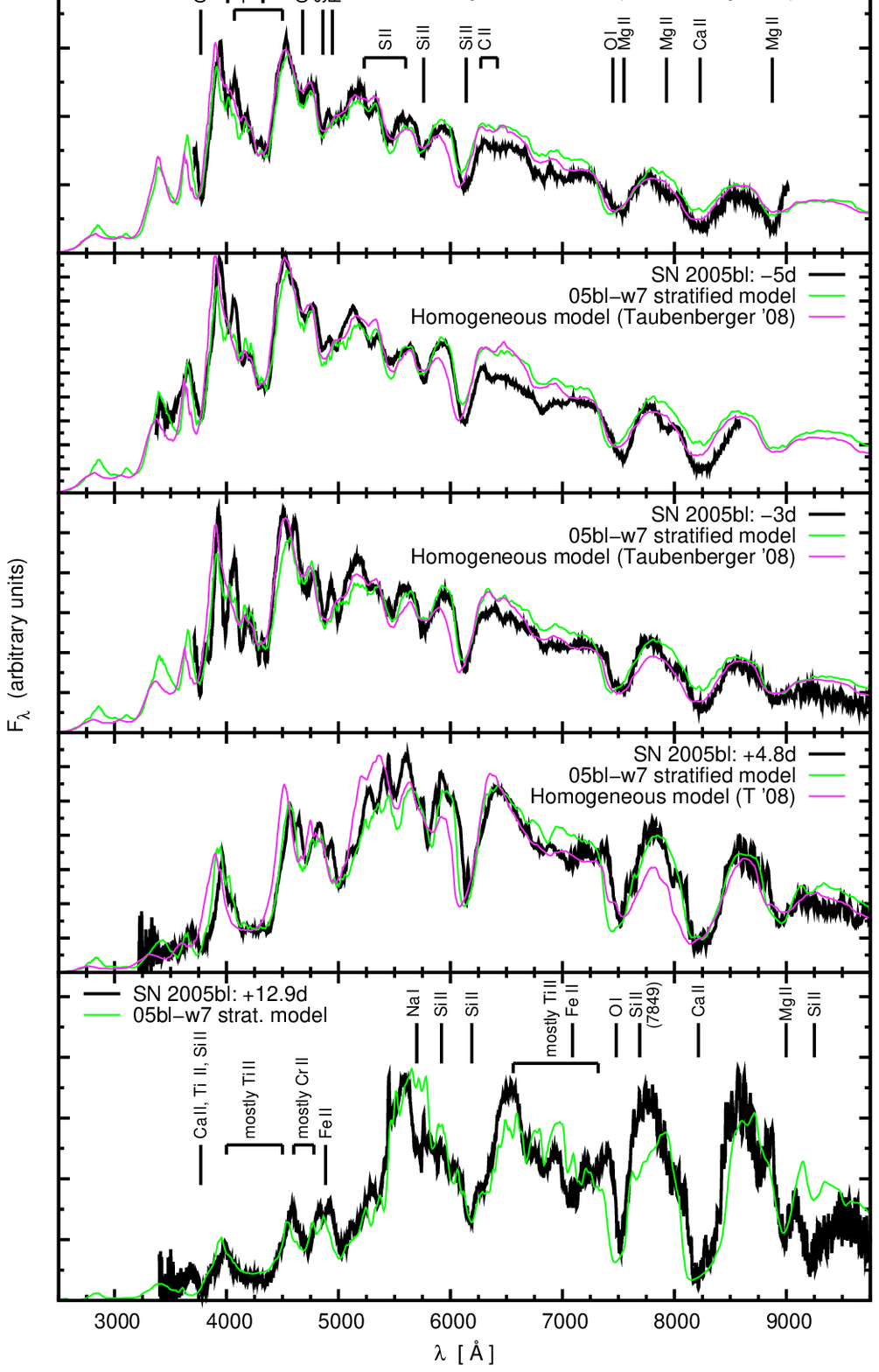}
   \caption{Model sequence based on the W7 density profile (\mbox{05bl-w7} sequence, green lines). Observed spectra (black lines) and one-zone model spectra (magenta lines) from \citet{tau08}, which only extend to \mbox{$+$$4.8$d}, are shown for comparison. Identifications of the most prominent features are given at the beginning and the end of the sequence. Especially in the \mbox{$+$$12.9$d} spectrum, many blends appear, so that the identifications are only approximate.}
   \label{fig:sequence-w7}
\end{figure*}

We analyse five spectra of SN~2005bl, taken at \mbox{$-$$6$d}, \mbox{$-$$5$d}, \mbox{$-$$3$d}, \mbox{$+$$4.8$d} and \mbox{$+$$12.9$d} with respect to $B$ maximum. Observational data and one-zone spectral models have already been presented in \citet{tau08}. As in that paper, we assume a total reddening of $E(B-V)$$=$$0.20$ and a $B$-band rise time of 17d to calculate the time $t$ from the onset of the explosion.

Later spectra were not modelled, as the photosphere has already receded to $v_\textrm{ph}$$<$$3500$\kms\ at \mbox{$+$$12.9$d} [for comparison, \citet{maz08eo} found $v_\textrm{ph}$$=$$4700$\kms\ at \mbox{$+$$14$d} and $v_\textrm{ph}$$=$$2800$\kms\ at \mbox{$+$$21$d} in SN~2004eo]. As the photosphere reaches the \Nifs-rich zone, some energy deposition should realistically take place above the photosphere itself. This is is not taken into account in our code. Thus, to explore the innermost layers one would need to model nebular spectra (which are not available for SN~2005bl) at least as a consistency check.

The outermost ejecta of SNe Ia may partly consist of unburned material \citep[cf.][]{maz07}. As the one-zone models for SN~2005bl \citep{tau08} showed too much absorption by burned material at high velocities, we introduced a zone with strongly reduced abundances of burning products above $v$$\gtrsim$$15000$\kms. This was done limiting the mass fractions of burning products in this zone to $\lesssim$$\frac{1}{10}$ their value at the photosphere at \mbox{$-$$6$d}. The unburned material at $v$$\gtrsim$$15000$\kms, which then constitutes $\gtrsim$$98$$\%$ by mass at these velocities, is assumed to consist of carbon and oxygen in a $\sim$$1$:$1$ ratio. In a preliminary stratified-abundance model, this was found clearly to improve the synthetic spectra, also with respect to the one-zone models (cf. Sec. \ref{sec:taubenbergercomparison}). Consequently, we implemented such a zone in all our stratified-abundance models (except when using the w7e0.35 density profile, which has negligible densities in the outer layers).

Below, we first discuss an abundance tomography experiment based on the original W7 density structure. We compare our synthetic spectra with the observed ones and with the one-zone model spectra of \citet{tau08}. After discussing the abundance profile, we then present models with different total mass and kinetic energy. Parameters (abundances, photospheric velocities, etc.) of all models are compiled in Appendix \ref{app:modelparameters}.

\subsection{Abundance tomography based on W7}

The spectral models discussed here are shown in Fig. \ref{fig:sequence-w7}, where the most important spectral features are marked.

\subsubsection{2005 April 16: -6d, $v_\textrm{ph}$$=$$8400\mathrm{km}\;\mathrm{s^{-1}}$}
\label{sec:firstspectrum-w7}

At this epoch, the supernova shows a spectrum dominated by singly-ionised species. In normal SNe, usually also doubly-ionised species are detected at such early epochs (e.g. \citealt{maz08eo}). The zone between $8400$ and $15000\textrm{\kms}$ is dominated by oxygen. The absence of the \SiII\ $\lambda6355$ emission peak suggests absorption by \CII\ $\lambda6580$. However, the mass fraction of C between $8400$ and $15000\textrm{\kms}$ must be $<$$10\%$; otherwise the \CII\ feature would become too deep. Burned material (oxygen as a burning product excluded) makes up for no more than $\sim$$15\%$ in mass according to the observed line depths.

While numerous lines of intermediate-mass elements (IME) are visible, there are no absorptions that can unambiguously be attributed to Fe. We determined an upper limit to the Fe abundance of $0.01\%$, avoiding the appearance of a spurious \FeII\ feature at $\sim$$4950\textrm{\AA}$. Yet, some burning products heavier than Si and S are seen in the spectra: some per mille of Ti and Cr are necessary to model the absorption trough at $\sim$$4100\textrm{\AA}$ and the feature at $\sim$$4700\textrm{\AA}$, respectively. These elements also contribute significantly to line blocking in the UV \citep{sau08}.

\subsubsection{2005 April 17: -5d, $v_\textrm{ph}$$=$$8100\mathrm{km}\;\mathrm{s^{-1}}$}

The April 17 spectrum is very similar to the previous one. As the material directly above the photosphere is highly ionised, many features in this spectrum  depend strongly on the abundances above $8400$\kms.

At $\sim$$4950\textrm{\AA}$, the stratified model has an absorption trough too deep. This is mostly due to the \SiII\ $\lambda5049$ line, whose strength largely depends on the Si abundance above $v$$=$$8400$\kms. We chose this abundance so as to match the \SiII\ $\lambda5972$ line of this and the previous spectrum, and a simultaneous match of the \SiII\ $\lambda5049$ line was not possible. Apart from this and some flux mismatch in the red, the observations are fitted well.

\subsubsection{2005 April 19: -3d, $v_\textrm{ph}$$=$$7500\mathrm{km}\;\mathrm{s^{-1}}$}

This model again matches the observed spectrum nicely in most regions. The Ti-dominated trough at $\sim$$4100\textrm{\AA}$ is now deeper than in the earlier spectra, and relatively hard to fit. A good model requires Ti abundances of the order of a few percent at the photosphere, and relatively large Ti abundances in the zones above. Thus, we set the Ti mass fraction to $1\%$ between $8100$ and $8400$\kms. At larger velocities, the abundances are sharply constrained to some per mille by the features in the \mbox{$-$$6$d} spectrum. 

There is still no evidence for significant amounts of Fe in the spectrum. Fe mass fractions of a few per cent in the layers between $7500$ and $8400$\kms\ are compatible with the observations, but not strictly required.

\subsubsection{2005 April 26: +4.8d, $v_\textrm{ph}$$=$$6600\mathrm{km}\;\mathrm{s^{-1}}$}
\label{sec:05blw7-p48}

In order to fit this spectrum with its low flux in the UV and blue, the model atmosphere must contain sufficient amounts of Ti, Cr and Fe. The layers at $v$$>$$8400$\kms\ contain relatively small amounts of these elements, as dictated by the pre-maximum spectral features and UV flux. To compensate for this, large amounts are needed close to the photosphere. While the flux-blocking in the UV is quite sensitive to the abundances close to the photosphere, the depth of individual features (such as \SiII\ $\lambda5972$) is still more strongly influenced by the composition at $>$$8400$\kms.

The most notable deviation the model from the observed spectrum occurs in the blue wing of \OI\ $\lambda 7773$, where there is too much absorption. In the outermost zone, O could only be replaced by C, but we already have a $\sim$1$:$1 C-O mixture there. If we wanted to reduce the \OI\ absorption strength by a factor of 2 in these layers, we would have to postulate a $\sim$3$:$1 C-O mixture, which would seem quite ad-hoc. In the layers between $8400$ and $15000\textrm{\kms}$, the amount of oxygen cannot be reduced (cf. Sec. \ref{sec:firstspectrum-w7}). In Sec. \ref{sec:w7e0.7spectra}, we will show that a reduction of the density in the outer layers can cure this problem.

There is some mismatch around $5700$\AA, which seems to be caused by a low pseudo-continuum. This impression is however also due to \NaI$\;\!$D absorption at the peak between the \SII\ trough and the \SiII\ $\lambda5972$ feature. We introduced a small amount of Na above $8100$\kms\ to obtain at least some \NaI$\;\!$D absorption at \mbox{$+$$12.9$d}. The spurious absorption appearing at \mbox{$+$$4.8$d} then indicates inaccuracies in the Na ionisation profile and its evolution with time, a common issue with synthetic spectra \citep{maz97bg}.

\subsubsection{2005 May 04: $+12.9\mathrm{d}$, $v_\textrm{ph}$$=$$3250\mathrm{km}\;\mathrm{s^{-1}}$}

This model carries some conceptual uncertainty, as a possible energy deposition by \Nifs\ above the photosphere is not simulated in our code. Yet, the overall fit is satisfactory. Some incompatibilities with the abundances inferred for the outer layers could not be resolved. It was, for example, impossible to get rid of the absorptions at $\sim$$6500$ and $\sim$$7700\textrm{\AA}$, which are due to \TiII\ $\lambda\lambda6680,6718,6785$ and \SiII\ $\lambda7849$, respectively. These lines were not visible in the earlier spectra.

The photosphere is now deep inside the Si-dominated zone. The extended red wing of the observed feature at $\sim$$9000$\AA, caused mostly by \SiII\ $\lambda 9242$, indicates a large Si mass fraction. On the other hand, the small flux in the blue and UV already demands a larger fraction of Fe-group elements. While the exact amounts of Fe, Co and Ni are somewhat uncertain, their sum can be estimated to be $\sim$$30\%$. The exact number depends on the abundances of other elements blocking UV flux (mostly Ti and Cr) between $3250$ and $6600$\kms, which are somewhat uncertain.

\begin{figure}   
   \centering
   \includegraphics[angle=270,width=8.0cm]{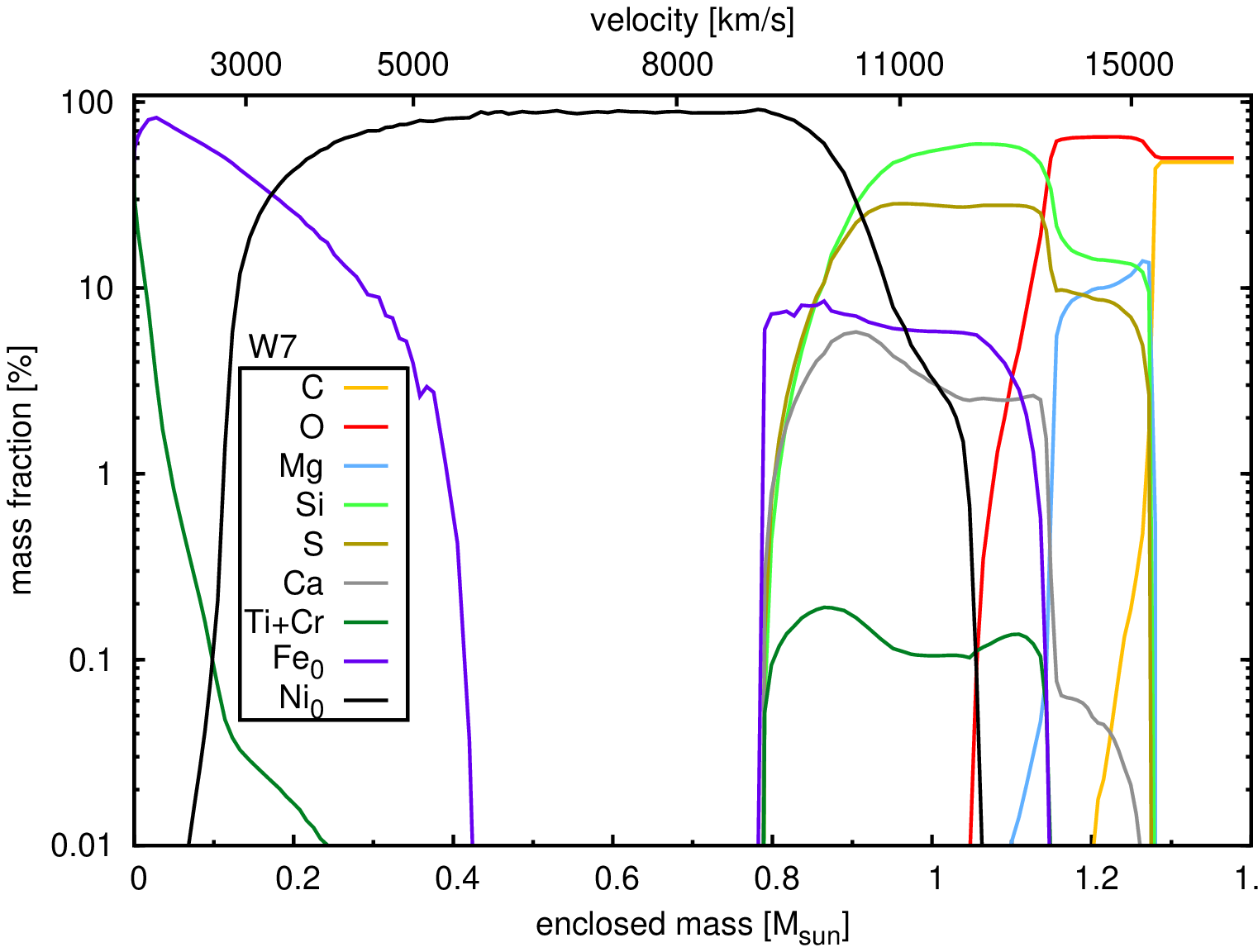}\\[0.35cm]
   \includegraphics[angle=270,width=8.0cm]{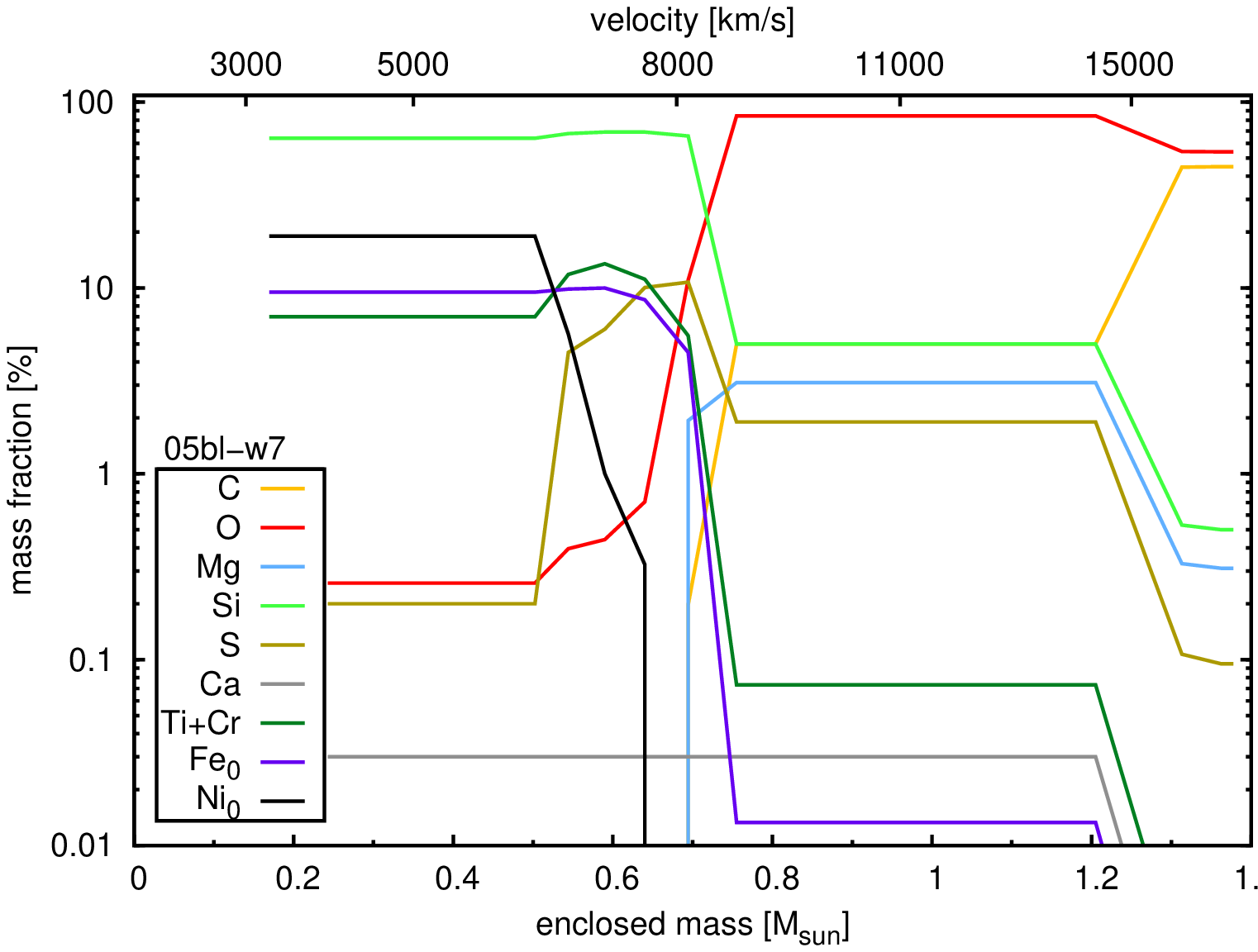}
   \caption{Abundances of W7 nucleosynthesis calculations \citep[][top panel]{iwa99} vs. abundance tomography of SN 2005bl, based on the original W7 density profile (\mbox{05bl-w7} model, bottom panel).}
   \label{fig:abundances-w7}
\end{figure}

\subsubsection{Comparison to one-zone models}
\label{sec:taubenbergercomparison}

Compared to one-zone models [\citet{tau08}, shown in Fig. \ref{fig:sequence-w7} as the magenta line], the stratified model sequence clearly constitutes an improvement in fitting the observations. The main reason for this is the C/O-dominated shell introduced at $v$$>$$15000\textrm{\kms}$, which makes spectral lines of burned material narrower. The changes with respect to the one-zone model are especially apparent in the pre-maximum spectra: the \CaII\ H\&K and \SiII\ $\lambda6355$ lines absorb less at high velocities, so that in the blue wings of the features only small mismatches are left. The \TiII-dominated trough around $4100$\AA\ now has more structure. Some deviations, even a bit more apparent then in the one-zone models, remain in the red wing of \SiII\ $\lambda6355$ in the earliest spectra. This is largely due to re-emission in this wavelength range, caused by elements such as Ti and Cr which block and redistribute UV flux. These elements are, however, necessary to model the spectral features (see Sec. \ref{sec:firstspectrum-w7}).

\subsubsection{Abundance profile}

In Fig. \ref{fig:abundances-w7}, we compare the abundance profile derived in our tomography experiment to the nucleosynthesis in W7 \citep{iwa99}, which approximately represents a normally-luminous SN Ia \citep{nom84}. In our models, unburned material (counting in all of the oxygen) constitutes a much larger fraction of the ejecta, almost the outer $\sim$$0.7\Msun$. Our analysis of the outer layers is still a bit coarse. A better-resolved analysis, yielding more exact results e.g. for the amount of IME between $8400$ and $15000$\kms, would be possible if spectra at earlier epochs were available (see Sec. \ref{sec:earlierspectra}).

Below the outer $\sim$$0.7\Msun$, the ejecta of SN~2005bl are dominated by IME. The transition happens in the zone between $6400$ and $8400$\kms. The exact transition velocity is difficult to infer, as the post-maximum spectra show only a limited sensitivity to the Si abundances below $8400$\kms. \citet{maz97bg} have conducted a fine analysis of the \OI\ $\lambda7773$ line profile in SN~1991bg, and found a lower cut-off velocity of $8600$\kms\ for O. We thus implemented a relatively sharp decrease of the O abundance in favour of Si below the \mbox{$-$$6$d} photosphere.

The layers between $6400$ and $8400$\kms\ already consist of $\sim$$100\%$ \Nifs\ in W7. We, in contrast, find (besides IME) comparatively large abundances of Ti and Cr as products of incomplete burning at these velocities (peak values in the order of some per cent). These elements contribute to the formation of the observed trough around $\sim$$4200$\AA\, which is characteristic of 91bg-like objects past maximum, but also to the line blocking in the UV. To some extent, their effects can also be mimicked by Fe, Co and Ni. With overly large amounts of Fe, however, individual lines in the optical may show up, and the flux distribution in the UV and blue may deviate from what is observed. Large abundances of \Nifs\ or its decay product \Cofs\ outside the centre would be in conflict with the nebular spectra of dim SNe Ia, which show very narrow lines \citep{maz97bg}.

The deepest zones that we reach with our analysis are still dominated by IME. However, there are signs of a transition to the NSE-burning zone: the large amount of line blocking and flux redistribution needed to fit the \mbox{$+$$12.9$d} spectrum clearly points towards Fe-group abundances of several $10\%$.

Compared to the one-zone models of \citet{tau08} epoch by epoch, the abundances of burning products at the respective photospheres are larger. In a model with homogeneous composition, the inferred abundances will always be some average between those at the photosphere and those further outwards, where less burning products are present.

\subsection{Models with modified density profiles}
\label{sec:modelcomparison}

\begin{figure*}   
   \centering
   \includegraphics[width=16cm]{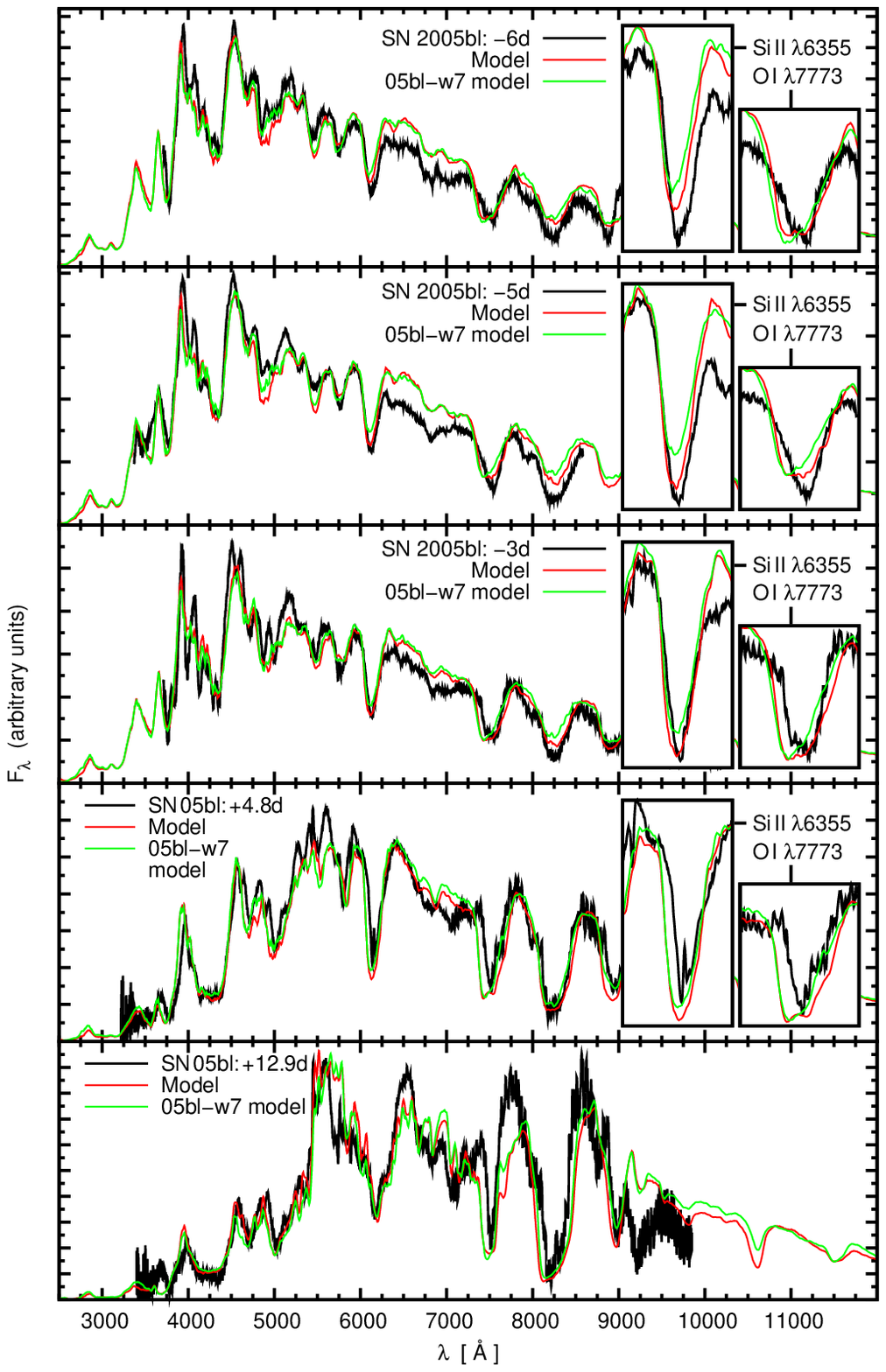}
   \caption{\mbox{05bl-w7e0.7m0.7} model sequence, based on the \mbox{w7e0.7m0.7} density profile (red lines). Observed spectra (black lines) and \mbox{05bl-w7} spectra from Fig. \ref{fig:sequence-w7} (green lines) are given for comparison. For all epochs except the last one, insets show the \SiII\ $\lambda 6355$ and \OI\ $\lambda 7773$ features in detail. Changes in the density profile usually manifest themselves most clearly in these two features (see also Sec. \ref{sec:assessment}).} 
   \label{fig:sequence-w7e0.7m0.7}
\end{figure*}

\begin{figure*}   
   \centering
   \includegraphics[width=16cm]{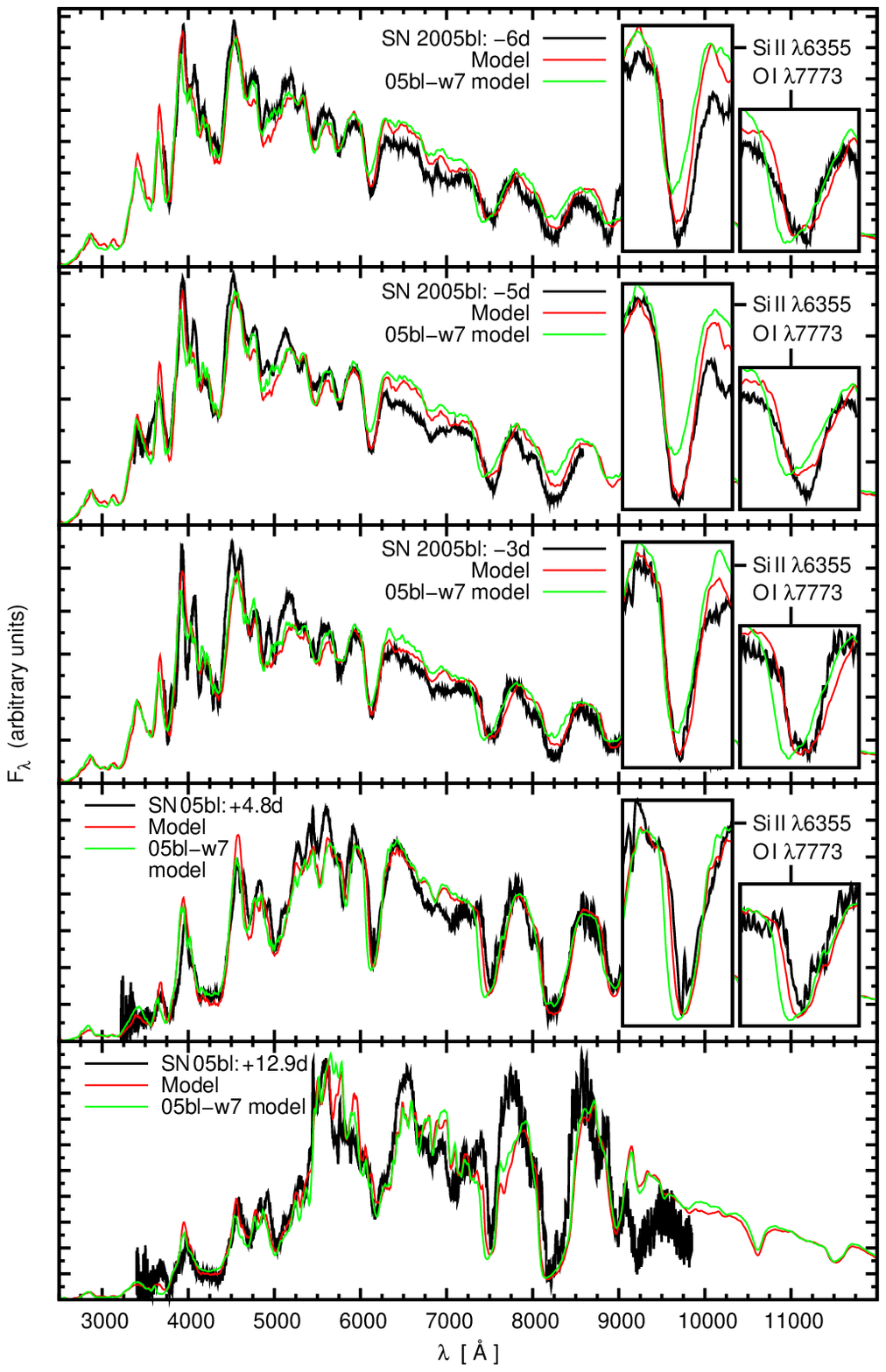}
   \caption{\mbox{05bl-w7e0.7} model sequence (red lines); plot analogous to Fig. \ref{fig:sequence-w7e0.7m0.7}.}
   \label{fig:sequence-w7e0.7}
\end{figure*}

We now show some representative spectral models based on modified density profiles (Sec. \ref{sec:densityprofile}). The reader interested in the abundances is referred to Section \ref{sec:w7e0.7discussion} and Appendix \ref{app:modelparameters}. Here, we focus on the differences in the spectra with respect to the W7-based models. To facilitate the understanding of these differences, we first discuss the properties of the scaled density models.

Our scaled density models span a range of masses and kinetic energies (see Table \ref{tab:scaledmodels}). Scaling the total mass and energy, the amplitude and/or form of the W7 density structure is changed. What exactly happens depends on the $\frac{E'_\textrm{k}}{M'}$ ratio of the final profile with respect to $\frac{E_{W7}}{M_{W7}}$. Here, we distinguish the following three cases, which result in three classes of scaled density profiles:

\begin{itemize}
\item[]$\!\!\!\!\!\!\!\!\!\!\frac{E'_\textrm{k}}{M'}\! = \! \frac{E_{W7}}{M_{W7}}$: In this case, the scaled velocity-density profile is obtained from W7 by reducing the density at each velocity by a uniform factor. The form of the density profile in velocity space is thus left unchanged.
\item[]$\!\!\!\!\!\!\!\!\!\!\frac{E'_\textrm{k}}{M'}\! < \! \frac{E_{W7}}{M_{W7}}$: Here, the energy per unit mass is reduced. This means that mass elements are ''shifted`` towards lower velocities. The density profile becomes steeper in velocity space, and the relative amount of mass at high velocities is smaller.
\item[]$\!\!\!\!\!\!\!\!\!\!\frac{E'_\textrm{k}}{M'}\! > \! \frac{E_{W7}}{M_{W7}}$: Increasing the energy per unit mass ''shifts`` material outwards, opposite to the case before. As the spectra of 91bg-like SNe Ia lack absorption at high velocities in \textit{all} lines, this is generally disfavoured. Thus, we calculated only one model sequence with such a density profile (w7m0.7).
\end{itemize}

The models we discuss below are exemplary for these three scaling types. They are named after the underlying density models (e.g. \mbox{05bl-w7m0.7} is based on w7m0.7).

\subsubsection{Reduced mass and energy, $\frac{E'_\textrm{k}}{M'}\! = \! \frac{E_{W7}}{M_{W7}}$: \mbox{05bl-w7e0.7m0.7}}

In these models, the density is decreased at all radii. This leads to a slight improvement of the spectra (Fig. \ref{fig:sequence-w7e0.7m0.7}), as the photospheres are deeper inside the ejecta, and the absorption velocities tend to be lower. Owing to the lower densities, larger mass fractions of burned material are necessary to fit the line depths. At high velocities, however, oxygen still dominates and the high-velocity absorption in the \OI\ $\lambda 7773$ line only becomes a bit weaker.

\subsubsection{Reduced energy, $\frac{E'_\textrm{k}}{M'}\! < \! \frac{E_{W7}}{M_{W7}}$: \mbox{05bl-w7e0.7}}
\label{sec:w7e0.7spectra}

In the w7e0.7 density profile (Fig. \ref{fig:sequence-w7e0.7}), the densities are significantly increased below $\sim$$6500$\kms\ and decreased above $\sim$$13000$\kms. Thus, the spectral features become narrower compared to the W7-based sequence. Line widths and positions now generally fit the structure of the observed spectra better. 

Owing to the lower densities in the outer part, there is less line blocking by heavy elements. This decreases the flux redistribution, so that the flux level in the red wing of \SiII\ $\lambda6355$ and redwards is matched better, especially at early times.

At the same time, the spurious high-velocity absorption is practically gone in Ca H\&K and \SiII\ $\lambda6355$, but even more importantly in \OI\ $\lambda7773$. The reason for this is again the decreased density in the outer layers. As the abundances in the outer layers, especially of oxygen, are not fundamentally changed with the density modification, the decrease in density translates into weaker absorption at high velocities.

\subsubsection{Increased mass, $\frac{E'_\textrm{k}}{M'}\! < \! \frac{E_{W7}}{M_{W7}}$: \mbox{05bl-w7m1.25}}

\begin{figure*}   
   \centering
   \includegraphics[width=16cm]{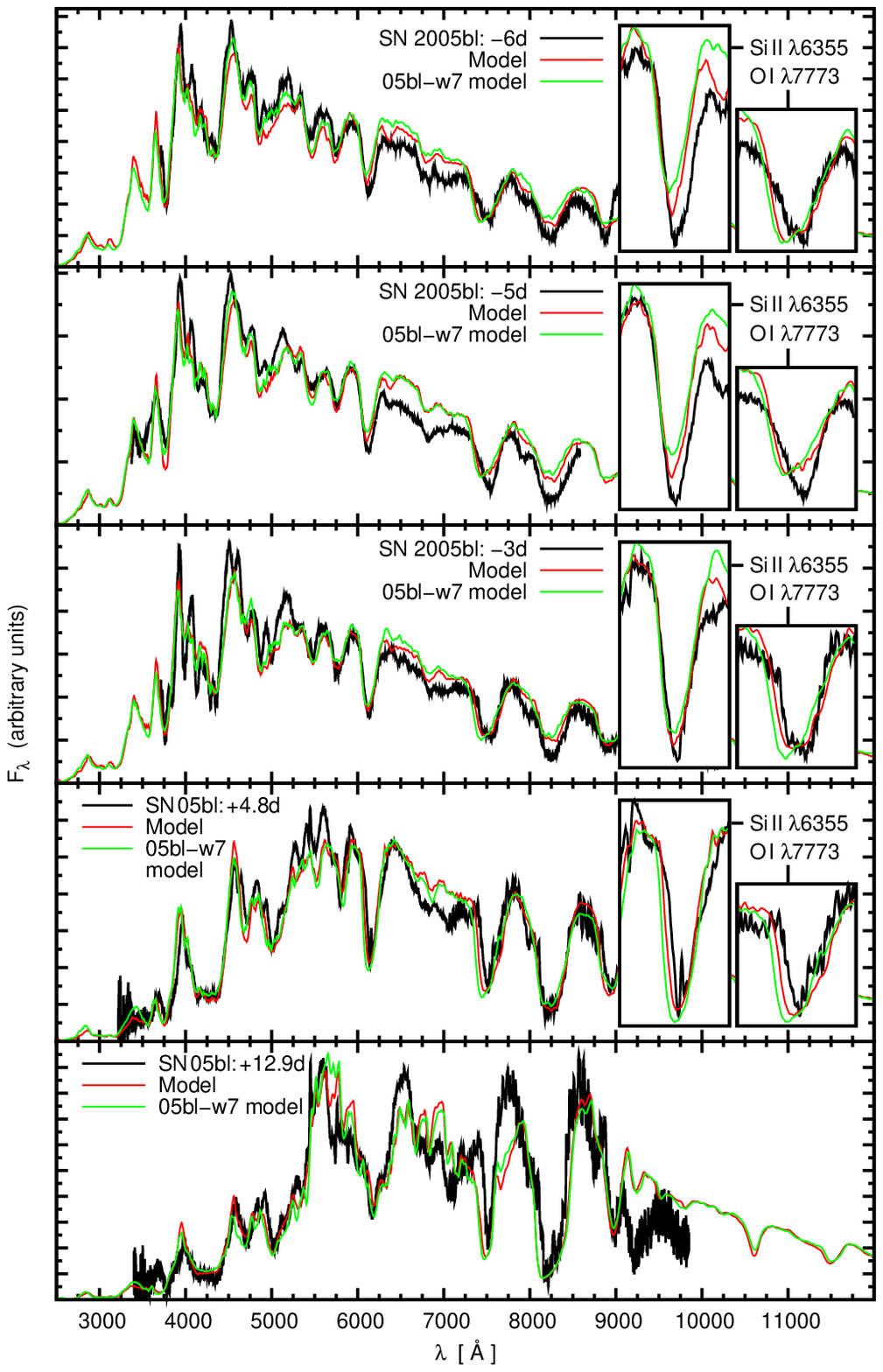}
   \caption{\mbox{05bl-w7m1.25} model sequence (red lines); plot analogous to Fig. \ref{fig:sequence-w7e0.7m0.7}.}
   \label{fig:sequence-w7m1.25}
\end{figure*}

Despite the larger mass, the \mbox{05bl-w7m1.25} spectra show a somewhat improved quality compared to W7 (Fig. \ref{fig:sequence-w7m1.25}, see also line velocity measurements in Sec. \ref{sec:assessment}). This illustrates that a super-Chandrasekhar total mass is not necessarily incompatible with the spectra of SN~2005bl. Remarkably, the improvement over the W7-based sequence is due to \textit{decreased} densities in the outermost layers ($v$$\gtrsim$$15000$\kms) of the warped density profile.

\subsubsection{Reduced mass, $\frac{E'_\textrm{k}}{M'}\! > \! \frac{E_{W7}}{M_{W7}}$: \mbox{05bl-w7m0.7}}

\begin{figure*}   
   \centering
   \includegraphics[width=16cm]{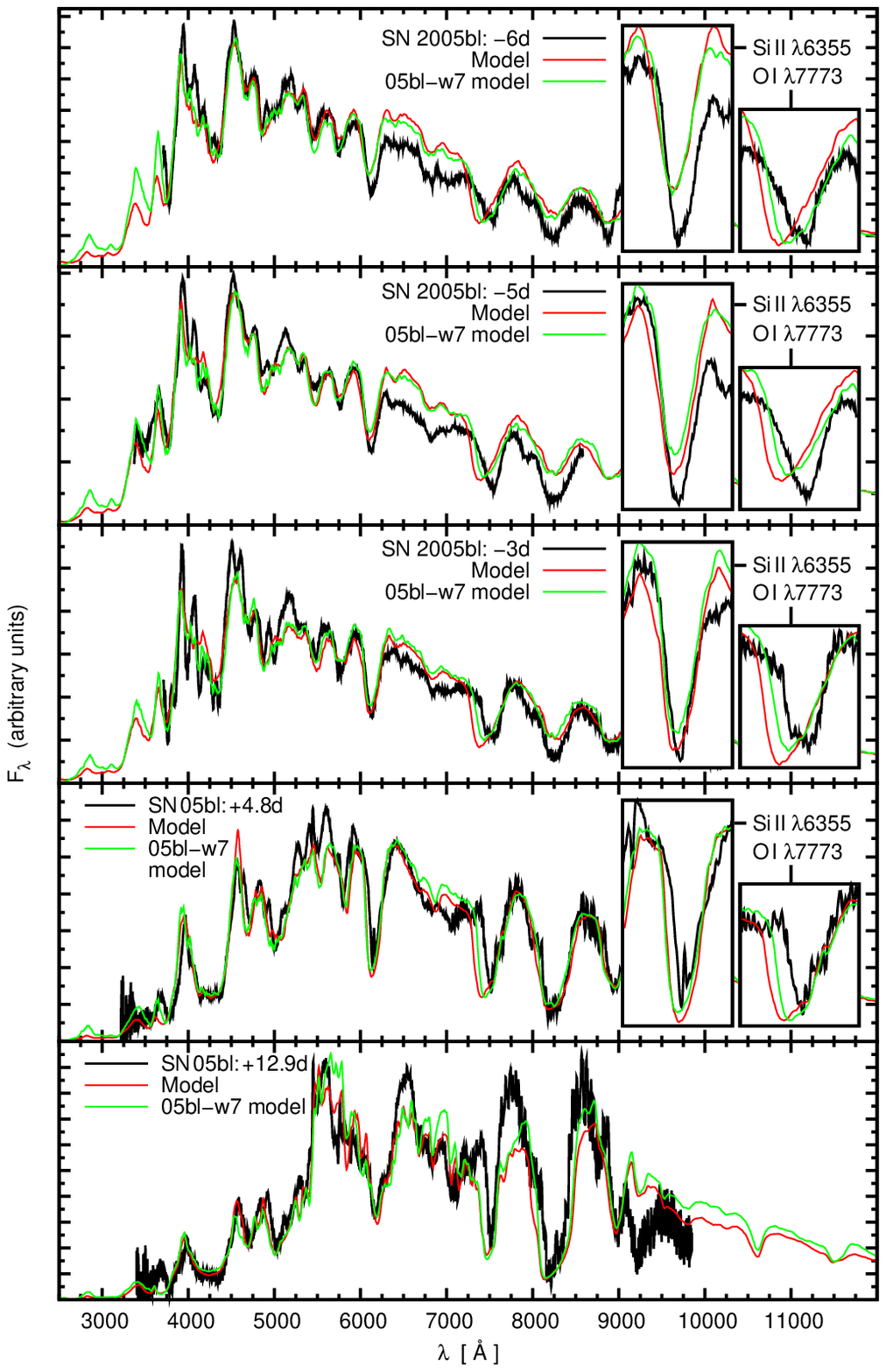}
   \caption{\mbox{05bl-w7m0.7} model sequence (red lines); plot analogous to Fig. \ref{fig:sequence-w7e0.7m0.7}.}
   \label{fig:sequence-w7m0.7}
\end{figure*}

Here, the densities in the outermost layers are increased with respect to W7. This can directly be seen in the spectra (Fig. \ref{fig:sequence-w7m0.7}): all the problems which are reduced in \mbox{05bl-w7e0.7} (compared to the original W7-based model sequence) are now exacerbated.

\section{Discussion}
\label{sec:discussion}

\subsection{Assessment of the models based on different density profiles -- mass and kinetic energy of dim SNe Ia.}
\label{sec:assessment}

Having discussed some representative cases in Section \ref{sec:modelcomparison}, we now systematically compare all models calculated on the basis of different density profiles. Our aim is to judge the quality of each model sequence in a simple and meaningful manner. To achieve this, we introduce three quality criteria:

\begin{enumerate}
 \item \textit{Consistence of spectra.} The main motivation to test modifications of the density were mismatches in the line velocities or widths remaining in the W7-based model sequence, especially in \OI\ $\lambda7773$ and \SiII\ $\lambda6355$. Other lines of the spectrum did not show deviations as apparent, apart from \CaII\ H\&K, which behaves quite similar to \SiII\ $\lambda6355$\footnote{\CaII\ H\&K, \SiII\ $\lambda6355$ and \OI\ $\lambda7773$ are usually the strongest lines in our spectra, which therefore have the highest probability of developing high-velocity absorptions.}. To assess if the lines are better fitted using different density profiles, we measured the velocities of \OI\ $\lambda7773$ and \SiII\ $\lambda6355$ in each synthetic and observed spectrum at \mbox{$-$$6$d}, \mbox{$-$$3$d} and \mbox{$+$$4.8$d}. Then we calculated, for each model sequence and line, the velocity difference between the observed and the synthetic spectra, averaged over the epochs.
 \item \textit{Consistence of kinetic energy.} We calculated a hypothetical kinetic energy ($E_{\textrm{k,hyp}}$) for each of the abundance profiles inferred. This is the nuclear energy release (assuming a pre-explosion composition of equal amounts C and O) minus the binding energy $|E_\textrm{bind}|$ of the WD (gravitational energy\footnote{By ''binding energy`` and ''gravitational energy`` we always mean the absolute values here, i.e. we treat them as positive numbers.} minus thermal and, in case of rotation, rotational energy). To judge the quality of a model sequence, we then compared the kinetic energy assumed in the density scaling ($E'_\textrm{k}$) to $E_{\textrm{k,hyp}}$. The calculation of $E_{\textrm{k,hyp}}$ depends on some assumptions, the first of which is that the mass fraction of IME in the obscured core below the \mbox{$+$$12.9$d} photosphere is $\frac{1}{2}$ of that above the \mbox{$+$$12.9$d} photosphere. Actually, this mass fraction may be between zero and the IME mass fraction above the \mbox{$+$$12.9$d} photosphere. The possible error due to this is given below. The binding energy $|E_\textrm{bind}|$ of the progenitors (except for the $0.69$$\Msun$ ones) was calculated following \citet{yoo05}, who assume a white dwarf rotation profile resulting from binary evolution. We used their ''$BE(M;\rho_c)$`` relation (eq. 33), assuming a central density $\rho_c$ of $2.0$$\cdot$$10^9$$\textrm{g}/\textrm{cm}^3$ (which is typical for WD ignition) for $M'$$\geq$$M_\textrm{Ch}$. For sub-Chandrasekhar WDs, the central densities are lower even in the absence of rotation. We assumed negligible rotation for these cases, and obtained the central density for a given mass inverting formula (22) of \citet{yoo05}\footnote{This formula cannot be applied for our lowest-mass models. Therefore, we inferred the binding energy of a $0.69$$\Msun$ progenitor from a WD model with constant temperature, which uses the Timmes equation of state \citep{tim99}. This equation of state takes into account a variable degree of electron degeneracy.}.
 \item \textit{Expected light-curve width.} For models with good consistence based on the first two criteria, we additionally can check whether the density and abundance structure implies a width of the bolometric light curve ($\tau_\textrm{LC}$) compatible with that of dim SNe Ia. We calculated an expected light curve width for each model sequence, following \citet{maz07}, from the respective kinetic energy $E_{\mathrm{k}}$, ejecta mass $M'$, and total masses of IME and NSE material $M_{\textrm{IME}},M_{\textrm{NSE}}$ as: 
\begin{equation*}
\tau_{\textrm{LC}}=\mathcal{N}\cdot \tilde{\kappa}^{\frac{1}{2}} E_{\mathrm{k}}^{-\frac{1}{4}} M'^{\frac{3}{4}}.
\end{equation*}
Here, $\tilde{\kappa}\!=\!(0.1M_{\textrm{IME}}+M_{\textrm{NSE}})/M'$ is proportional to the opacity estimate of \citet{maz07}, and $\mathcal{N}$ is a normalisation factor chosen so as to agree with their estimates of light-curve widths. In order to calculate $\tilde{\kappa}$, we can assume different burning efficiencies in the core, as above; additionally, we may adopt as $E_{\mathrm{k}}$ either the hypothetical value $E_{\textrm{k,hyp}}$  or the value $E'_{\textrm{k}}$ from the density scaling. We thus calculated again an average $\tau_{\textrm{LC}}$ and an estimate of the error introduced by these degrees of freedom. In order to judge the models, the values $\tau_{\textrm{LC}}$ were compared to $\tau_{\textrm{LC,dim}}$$=$$13.9$d, which is the average expected light curve width for the similarly dim SNe 1991bg and 1999by \citep{maz07}.
\end{enumerate}

We now discuss the quality of the models in terms of the three criteria.

\subsubsection{Line velocities -- consistence of spectra}

The differences in Doppler velocity of the \OI\ $\lambda7773$ and \SiII\ $\lambda6355$ lines between observed and synthetic spectra are shown in Table \ref{tab:velocities}. In this table, the models are ranked according to the absolute value of the ''mean velocity difference``, which is the average over both lines and all epochs.

\begin{table*}
\scriptsize
\caption{Time-averaged line velocity differences in \SiII\ $\lambda 6355$ and \OI\ $\lambda 7773$ from models to observed spectra (denoted by $\langle\Delta{}v(\textrm{\SiII\ }\lambda 6355)\rangle$ and $\langle\Delta{}v(\textrm{\OI}\lambda 7773)\rangle$, respectively). Positive differences mean that the lines are too fast (blue) in the synthetic spectra. The models are sorted according to the mean velocity difference averaged among both lines (ascending in absolute value).}
\label{tab:velocities}
\centering
\begin{tabular}{lccrrr}
Model $\!\!\!\!$ & $\!\!\!\!$ $E'_{\textrm{k}} / E_{\textrm{k},W7} $ $\!\!\!$ & $\!\!\!\!$ $M' / M_{W7}$ $\!\!\!\!$ &  $\langle\Delta{}v(\textrm{\SiII\ }\lambda 6355)\rangle$ $\!\!\!\!\!\!\!\!$ & $\langle\Delta{}v(\textrm{\OI\ }\lambda 7773)\rangle$  $\!\!\!\!\!\!\!\!$ & $\!\!\!\!\!\!$ $\langle\Delta{}v\rangle$  $\!\!$ \\ \hline
05bl-w7e0.7 $\!\!\!\!$ & $\!\!\!\!$ 0.70 $\!\!\!\!$ & $\!\!\!\!$ 1.00 $\!\!\!\!$ & $\!\!\!\!$ 64.1 $\quad\ $ & $\!\!\!\!$ -54.7 $\quad\ $ & $\!\!\!\!$ 4.7 \\ 
05bl-w7m1.45 $\!\!\!\!$ & $\!\!\!\!$ 1.00 $\!\!\!\!$ & $\!\!\!\!$ 1.45 $\!\!\!\!$ & $\!\!\!\!$ 206.0 $\quad\ $ & $\!\!\!\!$ 86.6 $\quad\ $ & $\!\!\!\!$ 146.3 \\ 
05bl-w7e0.5m0.7 $\!\!\!\!$ & $\!\!\!\!$ 0.50 $\!\!\!\!$ & $\!\!\!\!$ 0.70 $\!\!\!\!$ & $\!\!\!\!$ -26.5 $\quad\ $ & $\!\!\!\!$ -283.8 $\quad\ $ & $\!\!\!\!$ -155.1 \\ 
05bl-w7e0.7m1.25 $\!\!\!\!$ & $\!\!\!\!$ 0.70 $\!\!\!\!$ & $\!\!\!\!$ 1.25 $\!\!\!\!$ & $\!\!\!\!$ -117.2 $\quad\ $ & $\!\!\!\!$ -386.0 $\quad\ $ & $\!\!\!\!$ -251.6 \\ 
05bl-w7e0.5m0.5 $\!\!\!\!$ & $\!\!\!\!$ 0.50 $\!\!\!\!$ & $\!\!\!\!$ 0.50 $\!\!\!\!$ & $\!\!\!\!$ 292.2 $\quad\ $ & $\!\!\!\!$ 222.2 $\quad\ $ & $\!\!\!\!$ 257.2 \\ 
05bl-w7m1.25 $\!\!\!\!$ & $\!\!\!\!$ 1.00 $\!\!\!\!$ & $\!\!\!\!$ 1.25 $\!\!\!\!$ & $\!\!\!\!$ 368.1 $\quad\ $ & $\!\!\!\!$ 359.5 $\quad\ $ & $\!\!\!\!$ 363.8 \\ 
05bl-w7e0.7m0.7 $\!\!\!\!$ & $\!\!\!\!$ 0.70 $\!\!\!\!$ & $\!\!\!\!$ 0.70 $\!\!\!\!$ & $\!\!\!\!$ 380.4 $\quad\ $ & $\!\!\!\!$ 454.9 $\quad\ $ & $\!\!\!\!$ 417.7 \\ 
05bl-w7e0.5 $\!\!\!\!$ & $\!\!\!\!$ 0.50 $\!\!\!\!$ & $\!\!\!\!$ 1.00 $\!\!\!\!$ & $\!\!\!\!$ -325.5 $\quad\ $ & $\!\!\!\!$ -618.0 $\quad\ $ & $\!\!\!\!$ -471.7 \\ 
05bl-w7e0.7m1.45 $\!\!\!\!$ & $\!\!\!\!$ 0.70 $\!\!\!\!$ & $\!\!\!\!$ 1.45 $\!\!\!\!$ & $\!\!\!\!$ -223.4 $\quad\ $ & $\!\!\!\!$ -798.2 $\quad\ $ & $\!\!\!\!$ -510.8 \\ 
05bl-w7 $\!\!\!\!$ & $\!\!\!\!$ 1.00 $\!\!\!\!$ & $\!\!\!\!$ 1.00 $\!\!\!\!$ & $\!\!\!\!$ 494.7 $\quad\ $ & $\!\!\!\!$ 1012.4 $\quad\ $ & $\!\!\!\!$ 753.5 \\ 
05bl-w7e1.45m1.45 $\!\!\!\!$ & $\!\!\!\!$ 1.45 $\!\!\!\!$ & $\!\!\!\!$ 1.45 $\!\!\!\!$ & $\!\!\!\!$ 594.9 $\quad\ $ & $\!\!\!\!$ 942.8 $\quad\ $ & $\!\!\!\!$ 768.8 \\ 
05bl-w7e0.5m1.25 $\!\!\!\!$ & $\!\!\!\!$ 0.50 $\!\!\!\!$ & $\!\!\!\!$ 1.25 $\!\!\!\!$ & $\!\!\!\!$ -592.6 $\quad\ $ & $\!\!\!\!$ -976.9 $\quad\ $ & $\!\!\!\!$ -784.7 \\ 
05bl-w7e0.35 $\!\!\!\!$ & $\!\!\!\!$ 0.35 $\!\!\!\!$ & $\!\!\!\!$ 1.00 $\!\!\!\!$ & $\!\!\!\!$ -846.7 $\quad\ $ & $\!\!\!\!$ -1131.7 $\quad\ $ & $\!\!\!\!$ -989.2 \\ 
05bl-w7m0.7 $\!\!\!\!$ & $\!\!\!\!$ 1.00 $\!\!\!\!$ & $\!\!\!\!$ 0.70 $\!\!\!\!$ & $\!\!\!\!$ 457.9 $\quad\ $ & $\!\!\!\!$ 1815.7 $\quad\ $ & $\!\!\!\!$ 1136.8 \\ 
\hline
\end{tabular} 
\begin{flushleft}
\end{flushleft}
\end{table*} 

A decent match of line velocity is obtained especially for the \mbox{05bl-w7e0.7} model, but also, for example, for some super-Chandrasekhar mass models with $E'_\textrm{k}/M'$ lower than W7. This shows that a reduced density in the outer layers is the key to a better fit in the lines. To fit the observed lines well, models near the Chandrasekhar mass need a $E'_\textrm{k}/M'$ smaller by $\sim$$30$$-$$40\%$ with respect to W7. With too large a reduction in energy, line velocities become too low (see e.g. negative velocity differences for the \mbox{05bl-w7e0.5} model). At low masses, generally a smaller reduction in $E'_\textrm{k}/M'$ suffices: \mbox{05bl-w7e0.5m0.5} ($M'$$=$$0.69\Msun$) as an extreme model still gives a satisfactory fit with $E'_\textrm{k}/M'$$=$$\left(E_\textrm{k}/M\right)_{W7}$.

Remarkably, for all mass values probed in this work, a reasonably good model can be obtained (judged by the line velocities). The kinetic energy $E'_\textrm{k}$ of all well-fitting models, however, is lower than $E_{\textrm{k},W7}$.

\subsubsection{Energetic consistence}

In Table \ref{tab:energetics} we show our hypothetical kinetic energy values, as well as the quantities from which they were calculated. We then judge the models by the ratio of $E_{\textrm{k,hyp}}$ to the kinetic energy assumed in the density scaling ($E'_\textrm{k}$). Ideally, this ratio should be equal to one; the larger the deviation, the lower the rank of a model.

\begin{table*}
\scriptsize
\caption{Energetic balance of the models (see text for a description of the quantities). The models are ordered according to the deviation of their $\frac{E_{\textrm{k,hyp}}}{E'_\textrm{k}}$ ratio from 1 (cf. last column).}
\label{tab:energetics}
\centering
\begin{tabular}{lccccccr}
Model $\!\!\!\!$ & $\!\!\!\!$ $E'_{\textrm{k}}/E_{\textrm{k},W7}$ $\!\!\!\!$ & $\!\!\!\!$ $M'/M_{W7}$ $\!\!\!\!$ & $\!\!\!\!$ $E'_\textrm{k}$ $\!\!\!\!$ & $\!\!\!\!$ $E_{\textrm{nucl}}$ $\!\!\!\!$ & $\!\!\!\!$ $E_{\textrm{bind}}$ $\!\!\!\!$ & $\!\!\!\!$ $E_\textrm{k,hyp}$$^\textrm{a}$ $\!\!\!\!$ & $\!\!\!\!$ $\!\!\!\!\frac{E_\textrm{k,hyp}}{E'_\textrm{k}}\!-\!1\!\!\!\!$ \\
\vspace{-0.85pt} & \vspace{-0.85pt} & \vspace{-0.85pt} & \vspace{-0.85pt} & \vspace{-0.85pt} & \vspace{-0.85pt} & \vspace{-0.85pt} & \vspace{-0.85pt} \\
{} $\!\!\!\!$ & $\!\!\!\!$ {} $\!\!\!\!$ & $\!\!\!\!$ {} $\!\!\!\!$ & $\!\!\!\!$ [$10^{51}$erg] $\!\!\!\!$ & $\!\!\!\!$ [$10^{51}$erg] $\!\!\!\!$ & $\!\!\!\!$ [$10^{51}$erg] $\!\!\!\!$ & $\!\!\!\!$ [$10^{51}$erg] $\!\!\!\!$ & $\!\!\!\!$ {} \\ \hline
05bl-w7e0.7m1.45 $\!\!\!\!$ & $\!\!\!\!$ 0.70 $\!\!\!\!$ & $\!\!\!\!$ 1.45 $\!\!\!\!$ & $\!\!\!\!$ 0.93 $\!\!\!\!$ & $\!\!\!\!$ 2.10 $\!\!\!\!$ & $\!\!\!\!$ 1.15 $\!\!\!\!$ & $\!\!\!\!$ 0.96$\,\pm\,$0.10 $\!\!\!\!$ & $\!\!\!\!$ 0.03 $\ $ \\
05bl-w7e0.7m1.25 $\!\!\!\!$ & $\!\!\!\!$ 0.70 $\!\!\!\!$ & $\!\!\!\!$ 1.25 $\!\!\!\!$ & $\!\!\!\!$ 0.93 $\!\!\!\!$ & $\!\!\!\!$ 1.73 $\!\!\!\!$ & $\!\!\!\!$ 0.85 $\!\!\!\!$ & $\!\!\!\!$ 0.88$\,\pm\,$0.06 $\!\!\!\!$ & $\!\!\!\!$ -0.05 $\ $ \\
05bl-w7e0.5m0.7 $\!\!\!\!$ & $\!\!\!\!$ 0.50 $\!\!\!\!$ & $\!\!\!\!$ 0.70 $\!\!\!\!$ & $\!\!\!\!$ 0.66 $\!\!\!\!$ & $\!\!\!\!$ 0.87 $\!\!\!\!$ & $\!\!\!\!$ 0.13 $\!\!\!\!$ & $\!\!\!\!$ 0.74$\,\pm\,$0.02 $\!\!\!\!$ & $\!\!\!\!$ 0.11 $\ $ \\
05bl-w7e0.7 $\!\!\!\!$ & $\!\!\!\!$ 0.70 $\!\!\!\!$ & $\!\!\!\!$ 1.00 $\!\!\!\!$ & $\!\!\!\!$ 0.93 $\!\!\!\!$ & $\!\!\!\!$ 1.26 $\!\!\!\!$ & $\!\!\!\!$ 0.49 $\!\!\!\!$ & $\!\!\!\!$ 0.77$\,\pm\,$0.03 $\!\!\!\!$ & $\!\!\!\!$ -0.17 $\ $ \\
05bl-w7e0.5m0.5 $\!\!\!\!$ & $\!\!\!\!$ 0.50 $\!\!\!\!$ & $\!\!\!\!$ 0.50 $\!\!\!\!$ & $\!\!\!\!$ 0.66 $\!\!\!\!$ & $\!\!\!\!$ 0.60 $\!\!\!\!$ & $\!\!\!\!$ 0.06 $\!\!\!\!$ & $\!\!\!\!$ 0.53$\,\pm\,$0.00 $\!\!\!\!$ & $\!\!\!\!$ -0.20 $\ $ \\
05bl-w7e0.5m1.25 $\!\!\!\!$ & $\!\!\!\!$ 0.50 $\!\!\!\!$ & $\!\!\!\!$ 1.25 $\!\!\!\!$ & $\!\!\!\!$ 0.66 $\!\!\!\!$ & $\!\!\!\!$ 1.89 $\!\!\!\!$ & $\!\!\!\!$ 0.85 $\!\!\!\!$ & $\!\!\!\!$ 1.04$\,\pm\,$0.07 $\!\!\!\!$ & $\!\!\!\!$ 0.23 $\ $ \\
05bl-w7e0.7m0.7 $\!\!\!\!$ & $\!\!\!\!$ 0.70 $\!\!\!\!$ & $\!\!\!\!$ 0.70 $\!\!\!\!$ & $\!\!\!\!$ 0.93 $\!\!\!\!$ & $\!\!\!\!$ 0.80 $\!\!\!\!$ & $\!\!\!\!$ 0.13 $\!\!\!\!$ & $\!\!\!\!$ 0.68$\,\pm\,$0.01 $\!\!\!\!$ & $\!\!\!\!$ -0.27 $\ $ \\
05bl-w7m1.45 $\!\!\!\!$ & $\!\!\!\!$ 1.00 $\!\!\!\!$ & $\!\!\!\!$ 1.45 $\!\!\!\!$ & $\!\!\!\!$ 1.33 $\!\!\!\!$ & $\!\!\!\!$ 1.90 $\!\!\!\!$ & $\!\!\!\!$ 1.15 $\!\!\!\!$ & $\!\!\!\!$ 0.75$\,\pm\,$0.06 $\!\!\!\!$ & $\!\!\!\!$ -0.43 $\ $ \\
05bl-w7e0.5 $\!\!\!\!$ & $\!\!\!\!$ 0.50 $\!\!\!\!$ & $\!\!\!\!$ 1.00 $\!\!\!\!$ & $\!\!\!\!$ 0.66 $\!\!\!\!$ & $\!\!\!\!$ 1.44 $\!\!\!\!$ & $\!\!\!\!$ 0.49 $\!\!\!\!$ & $\!\!\!\!$ 0.95$\,\pm\,$0.05 $\!\!\!\!$ & $\!\!\!\!$ 0.43 $\ $ \\
05bl-w7m1.25 $\!\!\!\!$ & $\!\!\!\!$ 1.00 $\!\!\!\!$ & $\!\!\!\!$ 1.25 $\!\!\!\!$ & $\!\!\!\!$ 1.33 $\!\!\!\!$ & $\!\!\!\!$ 1.56 $\!\!\!\!$ & $\!\!\!\!$ 0.85 $\!\!\!\!$ & $\!\!\!\!$ 0.71$\,\pm\,$0.04 $\!\!\!\!$ & $\!\!\!\!$ -0.47 $\ $ \\
05bl-w7 $\!\!\!\!$ & $\!\!\!\!$ 1.00 $\!\!\!\!$ & $\!\!\!\!$ 1.00 $\!\!\!\!$ & $\!\!\!\!$ 1.33 $\!\!\!\!$ & $\!\!\!\!$ 1.14 $\!\!\!\!$ & $\!\!\!\!$ 0.49 $\!\!\!\!$ & $\!\!\!\!$ 0.65$\,\pm\,$0.02 $\!\!\!\!$ & $\!\!\!\!$ -0.51 $\ $ \\
05bl-w7m0.7 $\!\!\!\!$ & $\!\!\!\!$ 1.00 $\!\!\!\!$ & $\!\!\!\!$ 0.70 $\!\!\!\!$ & $\!\!\!\!$ 1.33 $\!\!\!\!$ & $\!\!\!\!$ 0.65 $\!\!\!\!$ & $\!\!\!\!$ 0.13 $\!\!\!\!$ & $\!\!\!\!$ 0.53$\,\pm\,$0.01 $\!\!\!\!$ & $\!\!\!\!$ -0.60 $\ $ \\
05bl-w7e1.45m1.45 $\!\!\!\!$ & $\!\!\!\!$ 1.45 $\!\!\!\!$ & $\!\!\!\!$ 1.45 $\!\!\!\!$ & $\!\!\!\!$ 1.93 $\!\!\!\!$ & $\!\!\!\!$ 1.69 $\!\!\!\!$ & $\!\!\!\!$ 1.15 $\!\!\!\!$ & $\!\!\!\!$ 0.54$\,\pm\,$0.04 $\!\!\!\!$ & $\!\!\!\!$ -0.72 $\ $ \\
05bl-w7e0.35 $\!\!\!\!$ & $\!\!\!\!$ 0.35 $\!\!\!\!$ & $\!\!\!\!$ 1.00 $\!\!\!\!$ & $\!\!\!\!$ 0.46 $\!\!\!\!$ & $\!\!\!\!$ 1.55 $\!\!\!\!$ & $\!\!\!\!$ 0.49 $\!\!\!\!$ & $\!\!\!\!$ 1.07$\,\pm\,$0.06 $\!\!\!\!$ & $\!\!\!\!$ 1.29 $\ $ \\
\hline
\end{tabular}
\begin{flushleft}
$^\textrm{a}$ The error estimate only reflects the error due to the unknown composition below the photosphere at \mbox{$+$$12.9$d}.
\end{flushleft}
\end{table*} 

For density profiles with the same mass, but different kinetic energy ${E'_\textrm{k}}$, the hypothetical kinetic energy ${E_{\textrm{k,hyp}}}$ usually varies systematically. In density models with smaller ${E'_\textrm{k}}$, densities are reduced in the high-velocity layers (see Sec. \ref{sec:w7e0.7spectra}), which contain mostly unburned material. At the same time, densities are increased in lower layers, where the material is mostly burned. The velocity at which the transition (between unburned and burned material) happens does not vary much from model to model as it is constrained by spectral features. Therefore, the change in the density profile results in a larger ratio of burned to unburned material and a larger ${E_{\textrm{k,hyp}}}$. Similarly, when ${E'_\textrm{k}}$ is increased, ${E_{\textrm{k,hyp}}}$ decreases. Equality, i.e. consistence, between $E_{\textrm{k,hyp}}$ and $E'_\textrm{k}$ is usually reached at a reduced value of $E'_\textrm{k}/M'$ with respect to W7. The required reduction varies with the mass of the models (see Sec. \ref{sec:modeldiagram}). 

In Table \ref{tab:energetics}, two supermassive models (\mbox{05bl-w7e0.7m1.25}, \mbox{05bl-w7e0.7m1.45}) rank top. However, it should be noted that the energetic quality criterion again does not single out a certain mass, but sets a point of energetic consistence for each mass. All models with larger ${E'_\textrm{k}}$ will then feature too little nucleosynthesis to explain the assumed kinetic energy. The opposite holds for models with lower ${E'_\textrm{k}}$.

\subsubsection{Expected light-curve width}

We calculated estimates of the width of the bolometric light curve for the models ranking best in spectroscopic and energetic consistence at each mass $M'$. The resulting values, and those of the quantities needed for the calculation, are given in Table \ref{tab:lcwidth}.

\begin{table*}
\scriptsize
\caption{Light-curve width estimates for the spectroscopically and energetically most consistent models at each mass $M'$. The models are ordered according to the deviation of the ratio $\frac{\tau_{\textrm{LC}}}{\tau_{\textrm{LC,dim}}}$ from 1; we assume $\tau_{\textrm{LC,dim}}$$=$$13.9$d (see text).}
\label{tab:lcwidth}
\centering
\begin{tabular}{lcccccccr}
Model $\!\!\!\!$ & $\!\!\!\!$ $E'_{\textrm{k}}/E_{\textrm{k},W7}$ $\!\!\!\!$ & $\!\!\!\!$ $M'/M_{W7}$ $\!\!\!\!$ & $\!\!\!\!$ $E'_\textrm{k}$ $\!\!\!\!$ & $\!\!\!\!$ $M'$ $\!\!\!\!$ & $\!\!\!\!$ $E_{\textrm{k,hyp}}$ $\!\!\!\!$ & $\!\!\!\!$ $\tilde{\kappa}$$^\textrm{a}$ $\!\!\!\!$ & $\!\!\!\!$ $\tau_{\textrm{LC}}$$^\textrm{a}$ $\!\!\!\!$ & $\!\!\!\!$ $\!\!\frac{\tau_{\textrm{LC}}}{\tau_{\textrm{LC,dim}}}\!-\!1\!\!\!\!$ \\
\vspace{-0.60pt} & \vspace{-0.60pt} & \vspace{-0.60pt} & \vspace{-0.60pt} & \vspace{-0.60pt} & \vspace{-0.60pt} & \vspace{-0.60pt} & \vspace{-0.60pt} & \vspace{-0.60pt} \\
{} $\!\!\!\!$ & $\!\!\!\!$ {} $\!\!\!\!$ & $\!\!\!\!$ {} $\!\!\!\!$ & $\!\!\!\!$ [$10^{51}$erg] $\!\!\!\!$ & $\!\!\!\!$ [$\Msun$] $\!\!\!\!$ & $\!\!\!\!$ [$10^{51}$erg] $\!\!\!\!$ & $\!\!\!\!$ {} $\!\!\!\!$ & $\!\!\!\!$ [d] $\!\!\!\!$ & $\!\!\!\!$ {} $\!\!\!\!$ \\ \hline
05bl-w7e0.7 $\!\!\!\!$ & $\!\!\!\!$ 0.70 $\!\!\!\!$ & $\!\!\!\!$ 1.00 $\!\!\!\!$ & $\!\!\!\!$ 0.93 $\!\!\!\!$ & $\!\!\!\!$ 1.31 $\!\!\!\!$ & $\!\!\!\!$ 0.77 $\!\!\!\!$ & $\!\!\!\!$ 0.31$\,\pm\,$0.10 $\!\!\!\!$ & $\!\!\!\!$ 13.8$\,\pm\,$2.5 $\!\!\!\!$ & $\!\!\!\!$ 0.00 $\quad\ $ \\
05bl-w7e0.5m0.7 $\!\!\!\!$ & $\!\!\!\!$ 0.50 $\!\!\!\!$ & $\!\!\!\!$ 0.70 $\!\!\!\!$ & $\!\!\!\!$ 0.66 $\!\!\!\!$ & $\!\!\!\!$ 0.97 $\!\!\!\!$ & $\!\!\!\!$ 0.74 $\!\!\!\!$ & $\!\!\!\!$ 0.34$\,\pm\,$0.10 $\!\!\!\!$ & $\!\!\!\!$ 11.7$\,\pm\,$1.9 $\!\!\!\!$ & $\!\!\!\!$ -0.16  $\quad\ $ \\
05bl-w7e0.7m1.25 $\!\!\!\!$ & $\!\!\!\!$ 0.70 $\!\!\!\!$ & $\!\!\!\!$ 1.25 $\!\!\!\!$ & $\!\!\!\!$ 0.93 $\!\!\!\!$ & $\!\!\!\!$ 1.73 $\!\!\!\!$ & $\!\!\!\!$ 0.88 $\!\!\!\!$ & $\!\!\!\!$ 0.32$\,\pm\,$0.12 $\!\!\!\!$ & $\!\!\!\!$ 16.1$\,\pm\,$3.4 $\!\!\!\!$ & $\!\!\!\!$ 0.16  $\quad\ $ \\
05bl-w7e0.5m0.5 $\!\!\!\!$ & $\!\!\!\!$ 0.50 $\!\!\!\!$ & $\!\!\!\!$ 0.50 $\!\!\!\!$ & $\!\!\!\!$ 0.66 $\!\!\!\!$ & $\!\!\!\!$ 0.69 $\!\!\!\!$ & $\!\!\!\!$ 0.53 $\!\!\!\!$ & $\!\!\!\!$ 0.47$\,\pm\,$0.07 $\!\!\!\!$ & $\!\!\!\!$ 11.2$\,\pm\,$1.2 $\!\!\!\!$ & $\!\!\!\!$ -0.19  $\quad\ $ \\
05bl-w7m1.45 $\!\!\!\!$ & $\!\!\!\!$ 1.00 $\!\!\!\!$ & $\!\!\!\!$ 1.45 $\!\!\!\!$ & $\!\!\!\!$ 1.33 $\!\!\!\!$ & $\!\!\!\!$ 2.00 $\!\!\!\!$ & $\!\!\!\!$ 0.75 $\!\!\!\!$ & $\!\!\!\!$ 0.28$\,\pm\,$0.11 $\!\!\!\!$ & $\!\!\!\!$ 16.9$\,\pm\,$4.6 $\!\!\!\!$ & $\!\!\!\!$ 0.21 $\quad\ $ \\
05bl-w7e0.7m1.45 $\!\!\!\!$ & $\!\!\!\!$ 0.70 $\!\!\!\!$ & $\!\!\!\!$ 1.45 $\!\!\!\!$ & $\!\!\!\!$ 0.93 $\!\!\!\!$ & $\!\!\!\!$ 2.00 $\!\!\!\!$ & $\!\!\!\!$ 0.96 $\!\!\!\!$ & $\!\!\!\!$ 0.35$\,\pm\,$0.16 $\!\!\!\!$ & $\!\!\!\!$ 18.5$\,\pm\,$4.5 $\!\!\!\!$ & $\!\!\!\!$ 0.34 $\quad\ $ \\
\hline
\end{tabular}
\begin{flushleft}
$^\textrm{a}$ The error estimate reflects the errors due to the unknown composition below the photosphere at \mbox{$+$$12.9$d}, and due to the uncertainties in $E_\textrm{k}$.
\end{flushleft}
\end{table*} 

The deviation of $\tau_{\textrm{LC}}$ from $\tau_{\textrm{LC,dim}}$$=$$13.9$d strongly depends on the mass $M'$. Models with larger mass clearly tend to have a larger light-curve width, although they often have lower values of $\tilde{\kappa}$, as relatively small abundances of burning products are needed to match the observed line strengths with the synthetic spectra.

Although our expected light-curve widths are quite rough estimates, one can clearly state that the criterion disfavours masses largely deviating from the Chandrasekhar mass. The least massive model, with a mass of $M'$$=$$0.5$$M_\textrm{Ch}$, presumably will not produce a broad enough light curve. Likewise, the models at $M'$$=$$1.45$$M_\textrm{Ch}$ will probably exhibit too broad a light curve, although these models are not strictly incompatible with $\tau_{\textrm{LC,dim}}$, as a large inaccuracy in $\tau_{\textrm{LC,dim}}$ results from the large mass in the obscured core.

\subsection{Location of consistent models in the $E'_\textrm{k}$$-$$M'$ plane}
\label{sec:modeldiagram}

\begin{figure*}   
   \centering
   \includegraphics[width=12.5cm]{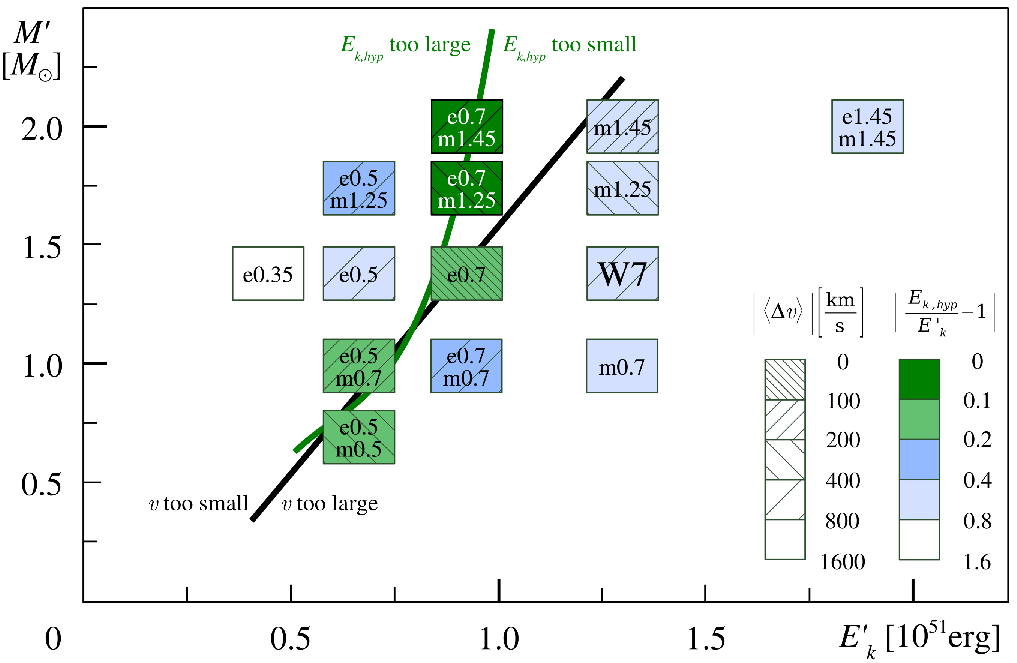}
   \caption{Overview of all models evaluated in this work. Colour and hatches mark the energetic and spectroscopic consistence of the models, as indicated by the quantities $\frac{E_\textrm{k,hyp}}{E'_\textrm{k}}\!-\!1$ and $\langle\Delta{}v\rangle$, respectively (darker colours / denser hatches meaning better consistence; numerical values see Tables \ref{tab:velocities} and \ref{tab:energetics}). The green line divides the regions where the models have too large and too small a hypothetical kinetic energy yield, compared to the kinetic energy assumed in the density scaling. To the left of the black line models show too low line velocities; to the right, the opposite holds.}
   \label{fig:modeldiagram}
\end{figure*}

Fig. \ref{fig:modeldiagram} gives an overview of all models in an $E'_\textrm{k}$$-$$M'$ plane. According to their quality in spectroscopic terms, the models are marked with different colours; the energetic consistence is indicated by hatches.

In the figure, we also indicate where spectroscopically and energetically consistent models can generally be expected in the plane: a black line is drawn approximately where the transition between too large and too small line velocities occurs. This line is straight and runs from massive models with low $E'_\textrm{k}/M'$ to submassive models with $E'_\textrm{k}/M'$$\approx$$\left(E_\textrm{k}/M\right)_{W7}$. A green line approximately divides the regions of too large and too small nucleosynthetic energy yields. It lies in the same region as the line of spectroscopic consistence, but is curved because the WD binding energy shows a disproportionately strong increase with WD mass. For models up to $M'$$\approx$$0.7 M_\textrm{Ch}$, the binding energy is negligible compared to the nuclear energy release, whereas at higher masses it is considerable, forcing the line of consistence towards smaller $E'_\textrm{k}$ and larger nucleosynthesis yields.

The two lines of consistence are especially close to one another for masses $M'$$\lesssim$$M_\textrm{Ch}$. Models at $M'$$=$$1.45$$M_\textrm{Ch}$ are either spectroscopically or energetically inconsistent, at least under the assumptions we made in this work. Additionally, the light-curve criterion indicates that the width of the light curve is too large for these most massive models. Our least massive models with $M'$$=$$0.5$$M_\textrm{Ch}$ are also disfavoured in this respect, as they would probably show too rapid a light-curve evolution.

Criteria like those used here could give stricter limits still, if the chemical composition in the inner layers was known. This requires studies of nebular spectra of dim SNe Ia.

\subsection{\mbox{05bl-w7e0.7} as a reference model}
\label{sec:w7e0.7discussion}

\begin{figure}   
   \centering
   \includegraphics[angle=270,width=8.0cm]{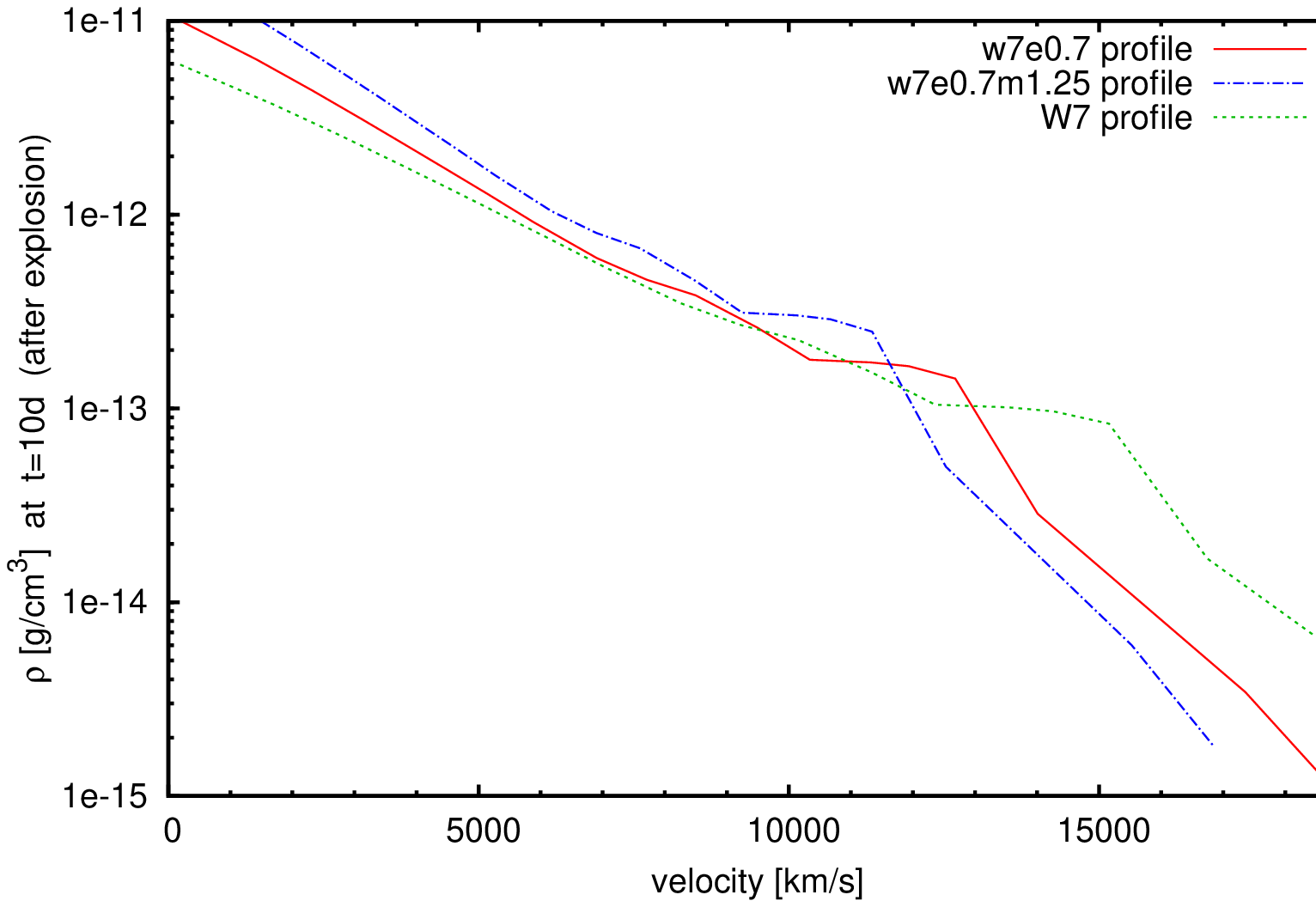}
   \caption{w7e0.7 compared to the standard W7 and the w7m1.25e0.7 density profiles.}
   \label{fig:density-reference}
\end{figure}   

\begin{figure*}   
   \centering
   \includegraphics[angle=270,width=8.0cm]{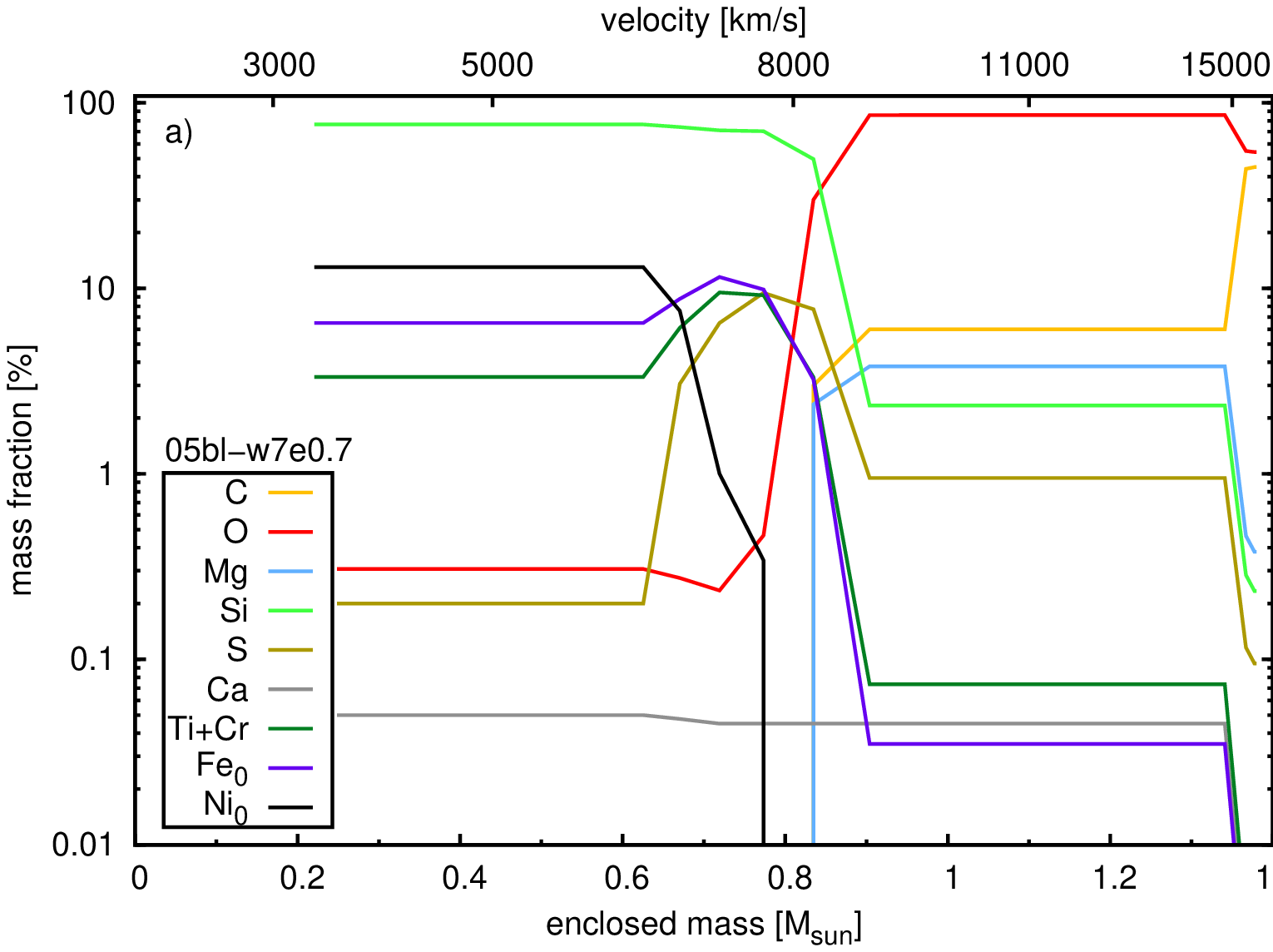} \hspace{1cm} \includegraphics[angle=270,width=8.0cm]{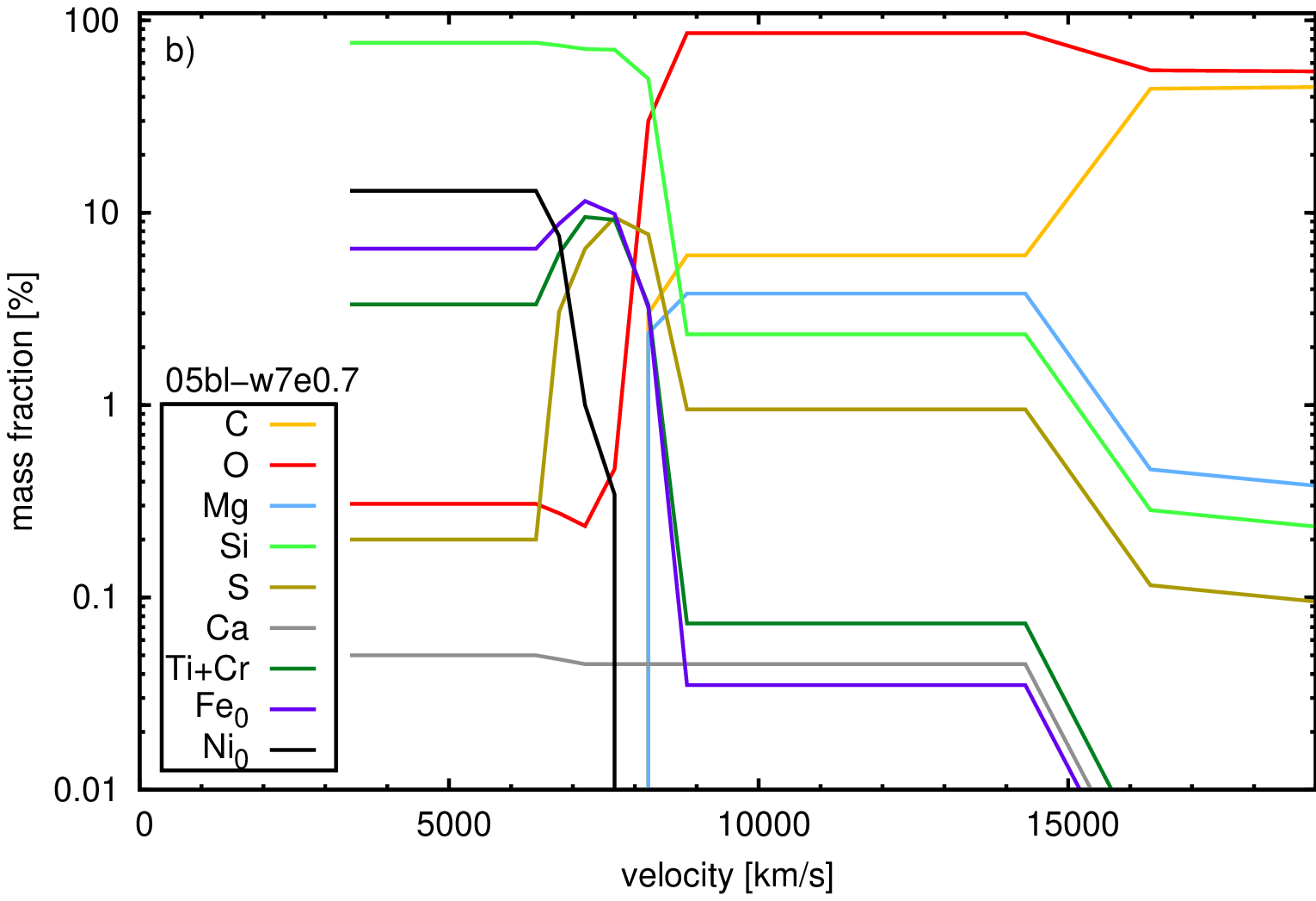} \\[0.5cm]
   \includegraphics[angle=270,width=8.0cm]{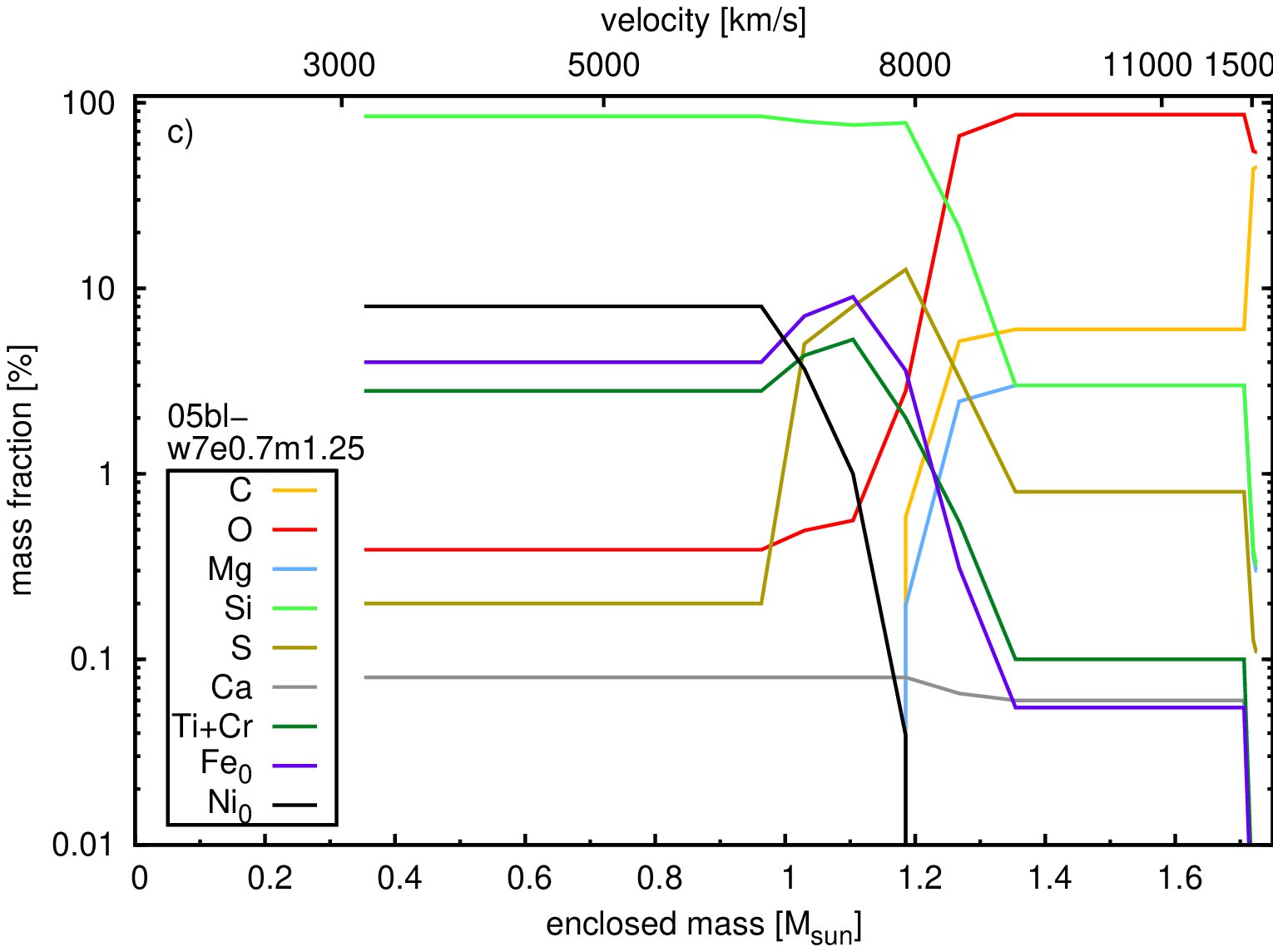} \hspace{1cm} \includegraphics[angle=270,width=8.0cm]{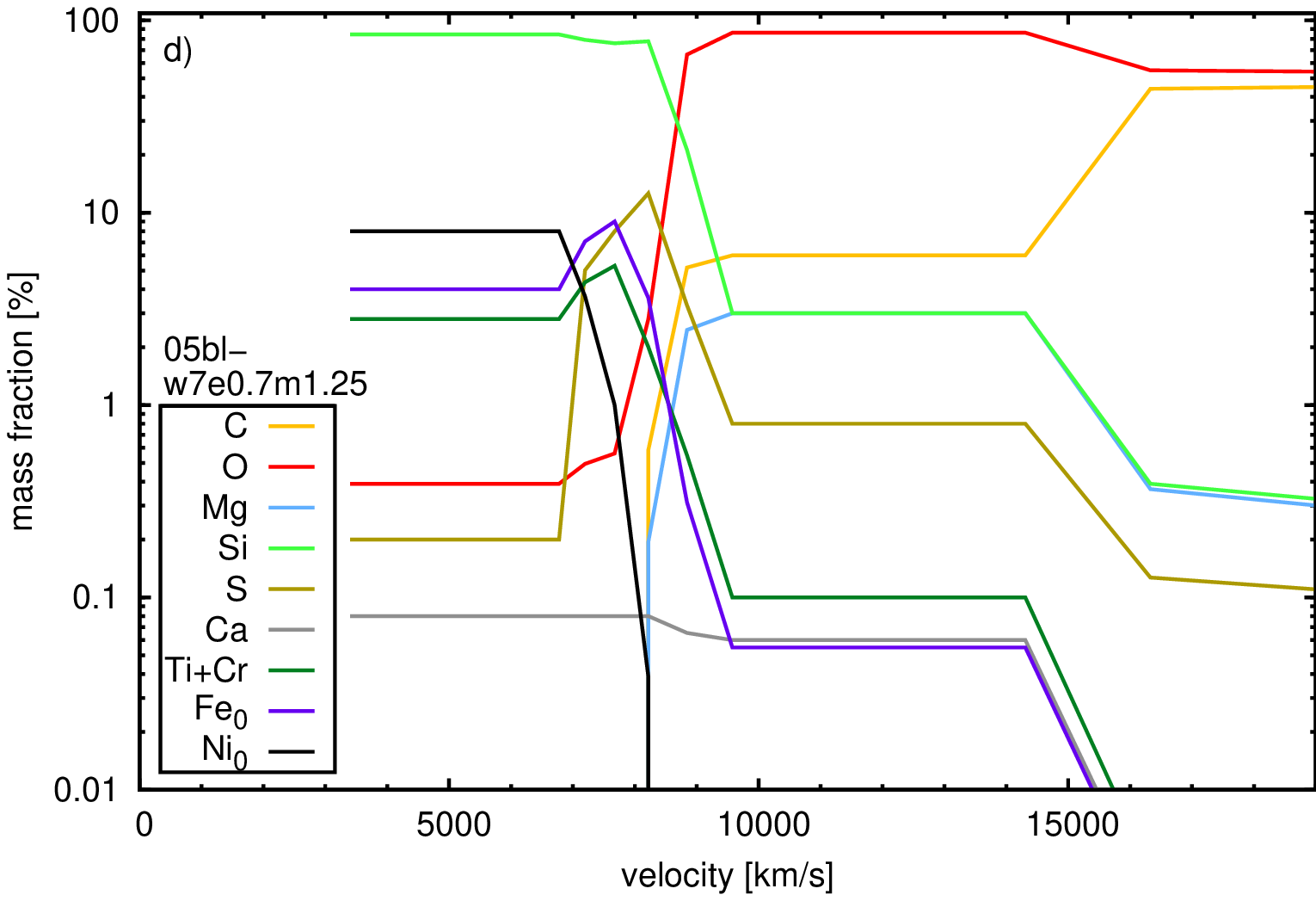} \\[0.5cm]
   \caption{Abundance tomography of SN 2005bl based on w7e0.7. The abundances are plotted versus enclosed mass (panel a) and velocity (panel b). For comparison, we also show the \mbox{05bl-w7m1.25e0.7} abundances (panels c and d). In velocity space, the patterns of the abundance profiles are very similar.}
   \label{fig:abundances-reference}
\end{figure*}

As discussed above, our models give no clear indication for a deviation from the Chandrasekhar mass. The simplest modification leading to better spectral fits and roughly consistent energetics is simply a moderate downscaling of the energy, as in the \mbox{05bl-w7e0.7} model. Thus, we consider the \mbox{05bl-w7e0.7} model a ``reference''. In Figures \ref{fig:density-reference} and \ref{fig:abundances-reference}, we show the density and abundance profiles of the \mbox{05bl-w7e0.7} model. Other spectroscopically consistent models show similar densities in the outer layers, and thus also similar abundances in that zone. This can be verified in Figures \ref{fig:density-reference} and \ref{fig:abundances-reference}, where the \mbox{05bl-w7e0.7m1.25} model is also plotted for comparison.

The \mbox{05bl-w7e0.7} model features $0.46$$\Msun$ of unburned material (including all oxygen; C constitutes $0.04$$\Msun$). IME are dominant, with a total abundance of $0.55$$\Msun$ above $3350$\kms, the velocity of the photosphere at \mbox{$+$$12.9$d}. Stable Fe is present in significant amounts ($0.05$$\Msun$). The mass of \Nifs\ (including decay products) above $3350$\kms\ is $0.06$$\Msun$, which is a bit higher than the $0.016$$-$$0.026\Msun$ found in \citet{maz97bg} above $3500$\kms\ for SN~1991bg. However, some of the \Nifs\ could be replaced by other UV-blocking elements without changing the quality of the fit. Some $0.23$$\Msun$ of material are still hidden below the \mbox{$+$$12.9$d} photosphere, where the IME abundances may still be significant (Si of the order of several $10\%$).

Alternative spectroscopically consistent models show similar patterns in the abundance profile in velocity space, but the exact densities and abundances below $\sim$$10000$\kms\ are somewhat different. In \mbox{05bl-w7e0.7m1.25}, as an example, the densities in the inner zones are larger. Thus, the abundances of Fe, Ti and Cr must be lower too keep UV opacities reasonable. For Si, moderate changes in the number density do not cause big changes in the spectra. Therefore, the smaller Fe, Ti and Cr abundances can be balanced by slightly larger Si abundances.

\subsection{The need for early time and nebular spectra of dim SNe~Ia}
\label{sec:earlierspectra}

The analysis presented here could still be refined for the outermost and innermost layers. The exact abundance stratification in the outer envelope cannot be inferred from the spectrum at \mbox{$-$$6$d}, whose photospheric velocity is already quite low. For the inner layers, especially the density structure and thus the abundance of Si is somewhat uncertain (see Sec. \ref{sec:w7e0.7discussion}). In order to make a more precise study of dim SNe Ia possible, additional spectra in the very early and in the nebular phase are needed.

The potential of an analysis of the nebular spectra has already been shown in \citet{maz97bg}. Here, we would like to illustrate the benefit of early time spectra, showing their sensitivity to the abundances in the outer envelope. We checked the influence of the abundances between $v$$\approx$$11000$\kms\  and $15000$\kms\ on the \mbox{$-$$6$d} spectrum, and found that these abundances have some effects difficult to distinguish from those of the chemical composition at lower velocities. Moreover, the (small) abundances of burned material at $\gtrsim$$15000$\kms\ cannot be exactly determined, as these only affect the extreme blue wings of the spectral features.

\begin{figure*}   
   \centering
   \includegraphics[angle=270,width=11.5cm]{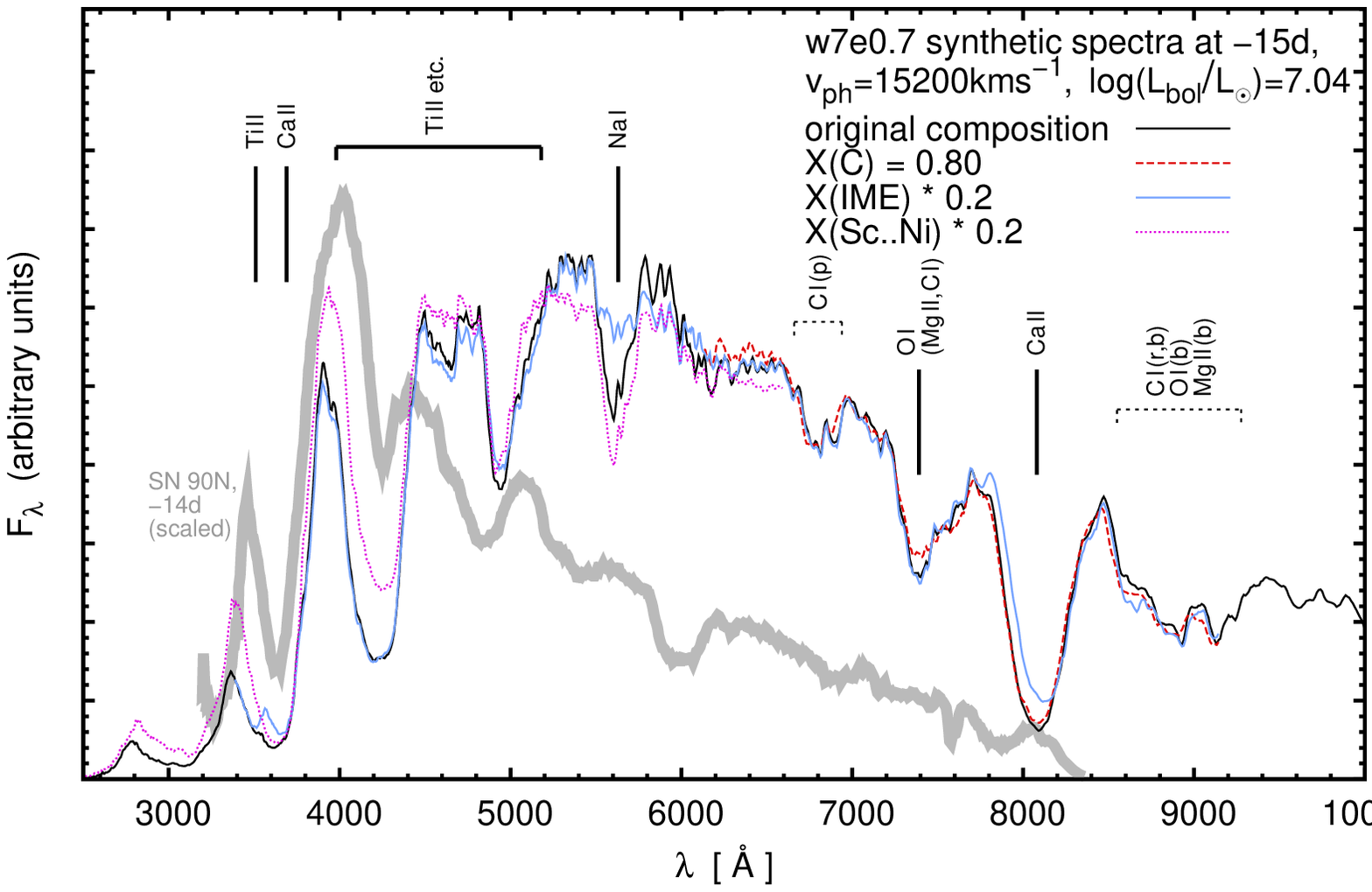}
   \includegraphics[angle=270,width=11.5cm]{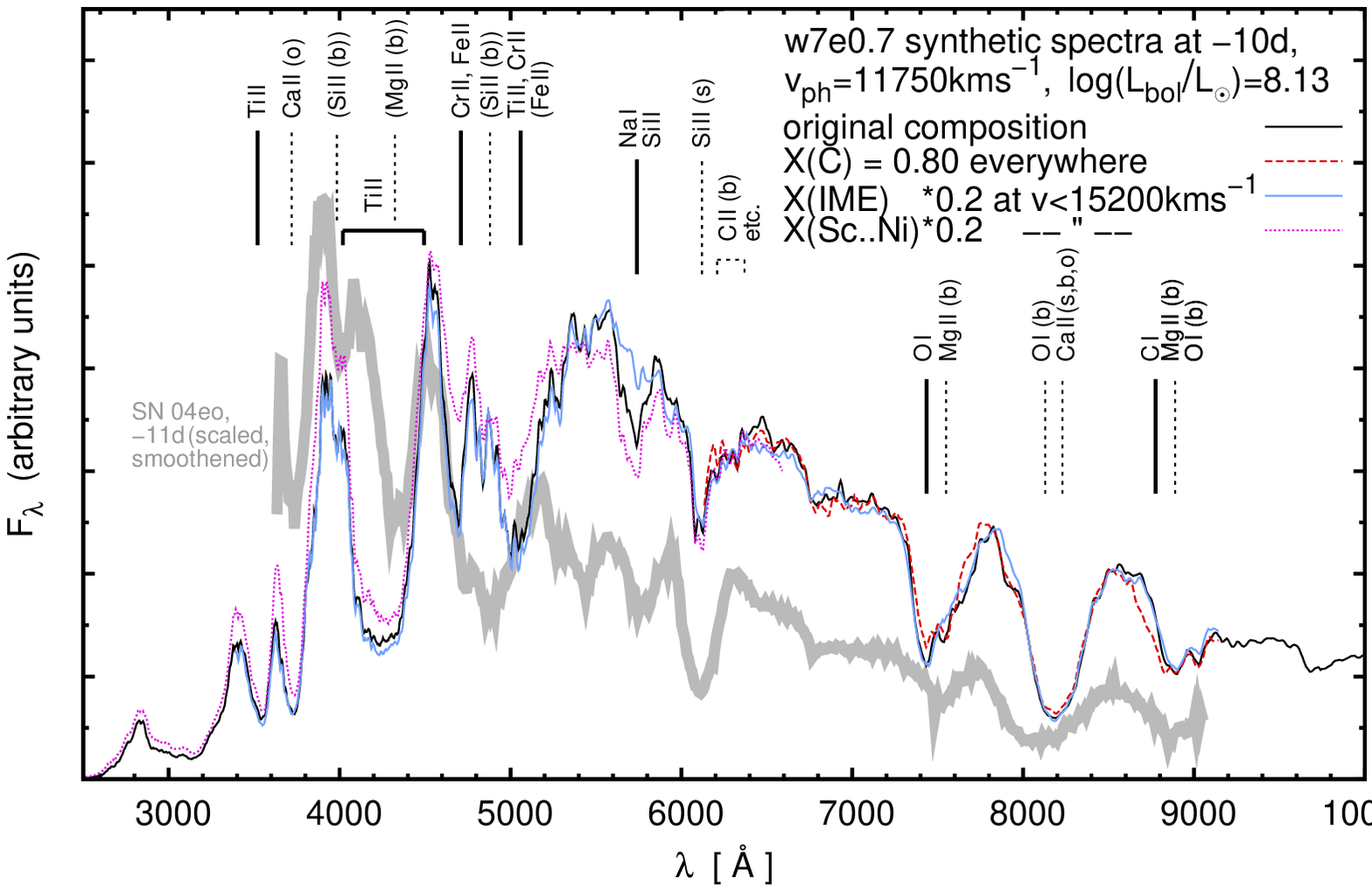}
   \caption{Early time synthetic spectra for 15d and 10d before $B$ maximum, based on \mbox{05bl-w7e0.7}. The black lines are spectra calculated with the original composition. The other lines illustrate the most notable changes which occur when setting $X(\textrm{C})$ to $80\%$ (red, dashed line) and when reducing IME (Mg up to Ca, blue, solid line) or heavier elements (Sc to Ni, magenta, dotted line) to 1/5 of their original abundances, respectively (in the velocity ranges indicated). For comparison, we plotted spectra of SN~1990N at -14d and of SN~2004eo at -11d as grey thick lines below the synthetic spectra. \protect{\newline}Approximate identifications are given for prominent features. Weak lines are given in parentheses; lines not reacting to changes in the abundances are shown with dotted marks, with the reasons for the insensitivity indicated as follows (based on a rough analysis): (b) -- line is heavily blended; (o) -- line is formed mainly at $>$$15000$\kms; (r) -- relative change of abundance is too small; (s) -- line is partially saturated.}
   \label{fig:earlytimespectra}
\end{figure*}

To explore the effect of the abundances in the outer envelope on early-time spectra, we calculated model spectra at \mbox{$-$$10$d} and \mbox{$-$$15$d} (Fig. \ref{fig:earlytimespectra}). The luminosities at these epochs were crudely estimated from the luminosity at \mbox{$-$$6$d} under the assumption of a quadratic light curve rise \citep[cf.][]{rie99}. We first calculated spectra assuming the \mbox{05bl-w7e0.7} density and abundance structure. For each of the two epochs, the photospheric position was shifted from its value at \mbox{$-$$6$d} to higher velocities, until the backscattering was reasonably reduced. This resulted in photospheric velocities of $11750$ and $15200$\kms, respectively.

After calculating these initial models, we explored the effect of changes in the chemical composition, performing three additional code runs for each epoch. In the first two runs, we reduced IME and heavier elements to $20\%$ of their original abundances, respectively. For the \mbox{$-$$15$d} model, these changes were applied to the whole atmosphere. In the \mbox{$-$$10$d} model, we kept the original composition at velocities $>$$15200$\kms constant, in order to show the sensitivity to the abundances in the zone not probed by the \mbox{$-$$15$d} spectrum. In the third code run, finally, we removed oxygen in favour of carbon (so that the mass fraction $X(\textrm{C})$$=$$80\%$). This change was applied to the whole atmosphere, also at -10d, as otherwise an inverted composition (larger C abundances further inwards) would have resulted.

In Fig. \ref{fig:earlytimespectra}, we show the resulting spectra and give line identifications to clarify the effect of the modified abundances. Moreover, it is indicated which lines do and which do not change significantly with the modifications. At \mbox{$-$$15$d}, the synthetic spectrum looks vastly different from the spectrum of a normal SN Ia. We illustrate this in the upper panel of Fig. \ref{fig:earlytimespectra} by additionally plotting the earliest SN~Ia spectrum ever observed (SN~1990N at \mbox{$-$$14$d}, \citealt{lei91}). Compared to spectra of normal SNe Ia, but also to the \mbox{$-$$6$d} spectrum, lines of less ionised species appear owing to the low temperatures, which result from the low luminosity. \SiII\ and \SII\ lines, which normally characterise SNe Ia, are absent. The spectrum is especially sensitive to the abundances of Na, Ca and Fe-group elements (with Na, uncertainties in the ionisation remain a caveat, see Sec. \ref{sec:05blw7-p48}). Furthermore, \CI\ features are present around $6800$\AA\ and $8700$\AA. As the C abundance is already quite large in the outermost layers of \mbox{05bl-w7e0.7}, these features do not react strongly to a further increase of $X($C$)$. However, if much less carbon was present, they should gradually disappear.

At \mbox{$-$$10$d}, the structure of the spectrum resembles somewhat more that at \mbox{$-$$6$d}. Yet, the spectrum has little in common with that of the moderately subluminous, spectroscopically rather normal SN~2004eo at \mbox{$-$$11$d} (\citealt{pas07}; plotted in the lower panel of Fig. \ref{fig:earlytimespectra}). Compared to \mbox{$-$$6$d}, the \mbox{$-$$10$d} spectrum still shows hints of lower temperatures: because of the scarce population of excited levels, the \SII\ ``W-trough'' does not show up. For the same reason, the \CII\ $\lambda 6580$ feature is weak. In the model with a larger C mass fraction, however, some of the strongest lines of \CI\ begin to absorb at $\sim8700$\AA. Furthermore, there are absorptions due to \OI, \NaI, \SiII, \TiII, \CrII\ and \FeII, which should allow for an analysis of the abundances in the outer layers as soon as observations are available.

The amount of extra information which can be inferred from early-time spectra will, of course, also depend on the actual luminosity of the SN at these epochs. Larger luminosities mean higher temperatures, making lines of different ions appear. However, our results already suggest that there are interesting possibilities to infer the chemical composition of the outermost ejecta.

\section{Conclusions}
\label{sec:conclusions}

We conducted an abundance tomography of SN~2005bl and confirmed that nuclear burning in dim, 91bg-like SNe Ia stops at less advanced stages compared to normal SNe Ia. The spectra indicate that the abundance of burned material above $\sim$$8500$\kms\ is much lower than even in moderately-luminous objects \citep{maz08eo}. From $\sim$$8500$\kms\ down to $\sim$$3300$\kms, IME dominate the ejecta. This points towards large-scale incomplete Si-burning or explosive O burning \citep[e.g.][]{woo73}. A detonation at low densities, as it proceeds in the outer layers of delayed-detonation models \citep{kho91}, may be responsible for the abundance pattern we find. Assuming this, we need to understand how low densities could prevail in such a large fraction of the envelope at the onset of the detonation. Up to now, all explosion models which pre-expand the star by a deflagration and then detonate (e.g. \citealt{hil00}, \citealt{bad03}, \citealt{gam04}, \citealt{roe07}, \citealt{bra09}) produce larger amounts of \Nifs. This indicates that either the pre-expansion is too weak or the amount of \Nifs\ produced in the deflagration stage is already too large. As it is uncertain whether a suitable single-degenerate model can be found, the possibility of a double-degenerate origin of dim SNe Ia deserves attention.

Besides the abundances, we have obtained information about the density profile of SN~2005bl. We showed that the spectra are incompatible with the presence of significant amounts of oxygen at $v$$\gtrsim$$13000$\kms. Together with the low abundances of burning products, this indicates a general lack of material at high velocities, albeit less extreme than in objects like SN~2005hk \citep{sah08}. We tested whether a good fit to the observed spectra is possible using a modified W7 model, scaled to a different total mass and/or energy. Indeed, a reduction of $\sim$$30\%$ in total kinetic energy yielded a spectroscopically, and also energetically consistent Chandrasekhar-mass model (\mbox{05bl-w7e0.7}). Such consistence can also be reached with somewhat super- or sub-Chandrasekhar mass density profiles, provided that they are similar to w7e0.7 at $v$$\gtrsim$$13000$\kms. Deviations of $\gtrsim$$30\%$ from the Chandrasekhar mass seem disfavoured. With our most massive models ($1.45$$M_\textrm{Ch}$), it proved impossible to obtain spectroscopic and energetic consistence at the same time. In addition, these models as well as the least massive ones ($0.5$$M_\textrm{Ch}$) most likely would yield a light curve not matching that of a dim SN~Ia.

Sharper constraints on density models, as well as on the abundance structure in the innermost and outermost layers may be obtained from very-early-epoch and nebular spectra of dim SNe~Ia. More extensive observations are needed in order to complete our picture of these objects, and of SNe Ia in general.

\section*{ACKNOWLEDGEMENTS}
This work was supported in part by the European Community's Human Potential Programme under contract HPRN-CT-2002-00303, `The Physics of Type Ia Supernovae', and by the DFG-TCRC 33 ``The Dark Universe''. We would like to thank the anonymous referee for constructive comments.

\appendix

\section{Parameters of the models}
\label{app:modelparameters}

Table \ref{tab:modelparameters} shows the code input parameters of all spectral models mentioned in the main paper. Apart from the abundances, the code takes as input the photospheric velocity $v_\textrm{ph}$, the time from explosion $t$ (see main text) and the bolometric luminosity $L_\textrm{bol}$. For different models of a given spectrum, these luminosities can differ a bit, depending on the model spectral energy distribution. In addition to the input values, Table \ref{tab:modelparameters} also gives the calculated temperature of the photospheric black body emission, $T_\textrm{BB}$, for each model.

\begin{table*}
\scriptsize
\caption{Parameters of the models. Abundances are given only for more significant elements.}
\label{tab:modelparameters}
\centering
\begin{tabular}{lrccrccccccccccccc}
Model	$\!\!\!\!$ & $\!\!$	epochs\vspace{0.2cm}	$\!\!\!\!\!\!$ & $\!$	$\mathrm{lg}\!\left(\frac{L_\textrm{bol}}{\Lsun}\right)$	$\!\!\!\!\!\!$ & $\!\!\!\!$	$v_{\textrm{ph}}$	$\!\!\!\!$ & $\!\!\!\!$	$T_\textrm{BB}$	$\!\!\!$ &	\multicolumn{12}{c}{Element abundances (mass fractions) }  \\					
\vspace{-0.80pt} & \vspace{-0.80pt} & \vspace{-0.80pt} & \vspace{-0.80pt} & \vspace{-0.80pt} & \vspace{-0.80pt} & \vspace{-0.80pt} & \vspace{-0.80pt} & \vspace{-0.80pt} & \vspace{-0.80pt} & \vspace{-0.80pt} & \vspace{-0.80pt} & \vspace{-0.80pt} & \vspace{-0.80pt} & \vspace{-0.80pt} & \vspace{-0.80pt} & \vspace{-0.80pt}  \\
$\!\!\!\!$ &  $\!\!\!\!$ [d] $\!\!$ & $\!\!\!\!$ $\!\!\!\!$ & $\!\!\!\!$	[\kms] $\!\!\!\!$  & $\!\!\!\!$ [K] $\!$ & $\!\!\!\!$	$X$(C)	$\!\!\!\!$ & $\!\!\!\!$	$X$(O)	$\!\!\!\!$ & $\!\!\!\!$	$X$(Na)	$\!\!\!\!$ & $\!\!\!\!$	$X$(Mg)	$\!\!\!\!$ & $\!\!\!\!$	$X$(Al)	$\!\!\!\!$ & $\!\!\!\!$	$X$(Si)	$\!\!\!\!$ & $\!\!\!\!$	$X$(S)	$\!\!\!\!$ & $\!\!\!\!$	$X$(Ca)	$\!\!\!\!$ & $\!\!\!\!$	$X$(Ti)	$\!\!\!\!$ & $\!\!\!\!$	$X$(Cr)	$\!\!\!\!$ & $\!\!\!\!$	$X$(Fe)$_{0}{}^{\textrm{a)}}$	$\!\!\!\!$ & $\!\!\!\!$	$X$(\Nifs)$_{0}{}^{\textrm{a)}}$ \\ 
\vspace{-0.60pt} & \vspace{-0.60pt} & \vspace{-0.60pt} & \vspace{-0.60pt} & \vspace{-0.60pt} & \vspace{-0.60pt} & \vspace{-0.60pt} & \vspace{-0.60pt} & \vspace{-0.60pt} & \vspace{-0.60pt} & \vspace{-0.60pt} & \vspace{-0.60pt} & \vspace{-0.60pt} & \vspace{-0.60pt} & \vspace{-0.60pt} & \vspace{-0.60pt} & \vspace{-0.60pt}  \\ \hline
05bl-w7e0.35 $\!\!\!\!$ & $\!\!\!\!$ -6 $\!\!\!\!$ & $\!\!\!\!$ 8.520 $\!\!\!\!$ & $\!\!\!\!$ 8200 $\!\!\!\!$ & $\!\!\!\!$ 9574.5 $\!\!\!\!$ & $\!\!\!\!$ 0.05 $\!\!\!\!$ & $\!\!\!\!$ 0.87 $\!\!\!\!$ & $\!\!\!\!$ 0.0080 $\!\!\!\!$ & $\!\!\!\!$ 0.03 $\!\!\!\!$ & $\!\!\!\!$ 0.0025 $\!\!\!\!$ & $\!\!\!\!$ 0.04 $\!\!\!\!$ & $\!\!\!\!$ 0.00 $\!\!\!\!$ & $\!\!\!\!$ 0.0009 $\!\!\!\!$ & $\!\!\!\!$ 0.0016 $\!\!\!\!$ & $\!\!\!\!$ 0.0008 $\!\!\!\!$ & $\!\!\!\!$ 0.0017 $\!\!\!\!$ & $\!\!\!\!$ 0.0000 \\ 
{} $\!\!\!\!$ & $\!\!\!\!$ -5 $\!\!\!\!$ & $\!\!\!\!$ 8.622 $\!\!\!\!$ & $\!\!\!\!$ 7975 $\!\!\!\!$ & $\!\!\!\!$ 10024.7 $\!\!\!\!$ & $\!\!\!\!$ 0.02 $\!\!\!\!$ & $\!\!\!\!$ 0.20 $\!\!\!\!$ & $\!\!\!\!$ 0.0050 $\!\!\!\!$ & $\!\!\!\!$ 0.03 $\!\!\!\!$ & $\!\!\!\!$ 0.0025 $\!\!\!\!$ & $\!\!\!\!$ 0.60 $\!\!\!\!$ & $\!\!\!\!$ 0.06 $\!\!\!\!$ & $\!\!\!\!$ 0.0009 $\!\!\!\!$ & $\!\!\!\!$ 0.0070 $\!\!\!\!$ & $\!\!\!\!$ 0.0035 $\!\!\!\!$ & $\!\!\!\!$ 0.0750 $\!\!\!\!$ & $\!\!\!\!$ 0.0000 \\ 
{} $\!\!\!\!$ & $\!\!\!\!$ -3 $\!\!\!\!$ & $\!\!\!\!$ 8.753 $\!\!\!\!$ & $\!\!\!\!$ 7600 $\!\!\!\!$ & $\!\!\!\!$ 10377.4 $\!\!\!\!$ & $\!\!\!\!$ 0.00 $\!\!\!\!$ & $\!\!\!\!$ 0.15 $\!\!\!\!$ & $\!\!\!\!$ 0.0000 $\!\!\!\!$ & $\!\!\!\!$ 0.04 $\!\!\!\!$ & $\!\!\!\!$ 0.0015 $\!\!\!\!$ & $\!\!\!\!$ 0.50 $\!\!\!\!$ & $\!\!\!\!$ 0.13 $\!\!\!\!$ & $\!\!\!\!$ 0.0009 $\!\!\!\!$ & $\!\!\!\!$ 0.0500 $\!\!\!\!$ & $\!\!\!\!$ 0.0250 $\!\!\!\!$ & $\!\!\!\!$ 0.1000 $\!\!\!\!$ & $\!\!\!\!$ 0.0000 \\ 
{} $\!\!\!\!$ & $\!\!\!\!$ 4.8 $\!\!\!\!$ & $\!\!\!\!$ 8.870 $\!\!\!\!$ & $\!\!\!\!$ 7050 $\!\!\!\!$ & $\!\!\!\!$ 8658.6 $\!\!\!\!$ & $\!\!\!\!$ 0.00 $\!\!\!\!$ & $\!\!\!\!$ 0.11 $\!\!\!\!$ & $\!\!\!\!$ 0.0000 $\!\!\!\!$ & $\!\!\!\!$ 0.00 $\!\!\!\!$ & $\!\!\!\!$ 0.0000 $\!\!\!\!$ & $\!\!\!\!$ 0.55 $\!\!\!\!$ & $\!\!\!\!$ 0.04 $\!\!\!\!$ & $\!\!\!\!$ 0.0009 $\!\!\!\!$ & $\!\!\!\!$ 0.0850 $\!\!\!\!$ & $\!\!\!\!$ 0.0850 $\!\!\!\!$ & $\!\!\!\!$ 0.1100 $\!\!\!\!$ & $\!\!\!\!$ 0.0100 \\ 
{} $\!\!\!\!$ & $\!\!\!\!$ 12.9 $\!\!\!\!$ & $\!\!\!\!$ 8.609 $\!\!\!\!$ & $\!\!\!\!$ 3350 $\!\!\!\!$ & $\!\!\!\!$ 10611.0 $\!\!\!\!$ & $\!\!\!\!$ 0.00 $\!\!\!\!$ & $\!\!\!\!$ 0.08 $\!\!\!\!$ & $\!\!\!\!$ 0.0000 $\!\!\!\!$ & $\!\!\!\!$ 0.00 $\!\!\!\!$ & $\!\!\!\!$ 0.0000 $\!\!\!\!$ & $\!\!\!\!$ 0.65 $\!\!\!\!$ & $\!\!\!\!$ 0.00 $\!\!\!\!$ & $\!\!\!\!$ 0.0009 $\!\!\!\!$ & $\!\!\!\!$ 0.0111 $\!\!\!\!$ & $\!\!\!\!$ 0.0111 $\!\!\!\!$ & $\!\!\!\!$ 0.0800 $\!\!\!\!$ & $\!\!\!\!$ 0.1600 \\ 
05bl-w7e0.5m0.5 $\!\!\!\!$ & $\!\!\!\!$ -6 $\!\!\!\!$ & $\!\!\!\!$ 8.510 $\!\!\!\!$ & $\!\!\!\!$ 7600 $\!\!\!\!$ & $\!\!\!\!$ 9626.0 $\!\!\!\!$ & $\!\!\!\!$ 0.06 $\!\!\!\!$ & $\!\!\!\!$ 0.72 $\!\!\!\!$ & $\!\!\!\!$ 0.0080 $\!\!\!\!$ & $\!\!\!\!$ 0.13 $\!\!\!\!$ & $\!\!\!\!$ 0.0070 $\!\!\!\!$ & $\!\!\!\!$ 0.06 $\!\!\!\!$ & $\!\!\!\!$ 0.01 $\!\!\!\!$ & $\!\!\!\!$ 0.0010 $\!\!\!\!$ & $\!\!\!\!$ 0.0005 $\!\!\!\!$ & $\!\!\!\!$ 0.0007 $\!\!\!\!$ & $\!\!\!\!$ 0.0003 $\!\!\!\!$ & $\!\!\!\!$ 0.0000 \\ 
{} $\!\!\!\!$ & $\!\!\!\!$ -5 $\!\!\!\!$ & $\!\!\!\!$ 8.618 $\!\!\!\!$ & $\!\!\!\!$ 7275 $\!\!\!\!$ & $\!\!\!\!$ 10220.3 $\!\!\!\!$ & $\!\!\!\!$ 0.03 $\!\!\!\!$ & $\!\!\!\!$ 0.17 $\!\!\!\!$ & $\!\!\!\!$ 0.0040 $\!\!\!\!$ & $\!\!\!\!$ 0.08 $\!\!\!\!$ & $\!\!\!\!$ 0.0070 $\!\!\!\!$ & $\!\!\!\!$ 0.60 $\!\!\!\!$ & $\!\!\!\!$ 0.01 $\!\!\!\!$ & $\!\!\!\!$ 0.0010 $\!\!\!\!$ & $\!\!\!\!$ 0.0300 $\!\!\!\!$ & $\!\!\!\!$ 0.0300 $\!\!\!\!$ & $\!\!\!\!$ 0.0400 $\!\!\!\!$ & $\!\!\!\!$ 0.0000 \\ 
{} $\!\!\!\!$ & $\!\!\!\!$ -3 $\!\!\!\!$ & $\!\!\!\!$ 8.736 $\!\!\!\!$ & $\!\!\!\!$ 7000 $\!\!\!\!$ & $\!\!\!\!$ 10242.6 $\!\!\!\!$ & $\!\!\!\!$ 0.00 $\!\!\!\!$ & $\!\!\!\!$ 0.01 $\!\!\!\!$ & $\!\!\!\!$ 0.0000 $\!\!\!\!$ & $\!\!\!\!$ 0.00 $\!\!\!\!$ & $\!\!\!\!$ 0.0030 $\!\!\!\!$ & $\!\!\!\!$ 0.52 $\!\!\!\!$ & $\!\!\!\!$ 0.12 $\!\!\!\!$ & $\!\!\!\!$ 0.0010 $\!\!\!\!$ & $\!\!\!\!$ 0.1100 $\!\!\!\!$ & $\!\!\!\!$ 0.1100 $\!\!\!\!$ & $\!\!\!\!$ 0.1250 $\!\!\!\!$ & $\!\!\!\!$ 0.0000 \\ 
{} $\!\!\!\!$ & $\!\!\!\!$ 4.8 $\!\!\!\!$ & $\!\!\!\!$ 8.837 $\!\!\!\!$ & $\!\!\!\!$ 6400 $\!\!\!\!$ & $\!\!\!\!$ 8613.5 $\!\!\!\!$ & $\!\!\!\!$ 0.00 $\!\!\!\!$ & $\!\!\!\!$ 0.00 $\!\!\!\!$ & $\!\!\!\!$ 0.0000 $\!\!\!\!$ & $\!\!\!\!$ 0.00 $\!\!\!\!$ & $\!\!\!\!$ 0.0000 $\!\!\!\!$ & $\!\!\!\!$ 0.17 $\!\!\!\!$ & $\!\!\!\!$ 0.04 $\!\!\!\!$ & $\!\!\!\!$ 0.0010 $\!\!\!\!$ & $\!\!\!\!$ 0.2550 $\!\!\!\!$ & $\!\!\!\!$ 0.2550 $\!\!\!\!$ & $\!\!\!\!$ 0.2550 $\!\!\!\!$ & $\!\!\!\!$ 0.0100 \\ 
{} $\!\!\!\!$ & $\!\!\!\!$ 12.9 $\!\!\!\!$ & $\!\!\!\!$ 8.570 $\!\!\!\!$ & $\!\!\!\!$ 3250 $\!\!\!\!$ & $\!\!\!\!$ 9195.7 $\!\!\!\!$ & $\!\!\!\!$ 0.00 $\!\!\!\!$ & $\!\!\!\!$ 0.00 $\!\!\!\!$ & $\!\!\!\!$ 0.0000 $\!\!\!\!$ & $\!\!\!\!$ 0.00 $\!\!\!\!$ & $\!\!\!\!$ 0.0000 $\!\!\!\!$ & $\!\!\!\!$ 0.17 $\!\!\!\!$ & $\!\!\!\!$ 0.00 $\!\!\!\!$ & $\!\!\!\!$ 0.0010 $\!\!\!\!$ & $\!\!\!\!$ 0.2150 $\!\!\!\!$ & $\!\!\!\!$ 0.2150 $\!\!\!\!$ & $\!\!\!\!$ 0.1300 $\!\!\!\!$ & $\!\!\!\!$ 0.2600 \\ 
05bl-w7e0.5m0.7 $\!\!\!\!$ & $\!\!\!\!$ -6 $\!\!\!\!$ & $\!\!\!\!$ 8.520 $\!\!\!\!$ & $\!\!\!\!$ 7900 $\!\!\!\!$ & $\!\!\!\!$ 9734.1 $\!\!\!\!$ & $\!\!\!\!$ 0.06 $\!\!\!\!$ & $\!\!\!\!$ 0.83 $\!\!\!\!$ & $\!\!\!\!$ 0.0060 $\!\!\!\!$ & $\!\!\!\!$ 0.06 $\!\!\!\!$ & $\!\!\!\!$ 0.0025 $\!\!\!\!$ & $\!\!\!\!$ 0.03 $\!\!\!\!$ & $\!\!\!\!$ 0.01 $\!\!\!\!$ & $\!\!\!\!$ 0.0008 $\!\!\!\!$ & $\!\!\!\!$ 0.0005 $\!\!\!\!$ & $\!\!\!\!$ 0.0004 $\!\!\!\!$ & $\!\!\!\!$ 0.0004 $\!\!\!\!$ & $\!\!\!\!$ 0.0000 \\ 
{} $\!\!\!\!$ & $\!\!\!\!$ -5 $\!\!\!\!$ & $\!\!\!\!$ 8.618 $\!\!\!\!$ & $\!\!\!\!$ 7600 $\!\!\!\!$ & $\!\!\!\!$ 10242.2 $\!\!\!\!$ & $\!\!\!\!$ 0.04 $\!\!\!\!$ & $\!\!\!\!$ 0.12 $\!\!\!\!$ & $\!\!\!\!$ 0.0030 $\!\!\!\!$ & $\!\!\!\!$ 0.04 $\!\!\!\!$ & $\!\!\!\!$ 0.0025 $\!\!\!\!$ & $\!\!\!\!$ 0.62 $\!\!\!\!$ & $\!\!\!\!$ 0.10 $\!\!\!\!$ & $\!\!\!\!$ 0.0008 $\!\!\!\!$ & $\!\!\!\!$ 0.0150 $\!\!\!\!$ & $\!\!\!\!$ 0.0150 $\!\!\!\!$ & $\!\!\!\!$ 0.0400 $\!\!\!\!$ & $\!\!\!\!$ 0.0000 \\ 
{} $\!\!\!\!$ & $\!\!\!\!$ -3 $\!\!\!\!$ & $\!\!\!\!$ 8.740 $\!\!\!\!$ & $\!\!\!\!$ 7100 $\!\!\!\!$ & $\!\!\!\!$ 10727.9 $\!\!\!\!$ & $\!\!\!\!$ 0.00 $\!\!\!\!$ & $\!\!\!\!$ 0.03 $\!\!\!\!$ & $\!\!\!\!$ 0.0000 $\!\!\!\!$ & $\!\!\!\!$ 0.00 $\!\!\!\!$ & $\!\!\!\!$ 0.0017 $\!\!\!\!$ & $\!\!\!\!$ 0.55 $\!\!\!\!$ & $\!\!\!\!$ 0.15 $\!\!\!\!$ & $\!\!\!\!$ 0.0008 $\!\!\!\!$ & $\!\!\!\!$ 0.0750 $\!\!\!\!$ & $\!\!\!\!$ 0.0750 $\!\!\!\!$ & $\!\!\!\!$ 0.1200 $\!\!\!\!$ & $\!\!\!\!$ 0.0000 \\ 
{} $\!\!\!\!$ & $\!\!\!\!$ 4.8 $\!\!\!\!$ & $\!\!\!\!$ 8.845 $\!\!\!\!$ & $\!\!\!\!$ 6550 $\!\!\!\!$ & $\!\!\!\!$ 8764.9 $\!\!\!\!$ & $\!\!\!\!$ 0.00 $\!\!\!\!$ & $\!\!\!\!$ 0.00 $\!\!\!\!$ & $\!\!\!\!$ 0.0000 $\!\!\!\!$ & $\!\!\!\!$ 0.00 $\!\!\!\!$ & $\!\!\!\!$ 0.0000 $\!\!\!\!$ & $\!\!\!\!$ 0.53 $\!\!\!\!$ & $\!\!\!\!$ 0.05 $\!\!\!\!$ & $\!\!\!\!$ 0.0008 $\!\!\!\!$ & $\!\!\!\!$ 0.1225 $\!\!\!\!$ & $\!\!\!\!$ 0.1225 $\!\!\!\!$ & $\!\!\!\!$ 0.1650 $\!\!\!\!$ & $\!\!\!\!$ 0.0100 \\ 
{} $\!\!\!\!$ & $\!\!\!\!$ 12.9 $\!\!\!\!$ & $\!\!\!\!$ 8.590 $\!\!\!\!$ & $\!\!\!\!$ 3350 $\!\!\!\!$ & $\!\!\!\!$ 9582.3 $\!\!\!\!$ & $\!\!\!\!$ 0.00 $\!\!\!\!$ & $\!\!\!\!$ 0.00 $\!\!\!\!$ & $\!\!\!\!$ 0.0000 $\!\!\!\!$ & $\!\!\!\!$ 0.00 $\!\!\!\!$ & $\!\!\!\!$ 0.0000 $\!\!\!\!$ & $\!\!\!\!$ 0.68 $\!\!\!\!$ & $\!\!\!\!$ 0.00 $\!\!\!\!$ & $\!\!\!\!$ 0.0008 $\!\!\!\!$ & $\!\!\!\!$ 0.0400 $\!\!\!\!$ & $\!\!\!\!$ 0.0400 $\!\!\!\!$ & $\!\!\!\!$ 0.0800 $\!\!\!\!$ & $\!\!\!\!$ 0.1550 \\ 
05bl-w7e0.5 $\!\!\!\!$ & $\!\!\!\!$ -6 $\!\!\!\!$ & $\!\!\!\!$ 8.525 $\!\!\!\!$ & $\!\!\!\!$ 8300 $\!\!\!\!$ & $\!\!\!\!$ 9705.2 $\!\!\!\!$ & $\!\!\!\!$ 0.05 $\!\!\!\!$ & $\!\!\!\!$ 0.88 $\!\!\!\!$ & $\!\!\!\!$ 0.0070 $\!\!\!\!$ & $\!\!\!\!$ 0.03 $\!\!\!\!$ & $\!\!\!\!$ 0.0020 $\!\!\!\!$ & $\!\!\!\!$ 0.03 $\!\!\!\!$ & $\!\!\!\!$ 0.01 $\!\!\!\!$ & $\!\!\!\!$ 0.0004 $\!\!\!\!$ & $\!\!\!\!$ 0.0008 $\!\!\!\!$ & $\!\!\!\!$ 0.0004 $\!\!\!\!$ & $\!\!\!\!$ 0.0008 $\!\!\!\!$ & $\!\!\!\!$ 0.0000 \\ 
{} $\!\!\!\!$ & $\!\!\!\!$ -5 $\!\!\!\!$ & $\!\!\!\!$ 8.625 $\!\!\!\!$ & $\!\!\!\!$ 8050 $\!\!\!\!$ & $\!\!\!\!$ 10185.4 $\!\!\!\!$ & $\!\!\!\!$ 0.02 $\!\!\!\!$ & $\!\!\!\!$ 0.12 $\!\!\!\!$ & $\!\!\!\!$ 0.0035 $\!\!\!\!$ & $\!\!\!\!$ 0.02 $\!\!\!\!$ & $\!\!\!\!$ 0.0020 $\!\!\!\!$ & $\!\!\!\!$ 0.70 $\!\!\!\!$ & $\!\!\!\!$ 0.07 $\!\!\!\!$ & $\!\!\!\!$ 0.0004 $\!\!\!\!$ & $\!\!\!\!$ 0.0300 $\!\!\!\!$ & $\!\!\!\!$ 0.0250 $\!\!\!\!$ & $\!\!\!\!$ 0.0100 $\!\!\!\!$ & $\!\!\!\!$ 0.0000 \\ 
{} $\!\!\!\!$ & $\!\!\!\!$ -3 $\!\!\!\!$ & $\!\!\!\!$ 8.748 $\!\!\!\!$ & $\!\!\!\!$ 7550 $\!\!\!\!$ & $\!\!\!\!$ 10745.4 $\!\!\!\!$ & $\!\!\!\!$ 0.00 $\!\!\!\!$ & $\!\!\!\!$ 0.01 $\!\!\!\!$ & $\!\!\!\!$ 0.0000 $\!\!\!\!$ & $\!\!\!\!$ 0.00 $\!\!\!\!$ & $\!\!\!\!$ 0.0010 $\!\!\!\!$ & $\!\!\!\!$ 0.62 $\!\!\!\!$ & $\!\!\!\!$ 0.14 $\!\!\!\!$ & $\!\!\!\!$ 0.0004 $\!\!\!\!$ & $\!\!\!\!$ 0.0625 $\!\!\!\!$ & $\!\!\!\!$ 0.0625 $\!\!\!\!$ & $\!\!\!\!$ 0.1000 $\!\!\!\!$ & $\!\!\!\!$ 0.0000 \\ 
{} $\!\!\!\!$ & $\!\!\!\!$ 4.8 $\!\!\!\!$ & $\!\!\!\!$ 8.866 $\!\!\!\!$ & $\!\!\!\!$ 6900 $\!\!\!\!$ & $\!\!\!\!$ 8959.2 $\!\!\!\!$ & $\!\!\!\!$ 0.00 $\!\!\!\!$ & $\!\!\!\!$ 0.00 $\!\!\!\!$ & $\!\!\!\!$ 0.0000 $\!\!\!\!$ & $\!\!\!\!$ 0.00 $\!\!\!\!$ & $\!\!\!\!$ 0.0000 $\!\!\!\!$ & $\!\!\!\!$ 0.62 $\!\!\!\!$ & $\!\!\!\!$ 0.06 $\!\!\!\!$ & $\!\!\!\!$ 0.0004 $\!\!\!\!$ & $\!\!\!\!$ 0.0675 $\!\!\!\!$ & $\!\!\!\!$ 0.0675 $\!\!\!\!$ & $\!\!\!\!$ 0.1600 $\!\!\!\!$ & $\!\!\!\!$ 0.0100 \\ 
{} $\!\!\!\!$ & $\!\!\!\!$ 12.9 $\!\!\!\!$ & $\!\!\!\!$ 8.602 $\!\!\!\!$ & $\!\!\!\!$ 3350 $\!\!\!\!$ & $\!\!\!\!$ 10432.1 $\!\!\!\!$ & $\!\!\!\!$ 0.00 $\!\!\!\!$ & $\!\!\!\!$ 0.00 $\!\!\!\!$ & $\!\!\!\!$ 0.0000 $\!\!\!\!$ & $\!\!\!\!$ 0.00 $\!\!\!\!$ & $\!\!\!\!$ 0.0000 $\!\!\!\!$ & $\!\!\!\!$ 0.78 $\!\!\!\!$ & $\!\!\!\!$ 0.00 $\!\!\!\!$ & $\!\!\!\!$ 0.0004 $\!\!\!\!$ & $\!\!\!\!$ 0.0190 $\!\!\!\!$ & $\!\!\!\!$ 0.0190 $\!\!\!\!$ & $\!\!\!\!$ 0.0600 $\!\!\!\!$ & $\!\!\!\!$ 0.1200 \\ 
05bl-w7e0.5m1.25 $\!\!\!\!$ & $\!\!\!\!$ -6 $\!\!\!\!$ & $\!\!\!\!$ 8.517 $\!\!\!\!$ & $\!\!\!\!$ 8250 $\!\!\!\!$ & $\!\!\!\!$ 10038.1 $\!\!\!\!$ & $\!\!\!\!$ 0.05 $\!\!\!\!$ & $\!\!\!\!$ 0.87 $\!\!\!\!$ & $\!\!\!\!$ 0.0045 $\!\!\!\!$ & $\!\!\!\!$ 0.04 $\!\!\!\!$ & $\!\!\!\!$ 0.0018 $\!\!\!\!$ & $\!\!\!\!$ 0.03 $\!\!\!\!$ & $\!\!\!\!$ 0.00 $\!\!\!\!$ & $\!\!\!\!$ 0.0006 $\!\!\!\!$ & $\!\!\!\!$ 0.0010 $\!\!\!\!$ & $\!\!\!\!$ 0.0008 $\!\!\!\!$ & $\!\!\!\!$ 0.0013 $\!\!\!\!$ & $\!\!\!\!$ 0.0000 \\ 
{} $\!\!\!\!$ & $\!\!\!\!$ -5 $\!\!\!\!$ & $\!\!\!\!$ 8.615 $\!\!\!\!$ & $\!\!\!\!$ 8000 $\!\!\!\!$ & $\!\!\!\!$ 10511.8 $\!\!\!\!$ & $\!\!\!\!$ 0.01 $\!\!\!\!$ & $\!\!\!\!$ 0.13 $\!\!\!\!$ & $\!\!\!\!$ 0.0045 $\!\!\!\!$ & $\!\!\!\!$ 0.04 $\!\!\!\!$ & $\!\!\!\!$ 0.0018 $\!\!\!\!$ & $\!\!\!\!$ 0.62 $\!\!\!\!$ & $\!\!\!\!$ 0.12 $\!\!\!\!$ & $\!\!\!\!$ 0.0006 $\!\!\!\!$ & $\!\!\!\!$ 0.0220 $\!\!\!\!$ & $\!\!\!\!$ 0.0220 $\!\!\!\!$ & $\!\!\!\!$ 0.0250 $\!\!\!\!$ & $\!\!\!\!$ 0.0000 \\ 
{} $\!\!\!\!$ & $\!\!\!\!$ -3 $\!\!\!\!$ & $\!\!\!\!$ 8.748 $\!\!\!\!$ & $\!\!\!\!$ 7650 $\!\!\!\!$ & $\!\!\!\!$ 10658.7 $\!\!\!\!$ & $\!\!\!\!$ 0.00 $\!\!\!\!$ & $\!\!\!\!$ 0.02 $\!\!\!\!$ & $\!\!\!\!$ 0.0020 $\!\!\!\!$ & $\!\!\!\!$ 0.00 $\!\!\!\!$ & $\!\!\!\!$ 0.0010 $\!\!\!\!$ & $\!\!\!\!$ 0.64 $\!\!\!\!$ & $\!\!\!\!$ 0.16 $\!\!\!\!$ & $\!\!\!\!$ 0.0006 $\!\!\!\!$ & $\!\!\!\!$ 0.0600 $\!\!\!\!$ & $\!\!\!\!$ 0.0600 $\!\!\!\!$ & $\!\!\!\!$ 0.0600 $\!\!\!\!$ & $\!\!\!\!$ 0.0000 \\ 
{} $\!\!\!\!$ & $\!\!\!\!$ 4.8 $\!\!\!\!$ & $\!\!\!\!$ 8.861 $\!\!\!\!$ & $\!\!\!\!$ 7250 $\!\!\!\!$ & $\!\!\!\!$ 8628.6 $\!\!\!\!$ & $\!\!\!\!$ 0.00 $\!\!\!\!$ & $\!\!\!\!$ 0.00 $\!\!\!\!$ & $\!\!\!\!$ 0.0000 $\!\!\!\!$ & $\!\!\!\!$ 0.00 $\!\!\!\!$ & $\!\!\!\!$ 0.0000 $\!\!\!\!$ & $\!\!\!\!$ 0.65 $\!\!\!\!$ & $\!\!\!\!$ 0.07 $\!\!\!\!$ & $\!\!\!\!$ 0.0006 $\!\!\!\!$ & $\!\!\!\!$ 0.0900 $\!\!\!\!$ & $\!\!\!\!$ 0.0900 $\!\!\!\!$ & $\!\!\!\!$ 0.0850 $\!\!\!\!$ & $\!\!\!\!$ 0.0085 \\ 
{} $\!\!\!\!$ & $\!\!\!\!$ 12.9 $\!\!\!\!$ & $\!\!\!\!$ 8.595 $\!\!\!\!$ & $\!\!\!\!$ 3350 $\!\!\!\!$ & $\!\!\!\!$ 10835.7 $\!\!\!\!$ & $\!\!\!\!$ 0.00 $\!\!\!\!$ & $\!\!\!\!$ 0.00 $\!\!\!\!$ & $\!\!\!\!$ 0.0000 $\!\!\!\!$ & $\!\!\!\!$ 0.00 $\!\!\!\!$ & $\!\!\!\!$ 0.0000 $\!\!\!\!$ & $\!\!\!\!$ 0.82 $\!\!\!\!$ & $\!\!\!\!$ 0.00 $\!\!\!\!$ & $\!\!\!\!$ 0.0006 $\!\!\!\!$ & $\!\!\!\!$ 0.0070 $\!\!\!\!$ & $\!\!\!\!$ 0.0070 $\!\!\!\!$ & $\!\!\!\!$ 0.0500 $\!\!\!\!$ & $\!\!\!\!$ 0.1100 \\ 
05bl-w7e0.7m0.7 $\!\!\!\!$ & $\!\!\!\!$ -6 $\!\!\!\!$ & $\!\!\!\!$ 8.514 $\!\!\!\!$ & $\!\!\!\!$ 7950 $\!\!\!\!$ & $\!\!\!\!$ 9749.2 $\!\!\!\!$ & $\!\!\!\!$ 0.07 $\!\!\!\!$ & $\!\!\!\!$ 0.75 $\!\!\!\!$ & $\!\!\!\!$ 0.0050 $\!\!\!\!$ & $\!\!\!\!$ 0.10 $\!\!\!\!$ & $\!\!\!\!$ 0.0035 $\!\!\!\!$ & $\!\!\!\!$ 0.04 $\!\!\!\!$ & $\!\!\!\!$ 0.03 $\!\!\!\!$ & $\!\!\!\!$ 0.0009 $\!\!\!\!$ & $\!\!\!\!$ 0.0004 $\!\!\!\!$ & $\!\!\!\!$ 0.0004 $\!\!\!\!$ & $\!\!\!\!$ 0.0003 $\!\!\!\!$ & $\!\!\!\!$ 0.0000 \\ 
{} $\!\!\!\!$ & $\!\!\!\!$ -5 $\!\!\!\!$ & $\!\!\!\!$ 8.619 $\!\!\!\!$ & $\!\!\!\!$ 7600 $\!\!\!\!$ & $\!\!\!\!$ 10346.2 $\!\!\!\!$ & $\!\!\!\!$ 0.02 $\!\!\!\!$ & $\!\!\!\!$ 0.10 $\!\!\!\!$ & $\!\!\!\!$ 0.0025 $\!\!\!\!$ & $\!\!\!\!$ 0.03 $\!\!\!\!$ & $\!\!\!\!$ 0.0035 $\!\!\!\!$ & $\!\!\!\!$ 0.70 $\!\!\!\!$ & $\!\!\!\!$ 0.07 $\!\!\!\!$ & $\!\!\!\!$ 0.0009 $\!\!\!\!$ & $\!\!\!\!$ 0.0200 $\!\!\!\!$ & $\!\!\!\!$ 0.0200 $\!\!\!\!$ & $\!\!\!\!$ 0.0300 $\!\!\!\!$ & $\!\!\!\!$ 0.0000 \\ 
{} $\!\!\!\!$ & $\!\!\!\!$ -3 $\!\!\!\!$ & $\!\!\!\!$ 8.744 $\!\!\!\!$ & $\!\!\!\!$ 7100 $\!\!\!\!$ & $\!\!\!\!$ 10760.9 $\!\!\!\!$ & $\!\!\!\!$ 0.00 $\!\!\!\!$ & $\!\!\!\!$ 0.02 $\!\!\!\!$ & $\!\!\!\!$ 0.0000 $\!\!\!\!$ & $\!\!\!\!$ 0.00 $\!\!\!\!$ & $\!\!\!\!$ 0.0018 $\!\!\!\!$ & $\!\!\!\!$ 0.60 $\!\!\!\!$ & $\!\!\!\!$ 0.09 $\!\!\!\!$ & $\!\!\!\!$ 0.0009 $\!\!\!\!$ & $\!\!\!\!$ 0.0700 $\!\!\!\!$ & $\!\!\!\!$ 0.0700 $\!\!\!\!$ & $\!\!\!\!$ 0.1500 $\!\!\!\!$ & $\!\!\!\!$ 0.0000 \\ 
{} $\!\!\!\!$ & $\!\!\!\!$ 4.8 $\!\!\!\!$ & $\!\!\!\!$ 8.843 $\!\!\!\!$ & $\!\!\!\!$ 6550 $\!\!\!\!$ & $\!\!\!\!$ 8730.7 $\!\!\!\!$ & $\!\!\!\!$ 0.00 $\!\!\!\!$ & $\!\!\!\!$ 0.01 $\!\!\!\!$ & $\!\!\!\!$ 0.0000 $\!\!\!\!$ & $\!\!\!\!$ 0.00 $\!\!\!\!$ & $\!\!\!\!$ 0.0000 $\!\!\!\!$ & $\!\!\!\!$ 0.51 $\!\!\!\!$ & $\!\!\!\!$ 0.07 $\!\!\!\!$ & $\!\!\!\!$ 0.0009 $\!\!\!\!$ & $\!\!\!\!$ 0.1200 $\!\!\!\!$ & $\!\!\!\!$ 0.1200 $\!\!\!\!$ & $\!\!\!\!$ 0.1600 $\!\!\!\!$ & $\!\!\!\!$ 0.0100 \\ 
{} $\!\!\!\!$ & $\!\!\!\!$ 12.9 $\!\!\!\!$ & $\!\!\!\!$ 8.587 $\!\!\!\!$ & $\!\!\!\!$ 3250 $\!\!\!\!$ & $\!\!\!\!$ 9539.3 $\!\!\!\!$ & $\!\!\!\!$ 0.00 $\!\!\!\!$ & $\!\!\!\!$ 0.00 $\!\!\!\!$ & $\!\!\!\!$ 0.0000 $\!\!\!\!$ & $\!\!\!\!$ 0.00 $\!\!\!\!$ & $\!\!\!\!$ 0.0000 $\!\!\!\!$ & $\!\!\!\!$ 0.50 $\!\!\!\!$ & $\!\!\!\!$ 0.00 $\!\!\!\!$ & $\!\!\!\!$ 0.0009 $\!\!\!\!$ & $\!\!\!\!$ 0.0590 $\!\!\!\!$ & $\!\!\!\!$ 0.0590 $\!\!\!\!$ & $\!\!\!\!$ 0.1250 $\!\!\!\!$ & $\!\!\!\!$ 0.2500 \\ 
05bl-w7e0.7 $\!\!\!\!$ & $\!\!\!\!$ -6 $\!\!\!\!$ & $\!\!\!\!$ 8.520 $\!\!\!\!$ & $\!\!\!\!$ 8350 $\!\!\!\!$ & $\!\!\!\!$ 9764.8 $\!\!\!\!$ & $\!\!\!\!$ 0.06 $\!\!\!\!$ & $\!\!\!\!$ 0.86 $\!\!\!\!$ & $\!\!\!\!$ 0.0060 $\!\!\!\!$ & $\!\!\!\!$ 0.04 $\!\!\!\!$ & $\!\!\!\!$ 0.0025 $\!\!\!\!$ & $\!\!\!\!$ 0.02 $\!\!\!\!$ & $\!\!\!\!$ 0.01 $\!\!\!\!$ & $\!\!\!\!$ 0.0004 $\!\!\!\!$ & $\!\!\!\!$ 0.0004 $\!\!\!\!$ & $\!\!\!\!$ 0.0003 $\!\!\!\!$ & $\!\!\!\!$ 0.0003 $\!\!\!\!$ & $\!\!\!\!$ 0.0000 \\ 
{} $\!\!\!\!$ & $\!\!\!\!$ -5 $\!\!\!\!$ & $\!\!\!\!$ 8.617 $\!\!\!\!$ & $\!\!\!\!$ 8100 $\!\!\!\!$ & $\!\!\!\!$ 10135.5 $\!\!\!\!$ & $\!\!\!\!$ 0.03 $\!\!\!\!$ & $\!\!\!\!$ 0.13 $\!\!\!\!$ & $\!\!\!\!$ 0.0030 $\!\!\!\!$ & $\!\!\!\!$ 0.03 $\!\!\!\!$ & $\!\!\!\!$ 0.0025 $\!\!\!\!$ & $\!\!\!\!$ 0.68 $\!\!\!\!$ & $\!\!\!\!$ 0.10 $\!\!\!\!$ & $\!\!\!\!$ 0.0004 $\!\!\!\!$ & $\!\!\!\!$ 0.0100 $\!\!\!\!$ & $\!\!\!\!$ 0.0070 $\!\!\!\!$ & $\!\!\!\!$ 0.0150 $\!\!\!\!$ & $\!\!\!\!$ 0.0000 \\ 
{} $\!\!\!\!$ & $\!\!\!\!$ -3 $\!\!\!\!$ & $\!\!\!\!$ 8.745 $\!\!\!\!$ & $\!\!\!\!$ 7600 $\!\!\!\!$ & $\!\!\!\!$ 10632.8 $\!\!\!\!$ & $\!\!\!\!$ 0.00 $\!\!\!\!$ & $\!\!\!\!$ 0.01 $\!\!\!\!$ & $\!\!\!\!$ 0.0000 $\!\!\!\!$ & $\!\!\!\!$ 0.00 $\!\!\!\!$ & $\!\!\!\!$ 0.0015 $\!\!\!\!$ & $\!\!\!\!$ 0.70 $\!\!\!\!$ & $\!\!\!\!$ 0.11 $\!\!\!\!$ & $\!\!\!\!$ 0.0004 $\!\!\!\!$ & $\!\!\!\!$ 0.0533 $\!\!\!\!$ & $\!\!\!\!$ 0.0367 $\!\!\!\!$ & $\!\!\!\!$ 0.0900 $\!\!\!\!$ & $\!\!\!\!$ 0.0000 \\ 
{} $\!\!\!\!$ & $\!\!\!\!$ 4.8 $\!\!\!\!$ & $\!\!\!\!$ 8.861 $\!\!\!\!$ & $\!\!\!\!$ 6800 $\!\!\!\!$ & $\!\!\!\!$ 8930.6 $\!\!\!\!$ & $\!\!\!\!$ 0.00 $\!\!\!\!$ & $\!\!\!\!$ 0.00 $\!\!\!\!$ & $\!\!\!\!$ 0.0000 $\!\!\!\!$ & $\!\!\!\!$ 0.00 $\!\!\!\!$ & $\!\!\!\!$ 0.0000 $\!\!\!\!$ & $\!\!\!\!$ 0.71 $\!\!\!\!$ & $\!\!\!\!$ 0.07 $\!\!\!\!$ & $\!\!\!\!$ 0.0004 $\!\!\!\!$ & $\!\!\!\!$ 0.0550 $\!\!\!\!$ & $\!\!\!\!$ 0.0400 $\!\!\!\!$ & $\!\!\!\!$ 0.1150 $\!\!\!\!$ & $\!\!\!\!$ 0.0100 \\ 
{} $\!\!\!\!$ & $\!\!\!\!$ 12.9 $\!\!\!\!$ & $\!\!\!\!$ 8.594 $\!\!\!\!$ & $\!\!\!\!$ 3350 $\!\!\!\!$ & $\!\!\!\!$ 10071.5 $\!\!\!\!$ & $\!\!\!\!$ 0.00 $\!\!\!\!$ & $\!\!\!\!$ 0.00 $\!\!\!\!$ & $\!\!\!\!$ 0.0000 $\!\!\!\!$ & $\!\!\!\!$ 0.00 $\!\!\!\!$ & $\!\!\!\!$ 0.0000 $\!\!\!\!$ & $\!\!\!\!$ 0.77 $\!\!\!\!$ & $\!\!\!\!$ 0.00 $\!\!\!\!$ & $\!\!\!\!$ 0.0005 $\!\!\!\!$ & $\!\!\!\!$ 0.0167 $\!\!\!\!$ & $\!\!\!\!$ 0.0167 $\!\!\!\!$ & $\!\!\!\!$ 0.0650 $\!\!\!\!$ & $\!\!\!\!$ 0.1300 \\ 
05bl-w7e0.7m1.25 $\!\!\!\!$ & $\!\!\!\!$ -6 $\!\!\!\!$ & $\!\!\!\!$ 8.523 $\!\!\!\!$ & $\!\!\!\!$ 8700 $\!\!\!\!$ & $\!\!\!\!$ 9667.9 $\!\!\!\!$ & $\!\!\!\!$ 0.06 $\!\!\!\!$ & $\!\!\!\!$ 0.86 $\!\!\!\!$ & $\!\!\!\!$ 0.0045 $\!\!\!\!$ & $\!\!\!\!$ 0.03 $\!\!\!\!$ & $\!\!\!\!$ 0.0020 $\!\!\!\!$ & $\!\!\!\!$ 0.03 $\!\!\!\!$ & $\!\!\!\!$ 0.01 $\!\!\!\!$ & $\!\!\!\!$ 0.0006 $\!\!\!\!$ & $\!\!\!\!$ 0.0005 $\!\!\!\!$ & $\!\!\!\!$ 0.0005 $\!\!\!\!$ & $\!\!\!\!$ 0.0006 $\!\!\!\!$ & $\!\!\!\!$ 0.0000 \\ 
{} $\!\!\!\!$ & $\!\!\!\!$ -5 $\!\!\!\!$ & $\!\!\!\!$ 8.623 $\!\!\!\!$ & $\!\!\!\!$ 8400 $\!\!\!\!$ & $\!\!\!\!$ 10101.1 $\!\!\!\!$ & $\!\!\!\!$ 0.03 $\!\!\!\!$ & $\!\!\!\!$ 0.13 $\!\!\!\!$ & $\!\!\!\!$ 0.0020 $\!\!\!\!$ & $\!\!\!\!$ 0.01 $\!\!\!\!$ & $\!\!\!\!$ 0.0020 $\!\!\!\!$ & $\!\!\!\!$ 0.70 $\!\!\!\!$ & $\!\!\!\!$ 0.10 $\!\!\!\!$ & $\!\!\!\!$ 0.0008 $\!\!\!\!$ & $\!\!\!\!$ 0.0100 $\!\!\!\!$ & $\!\!\!\!$ 0.0075 $\!\!\!\!$ & $\!\!\!\!$ 0.0100 $\!\!\!\!$ & $\!\!\!\!$ 0.0000 \\ 
{} $\!\!\!\!$ & $\!\!\!\!$ -3 $\!\!\!\!$ & $\!\!\!\!$ 8.750 $\!\!\!\!$ & $\!\!\!\!$ 7950 $\!\!\!\!$ & $\!\!\!\!$ 10346.8 $\!\!\!\!$ & $\!\!\!\!$ 0.00 $\!\!\!\!$ & $\!\!\!\!$ 0.00 $\!\!\!\!$ & $\!\!\!\!$ 0.0000 $\!\!\!\!$ & $\!\!\!\!$ 0.00 $\!\!\!\!$ & $\!\!\!\!$ 0.0010 $\!\!\!\!$ & $\!\!\!\!$ 0.80 $\!\!\!\!$ & $\!\!\!\!$ 0.14 $\!\!\!\!$ & $\!\!\!\!$ 0.0008 $\!\!\!\!$ & $\!\!\!\!$ 0.0110 $\!\!\!\!$ & $\!\!\!\!$ 0.0080 $\!\!\!\!$ & $\!\!\!\!$ 0.0400 $\!\!\!\!$ & $\!\!\!\!$ 0.0000 \\ 
{} $\!\!\!\!$ & $\!\!\!\!$ 4.8 $\!\!\!\!$ & $\!\!\!\!$ 8.858 $\!\!\!\!$ & $\!\!\!\!$ 7150 $\!\!\!\!$ & $\!\!\!\!$ 8750.8 $\!\!\!\!$ & $\!\!\!\!$ 0.00 $\!\!\!\!$ & $\!\!\!\!$ 0.01 $\!\!\!\!$ & $\!\!\!\!$ 0.0000 $\!\!\!\!$ & $\!\!\!\!$ 0.00 $\!\!\!\!$ & $\!\!\!\!$ 0.0000 $\!\!\!\!$ & $\!\!\!\!$ 0.76 $\!\!\!\!$ & $\!\!\!\!$ 0.08 $\!\!\!\!$ & $\!\!\!\!$ 0.0008 $\!\!\!\!$ & $\!\!\!\!$ 0.0290 $\!\!\!\!$ & $\!\!\!\!$ 0.0240 $\!\!\!\!$ & $\!\!\!\!$ 0.0900 $\!\!\!\!$ & $\!\!\!\!$ 0.0100 \\ 
{} $\!\!\!\!$ & $\!\!\!\!$ 12.9 $\!\!\!\!$ & $\!\!\!\!$ 8.594 $\!\!\!\!$ & $\!\!\!\!$ 3350 $\!\!\!\!$ & $\!\!\!\!$ 10598.5 $\!\!\!\!$ & $\!\!\!\!$ 0.00 $\!\!\!\!$ & $\!\!\!\!$ 0.00 $\!\!\!\!$ & $\!\!\!\!$ 0.0000 $\!\!\!\!$ & $\!\!\!\!$ 0.00 $\!\!\!\!$ & $\!\!\!\!$ 0.0000 $\!\!\!\!$ & $\!\!\!\!$ 0.84 $\!\!\!\!$ & $\!\!\!\!$ 0.00 $\!\!\!\!$ & $\!\!\!\!$ 0.0008 $\!\!\!\!$ & $\!\!\!\!$ 0.0140 $\!\!\!\!$ & $\!\!\!\!$ 0.0140 $\!\!\!\!$ & $\!\!\!\!$ 0.0400 $\!\!\!\!$ & $\!\!\!\!$ 0.0800 \\ 
05bl-w7e0.7m1.45 $\!\!\!\!$ & $\!\!\!\!$ -6 $\!\!\!\!$ & $\!\!\!\!$ 8.519 $\!\!\!\!$ & $\!\!\!\!$ 8700 $\!\!\!\!$ & $\!\!\!\!$ 9896.0 $\!\!\!\!$ & $\!\!\!\!$ 0.09 $\!\!\!\!$ & $\!\!\!\!$ 0.82 $\!\!\!\!$ & $\!\!\!\!$ 0.0055 $\!\!\!\!$ & $\!\!\!\!$ 0.04 $\!\!\!\!$ & $\!\!\!\!$ 0.0020 $\!\!\!\!$ & $\!\!\!\!$ 0.03 $\!\!\!\!$ & $\!\!\!\!$ 0.01 $\!\!\!\!$ & $\!\!\!\!$ 0.0006 $\!\!\!\!$ & $\!\!\!\!$ 0.0006 $\!\!\!\!$ & $\!\!\!\!$ 0.0006 $\!\!\!\!$ & $\!\!\!\!$ 0.0009 $\!\!\!\!$ & $\!\!\!\!$ 0.0000 \\ 
{} $\!\!\!\!$ & $\!\!\!\!$ -5 $\!\!\!\!$ & $\!\!\!\!$ 8.620 $\!\!\!\!$ & $\!\!\!\!$ 8450 $\!\!\!\!$ & $\!\!\!\!$ 10296.5 $\!\!\!\!$ & $\!\!\!\!$ 0.07 $\!\!\!\!$ & $\!\!\!\!$ 0.13 $\!\!\!\!$ & $\!\!\!\!$ 0.0030 $\!\!\!\!$ & $\!\!\!\!$ 0.04 $\!\!\!\!$ & $\!\!\!\!$ 0.0020 $\!\!\!\!$ & $\!\!\!\!$ 0.60 $\!\!\!\!$ & $\!\!\!\!$ 0.10 $\!\!\!\!$ & $\!\!\!\!$ 0.0006 $\!\!\!\!$ & $\!\!\!\!$ 0.0220 $\!\!\!\!$ & $\!\!\!\!$ 0.0220 $\!\!\!\!$ & $\!\!\!\!$ 0.0150 $\!\!\!\!$ & $\!\!\!\!$ 0.0000 \\ 
{} $\!\!\!\!$ & $\!\!\!\!$ -3 $\!\!\!\!$ & $\!\!\!\!$ 8.752 $\!\!\!\!$ & $\!\!\!\!$ 8200 $\!\!\!\!$ & $\!\!\!\!$ 10222.8 $\!\!\!\!$ & $\!\!\!\!$ 0.00 $\!\!\!\!$ & $\!\!\!\!$ 0.01 $\!\!\!\!$ & $\!\!\!\!$ 0.0000 $\!\!\!\!$ & $\!\!\!\!$ 0.00 $\!\!\!\!$ & $\!\!\!\!$ 0.0010 $\!\!\!\!$ & $\!\!\!\!$ 0.76 $\!\!\!\!$ & $\!\!\!\!$ 0.12 $\!\!\!\!$ & $\!\!\!\!$ 0.0013 $\!\!\!\!$ & $\!\!\!\!$ 0.0275 $\!\!\!\!$ & $\!\!\!\!$ 0.0275 $\!\!\!\!$ & $\!\!\!\!$ 0.0450 $\!\!\!\!$ & $\!\!\!\!$ 0.0000 \\ 
{} $\!\!\!\!$ & $\!\!\!\!$ 4.8 $\!\!\!\!$ & $\!\!\!\!$ 8.864 $\!\!\!\!$ & $\!\!\!\!$ 7350 $\!\!\!\!$ & $\!\!\!\!$ 8761.1 $\!\!\!\!$ & $\!\!\!\!$ 0.00 $\!\!\!\!$ & $\!\!\!\!$ 0.01 $\!\!\!\!$ & $\!\!\!\!$ 0.0000 $\!\!\!\!$ & $\!\!\!\!$ 0.00 $\!\!\!\!$ & $\!\!\!\!$ 0.0000 $\!\!\!\!$ & $\!\!\!\!$ 0.78 $\!\!\!\!$ & $\!\!\!\!$ 0.07 $\!\!\!\!$ & $\!\!\!\!$ 0.0006 $\!\!\!\!$ & $\!\!\!\!$ 0.0333 $\!\!\!\!$ & $\!\!\!\!$ 0.0333 $\!\!\!\!$ & $\!\!\!\!$ 0.0700 $\!\!\!\!$ & $\!\!\!\!$ 0.0070 \\ 
{} $\!\!\!\!$ & $\!\!\!\!$ 12.9 $\!\!\!\!$ & $\!\!\!\!$ 8.605 $\!\!\!\!$ & $\!\!\!\!$ 3525 $\!\!\!\!$ & $\!\!\!\!$ 10682.2 $\!\!\!\!$ & $\!\!\!\!$ 0.00 $\!\!\!\!$ & $\!\!\!\!$ 0.00 $\!\!\!\!$ & $\!\!\!\!$ 0.0000 $\!\!\!\!$ & $\!\!\!\!$ 0.00 $\!\!\!\!$ & $\!\!\!\!$ 0.0000 $\!\!\!\!$ & $\!\!\!\!$ 0.87 $\!\!\!\!$ & $\!\!\!\!$ 0.00 $\!\!\!\!$ & $\!\!\!\!$ 0.0006 $\!\!\!\!$ & $\!\!\!\!$ 0.0170 $\!\!\!\!$ & $\!\!\!\!$ 0.0170 $\!\!\!\!$ & $\!\!\!\!$ 0.0250 $\!\!\!\!$ & $\!\!\!\!$ 0.0650 \\ 
05bl-w7m0.7 $\!\!\!\!$ & $\!\!\!\!$ -6 $\!\!\!\!$ & $\!\!\!\!$ 8.520 $\!\!\!\!$ & $\!\!\!\!$ 8150 $\!\!\!\!$ & $\!\!\!\!$ 9593.4 $\!\!\!\!$ & $\!\!\!\!$ 0.06 $\!\!\!\!$ & $\!\!\!\!$ 0.80 $\!\!\!\!$ & $\!\!\!\!$ 0.0060 $\!\!\!\!$ & $\!\!\!\!$ 0.05 $\!\!\!\!$ & $\!\!\!\!$ 0.0060 $\!\!\!\!$ & $\!\!\!\!$ 0.05 $\!\!\!\!$ & $\!\!\!\!$ 0.02 $\!\!\!\!$ & $\!\!\!\!$ 0.0010 $\!\!\!\!$ & $\!\!\!\!$ 0.0004 $\!\!\!\!$ & $\!\!\!\!$ 0.0008 $\!\!\!\!$ & $\!\!\!\!$ 0.0003 $\!\!\!\!$ & $\!\!\!\!$ 0.0000 \\ 
{} $\!\!\!\!$ & $\!\!\!\!$ -5 $\!\!\!\!$ & $\!\!\!\!$ 8.630 $\!\!\!\!$ & $\!\!\!\!$ 7850 $\!\!\!\!$ & $\!\!\!\!$ 9927.4 $\!\!\!\!$ & $\!\!\!\!$ 0.03 $\!\!\!\!$ & $\!\!\!\!$ 0.11 $\!\!\!\!$ & $\!\!\!\!$ 0.0020 $\!\!\!\!$ & $\!\!\!\!$ 0.04 $\!\!\!\!$ & $\!\!\!\!$ 0.0050 $\!\!\!\!$ & $\!\!\!\!$ 0.72 $\!\!\!\!$ & $\!\!\!\!$ 0.07 $\!\!\!\!$ & $\!\!\!\!$ 0.0010 $\!\!\!\!$ & $\!\!\!\!$ 0.0050 $\!\!\!\!$ & $\!\!\!\!$ 0.0050 $\!\!\!\!$ & $\!\!\!\!$ 0.0050 $\!\!\!\!$ & $\!\!\!\!$ 0.0000 \\ 
{} $\!\!\!\!$ & $\!\!\!\!$ -3 $\!\!\!\!$ & $\!\!\!\!$ 8.745 $\!\!\!\!$ & $\!\!\!\!$ 7475 $\!\!\!\!$ & $\!\!\!\!$ 9998.2 $\!\!\!\!$ & $\!\!\!\!$ 0.00 $\!\!\!\!$ & $\!\!\!\!$ 0.00 $\!\!\!\!$ & $\!\!\!\!$ 0.0000 $\!\!\!\!$ & $\!\!\!\!$ 0.00 $\!\!\!\!$ & $\!\!\!\!$ 0.0020 $\!\!\!\!$ & $\!\!\!\!$ 0.72 $\!\!\!\!$ & $\!\!\!\!$ 0.15 $\!\!\!\!$ & $\!\!\!\!$ 0.0010 $\!\!\!\!$ & $\!\!\!\!$ 0.0375 $\!\!\!\!$ & $\!\!\!\!$ 0.0375 $\!\!\!\!$ & $\!\!\!\!$ 0.0475 $\!\!\!\!$ & $\!\!\!\!$ 0.0000 \\ 
{} $\!\!\!\!$ & $\!\!\!\!$ 4.8 $\!\!\!\!$ & $\!\!\!\!$ 8.844 $\!\!\!\!$ & $\!\!\!\!$ 6550 $\!\!\!\!$ & $\!\!\!\!$ 8654.3 $\!\!\!\!$ & $\!\!\!\!$ 0.00 $\!\!\!\!$ & $\!\!\!\!$ 0.01 $\!\!\!\!$ & $\!\!\!\!$ 0.0000 $\!\!\!\!$ & $\!\!\!\!$ 0.00 $\!\!\!\!$ & $\!\!\!\!$ 0.0000 $\!\!\!\!$ & $\!\!\!\!$ 0.60 $\!\!\!\!$ & $\!\!\!\!$ 0.05 $\!\!\!\!$ & $\!\!\!\!$ 0.0010 $\!\!\!\!$ & $\!\!\!\!$ 0.0900 $\!\!\!\!$ & $\!\!\!\!$ 0.0900 $\!\!\!\!$ & $\!\!\!\!$ 0.1500 $\!\!\!\!$ & $\!\!\!\!$ 0.0100 \\ 
{} $\!\!\!\!$ & $\!\!\!\!$ 12.9 $\!\!\!\!$ & $\!\!\!\!$ 8.566 $\!\!\!\!$ & $\!\!\!\!$ 3250 $\!\!\!\!$ & $\!\!\!\!$ 9133.8 $\!\!\!\!$ & $\!\!\!\!$ 0.00 $\!\!\!\!$ & $\!\!\!\!$ 0.00 $\!\!\!\!$ & $\!\!\!\!$ 0.0000 $\!\!\!\!$ & $\!\!\!\!$ 0.00 $\!\!\!\!$ & $\!\!\!\!$ 0.0000 $\!\!\!\!$ & $\!\!\!\!$ 0.46 $\!\!\!\!$ & $\!\!\!\!$ 0.00 $\!\!\!\!$ & $\!\!\!\!$ 0.0005 $\!\!\!\!$ & $\!\!\!\!$ 0.0800 $\!\!\!\!$ & $\!\!\!\!$ 0.0800 $\!\!\!\!$ & $\!\!\!\!$ 0.1250 $\!\!\!\!$ & $\!\!\!\!$ 0.2500 \\ 
05bl-w7 $\!\!\!\!$ & $\!\!\!\!$ -6 $\!\!\!\!$ & $\!\!\!\!$ 8.525 $\!\!\!\!$ & $\!\!\!\!$ 8400 $\!\!\!\!$ & $\!\!\!\!$ 9845.6 $\!\!\!\!$ & $\!\!\!\!$ 0.05 $\!\!\!\!$ & $\!\!\!\!$ 0.84 $\!\!\!\!$ & $\!\!\!\!$ 0.0025 $\!\!\!\!$ & $\!\!\!\!$ 0.03 $\!\!\!\!$ & $\!\!\!\!$ 0.0032 $\!\!\!\!$ & $\!\!\!\!$ 0.05 $\!\!\!\!$ & $\!\!\!\!$ 0.02 $\!\!\!\!$ & $\!\!\!\!$ 0.0003 $\!\!\!\!$ & $\!\!\!\!$ 0.0004 $\!\!\!\!$ & $\!\!\!\!$ 0.0004 $\!\!\!\!$ & $\!\!\!\!$ 0.0001 $\!\!\!\!$ & $\!\!\!\!$ 0.0000 \\ 
{} $\!\!\!\!$ & $\!\!\!\!$ -5 $\!\!\!\!$ & $\!\!\!\!$ 8.627 $\!\!\!\!$ & $\!\!\!\!$ 8100 $\!\!\!\!$ & $\!\!\!\!$ 10291.1 $\!\!\!\!$ & $\!\!\!\!$ 0.00 $\!\!\!\!$ & $\!\!\!\!$ 0.17 $\!\!\!\!$ & $\!\!\!\!$ 0.0013 $\!\!\!\!$ & $\!\!\!\!$ 0.04 $\!\!\!\!$ & $\!\!\!\!$ 0.0032 $\!\!\!\!$ & $\!\!\!\!$ 0.65 $\!\!\!\!$ & $\!\!\!\!$ 0.10 $\!\!\!\!$ & $\!\!\!\!$ 0.0003 $\!\!\!\!$ & $\!\!\!\!$ 0.0100 $\!\!\!\!$ & $\!\!\!\!$ 0.0100 $\!\!\!\!$ & $\!\!\!\!$ 0.0175 $\!\!\!\!$ & $\!\!\!\!$ 0.0000 \\ 
{} $\!\!\!\!$ & $\!\!\!\!$ -3 $\!\!\!\!$ & $\!\!\!\!$ 8.754 $\!\!\!\!$ & $\!\!\!\!$ 7500 $\!\!\!\!$ & $\!\!\!\!$ 10859.7 $\!\!\!\!$ & $\!\!\!\!$ 0.00 $\!\!\!\!$ & $\!\!\!\!$ 0.01 $\!\!\!\!$ & $\!\!\!\!$ 0.0000 $\!\!\!\!$ & $\!\!\!\!$ 0.00 $\!\!\!\!$ & $\!\!\!\!$ 0.0011 $\!\!\!\!$ & $\!\!\!\!$ 0.69 $\!\!\!\!$ & $\!\!\!\!$ 0.12 $\!\!\!\!$ & $\!\!\!\!$ 0.0003 $\!\!\!\!$ & $\!\!\!\!$ 0.0500 $\!\!\!\!$ & $\!\!\!\!$ 0.0500 $\!\!\!\!$ & $\!\!\!\!$ 0.0800 $\!\!\!\!$ & $\!\!\!\!$ 0.0000 \\ 
{} $\!\!\!\!$ & $\!\!\!\!$ 4.8 $\!\!\!\!$ & $\!\!\!\!$ 8.859 $\!\!\!\!$ & $\!\!\!\!$ 6600 $\!\!\!\!$ & $\!\!\!\!$ 9110.3 $\!\!\!\!$ & $\!\!\!\!$ 0.00 $\!\!\!\!$ & $\!\!\!\!$ 0.00 $\!\!\!\!$ & $\!\!\!\!$ 0.0000 $\!\!\!\!$ & $\!\!\!\!$ 0.00 $\!\!\!\!$ & $\!\!\!\!$ 0.0000 $\!\!\!\!$ & $\!\!\!\!$ 0.69 $\!\!\!\!$ & $\!\!\!\!$ 0.06 $\!\!\!\!$ & $\!\!\!\!$ 0.0003 $\!\!\!\!$ & $\!\!\!\!$ 0.0675 $\!\!\!\!$ & $\!\!\!\!$ 0.0675 $\!\!\!\!$ & $\!\!\!\!$ 0.1000 $\!\!\!\!$ & $\!\!\!\!$ 0.0100 \\ 
{} $\!\!\!\!$ & $\!\!\!\!$ 12.9 $\!\!\!\!$ & $\!\!\!\!$ 8.594 $\!\!\!\!$ & $\!\!\!\!$ 3300 $\!\!\!\!$ & $\!\!\!\!$ 9957.4 $\!\!\!\!$ & $\!\!\!\!$ 0.00 $\!\!\!\!$ & $\!\!\!\!$ 0.00 $\!\!\!\!$ & $\!\!\!\!$ 0.0000 $\!\!\!\!$ & $\!\!\!\!$ 0.00 $\!\!\!\!$ & $\!\!\!\!$ 0.0000 $\!\!\!\!$ & $\!\!\!\!$ 0.64 $\!\!\!\!$ & $\!\!\!\!$ 0.00 $\!\!\!\!$ & $\!\!\!\!$ 0.0003 $\!\!\!\!$ & $\!\!\!\!$ 0.0350 $\!\!\!\!$ & $\!\!\!\!$ 0.0350 $\!\!\!\!$ & $\!\!\!\!$ 0.0950 $\!\!\!\!$ & $\!\!\!\!$ 0.1900 \\ 
05bl-w7m1.25 $\!\!\!\!$ & $\!\!\!\!$ -6 $\!\!\!\!$ & $\!\!\!\!$ 8.515 $\!\!\!\!$ & $\!\!\!\!$ 8800 $\!\!\!\!$ & $\!\!\!\!$ 9685.6 $\!\!\!\!$ & $\!\!\!\!$ 0.08 $\!\!\!\!$ & $\!\!\!\!$ 0.81 $\!\!\!\!$ & $\!\!\!\!$ 0.0040 $\!\!\!\!$ & $\!\!\!\!$ 0.05 $\!\!\!\!$ & $\!\!\!\!$ 0.0023 $\!\!\!\!$ & $\!\!\!\!$ 0.04 $\!\!\!\!$ & $\!\!\!\!$ 0.02 $\!\!\!\!$ & $\!\!\!\!$ 0.0003 $\!\!\!\!$ & $\!\!\!\!$ 0.0002 $\!\!\!\!$ & $\!\!\!\!$ 0.0002 $\!\!\!\!$ & $\!\!\!\!$ 0.0002 $\!\!\!\!$ & $\!\!\!\!$ 0.0000 \\ 
{} $\!\!\!\!$ & $\!\!\!\!$ -5 $\!\!\!\!$ & $\!\!\!\!$ 8.622 $\!\!\!\!$ & $\!\!\!\!$ 8500 $\!\!\!\!$ & $\!\!\!\!$ 10207.5 $\!\!\!\!$ & $\!\!\!\!$ 0.05 $\!\!\!\!$ & $\!\!\!\!$ 0.12 $\!\!\!\!$ & $\!\!\!\!$ 0.0040 $\!\!\!\!$ & $\!\!\!\!$ 0.05 $\!\!\!\!$ & $\!\!\!\!$ 0.0023 $\!\!\!\!$ & $\!\!\!\!$ 0.65 $\!\!\!\!$ & $\!\!\!\!$ 0.08 $\!\!\!\!$ & $\!\!\!\!$ 0.0003 $\!\!\!\!$ & $\!\!\!\!$ 0.0160 $\!\!\!\!$ & $\!\!\!\!$ 0.0120 $\!\!\!\!$ & $\!\!\!\!$ 0.0190 $\!\!\!\!$ & $\!\!\!\!$ 0.0000 \\ 
{} $\!\!\!\!$ & $\!\!\!\!$ -3 $\!\!\!\!$ & $\!\!\!\!$ 8.756 $\!\!\!\!$ & $\!\!\!\!$ 7950 $\!\!\!\!$ & $\!\!\!\!$ 10729.3 $\!\!\!\!$ & $\!\!\!\!$ 0.00 $\!\!\!\!$ & $\!\!\!\!$ 0.01 $\!\!\!\!$ & $\!\!\!\!$ 0.0000 $\!\!\!\!$ & $\!\!\!\!$ 0.00 $\!\!\!\!$ & $\!\!\!\!$ 0.0010 $\!\!\!\!$ & $\!\!\!\!$ 0.72 $\!\!\!\!$ & $\!\!\!\!$ 0.15 $\!\!\!\!$ & $\!\!\!\!$ 0.0003 $\!\!\!\!$ & $\!\!\!\!$ 0.0425 $\!\!\!\!$ & $\!\!\!\!$ 0.0400 $\!\!\!\!$ & $\!\!\!\!$ 0.0400 $\!\!\!\!$ & $\!\!\!\!$ 0.0000 \\ 
{} $\!\!\!\!$ & $\!\!\!\!$ 4.8 $\!\!\!\!$ & $\!\!\!\!$ 8.862 $\!\!\!\!$ & $\!\!\!\!$ 7000 $\!\!\!\!$ & $\!\!\!\!$ 8996.5 $\!\!\!\!$ & $\!\!\!\!$ 0.00 $\!\!\!\!$ & $\!\!\!\!$ 0.00 $\!\!\!\!$ & $\!\!\!\!$ 0.0000 $\!\!\!\!$ & $\!\!\!\!$ 0.00 $\!\!\!\!$ & $\!\!\!\!$ 0.0000 $\!\!\!\!$ & $\!\!\!\!$ 0.78 $\!\!\!\!$ & $\!\!\!\!$ 0.09 $\!\!\!\!$ & $\!\!\!\!$ 0.0003 $\!\!\!\!$ & $\!\!\!\!$ 0.0450 $\!\!\!\!$ & $\!\!\!\!$ 0.0450 $\!\!\!\!$ & $\!\!\!\!$ 0.0400 $\!\!\!\!$ & $\!\!\!\!$ 0.0050 \\ 
{} $\!\!\!\!$ & $\!\!\!\!$ 12.9 $\!\!\!\!$ & $\!\!\!\!$ 8.598 $\!\!\!\!$ & $\!\!\!\!$ 3325 $\!\!\!\!$ & $\!\!\!\!$ 10378.3 $\!\!\!\!$ & $\!\!\!\!$ 0.00 $\!\!\!\!$ & $\!\!\!\!$ 0.00 $\!\!\!\!$ & $\!\!\!\!$ 0.0000 $\!\!\!\!$ & $\!\!\!\!$ 0.00 $\!\!\!\!$ & $\!\!\!\!$ 0.0000 $\!\!\!\!$ & $\!\!\!\!$ 0.80 $\!\!\!\!$ & $\!\!\!\!$ 0.00 $\!\!\!\!$ & $\!\!\!\!$ 0.0003 $\!\!\!\!$ & $\!\!\!\!$ 0.0135 $\!\!\!\!$ & $\!\!\!\!$ 0.0135 $\!\!\!\!$ & $\!\!\!\!$ 0.0600 $\!\!\!\!$ & $\!\!\!\!$ 0.1100 \\ 
05bl-w7m1.45 $\!\!\!\!$ & $\!\!\!\!$ -6 $\!\!\!\!$ & $\!\!\!\!$ 8.520 $\!\!\!\!$ & $\!\!\!\!$ 8800 $\!\!\!\!$ & $\!\!\!\!$ 9990.6 $\!\!\!\!$ & $\!\!\!\!$ 0.06 $\!\!\!\!$ & $\!\!\!\!$ 0.83 $\!\!\!\!$ & $\!\!\!\!$ 0.0040 $\!\!\!\!$ & $\!\!\!\!$ 0.06 $\!\!\!\!$ & $\!\!\!\!$ 0.0020 $\!\!\!\!$ & $\!\!\!\!$ 0.04 $\!\!\!\!$ & $\!\!\!\!$ 0.01 $\!\!\!\!$ & $\!\!\!\!$ 0.0005 $\!\!\!\!$ & $\!\!\!\!$ 0.0003 $\!\!\!\!$ & $\!\!\!\!$ 0.0003 $\!\!\!\!$ & $\!\!\!\!$ 0.0006 $\!\!\!\!$ & $\!\!\!\!$ 0.0000 \\ 
{} $\!\!\!\!$ & $\!\!\!\!$ -5 $\!\!\!\!$ & $\!\!\!\!$ 8.628 $\!\!\!\!$ & $\!\!\!\!$ 8550 $\!\!\!\!$ & $\!\!\!\!$ 10483.5 $\!\!\!\!$ & $\!\!\!\!$ 0.03 $\!\!\!\!$ & $\!\!\!\!$ 0.10 $\!\!\!\!$ & $\!\!\!\!$ 0.0020 $\!\!\!\!$ & $\!\!\!\!$ 0.04 $\!\!\!\!$ & $\!\!\!\!$ 0.0020 $\!\!\!\!$ & $\!\!\!\!$ 0.65 $\!\!\!\!$ & $\!\!\!\!$ 0.12 $\!\!\!\!$ & $\!\!\!\!$ 0.0005 $\!\!\!\!$ & $\!\!\!\!$ 0.0150 $\!\!\!\!$ & $\!\!\!\!$ 0.0150 $\!\!\!\!$ & $\!\!\!\!$ 0.0200 $\!\!\!\!$ & $\!\!\!\!$ 0.0000 \\ 
{} $\!\!\!\!$ & $\!\!\!\!$ -3 $\!\!\!\!$ & $\!\!\!\!$ 8.758 $\!\!\!\!$ & $\!\!\!\!$ 8200 $\!\!\!\!$ & $\!\!\!\!$ 10562.5 $\!\!\!\!$ & $\!\!\!\!$ 0.00 $\!\!\!\!$ & $\!\!\!\!$ 0.01 $\!\!\!\!$ & $\!\!\!\!$ 0.0000 $\!\!\!\!$ & $\!\!\!\!$ 0.00 $\!\!\!\!$ & $\!\!\!\!$ 0.0010 $\!\!\!\!$ & $\!\!\!\!$ 0.78 $\!\!\!\!$ & $\!\!\!\!$ 0.10 $\!\!\!\!$ & $\!\!\!\!$ 0.0005 $\!\!\!\!$ & $\!\!\!\!$ 0.0275 $\!\!\!\!$ & $\!\!\!\!$ 0.0275 $\!\!\!\!$ & $\!\!\!\!$ 0.0525 $\!\!\!\!$ & $\!\!\!\!$ 0.0000 \\ 
{} $\!\!\!\!$ & $\!\!\!\!$ 4.8 $\!\!\!\!$ & $\!\!\!\!$ 8.861 $\!\!\!\!$ & $\!\!\!\!$ 7175 $\!\!\!\!$ & $\!\!\!\!$ 9017.1 $\!\!\!\!$ & $\!\!\!\!$ 0.00 $\!\!\!\!$ & $\!\!\!\!$ 0.01 $\!\!\!\!$ & $\!\!\!\!$ 0.0000 $\!\!\!\!$ & $\!\!\!\!$ 0.00 $\!\!\!\!$ & $\!\!\!\!$ 0.0000 $\!\!\!\!$ & $\!\!\!\!$ 0.78 $\!\!\!\!$ & $\!\!\!\!$ 0.08 $\!\!\!\!$ & $\!\!\!\!$ 0.0005 $\!\!\!\!$ & $\!\!\!\!$ 0.0300 $\!\!\!\!$ & $\!\!\!\!$ 0.0300 $\!\!\!\!$ & $\!\!\!\!$ 0.0700 $\!\!\!\!$ & $\!\!\!\!$ 0.0070 \\ 
{} $\!\!\!\!$ & $\!\!\!\!$ 12.9 $\!\!\!\!$ & $\!\!\!\!$ 8.605 $\!\!\!\!$ & $\!\!\!\!$ 3475 $\!\!\!\!$ & $\!\!\!\!$ 10418.1 $\!\!\!\!$ & $\!\!\!\!$ 0.00 $\!\!\!\!$ & $\!\!\!\!$ 0.00 $\!\!\!\!$ & $\!\!\!\!$ 0.0000 $\!\!\!\!$ & $\!\!\!\!$ 0.00 $\!\!\!\!$ & $\!\!\!\!$ 0.0000 $\!\!\!\!$ & $\!\!\!\!$ 0.87 $\!\!\!\!$ & $\!\!\!\!$ 0.00 $\!\!\!\!$ & $\!\!\!\!$ 0.0005 $\!\!\!\!$ & $\!\!\!\!$ 0.0097 $\!\!\!\!$ & $\!\!\!\!$ 0.0097 $\!\!\!\!$ & $\!\!\!\!$ 0.0350 $\!\!\!\!$ & $\!\!\!\!$ 0.0700 \\ 
05bl-w7e1.45m1.45 $\!\!\!\!$ & $\!\!\!\!$ -6 $\!\!\!\!$ & $\!\!\!\!$ 8.524 $\!\!\!\!$ & $\!\!\!\!$ 9000 $\!\!\!\!$ & $\!\!\!\!$ 10051.5 $\!\!\!\!$ & $\!\!\!\!$ 0.08 $\!\!\!\!$ & $\!\!\!\!$ 0.71 $\!\!\!\!$ & $\!\!\!\!$ 0.0043 $\!\!\!\!$ & $\!\!\!\!$ 0.10 $\!\!\!\!$ & $\!\!\!\!$ 0.0035 $\!\!\!\!$ & $\!\!\!\!$ 0.07 $\!\!\!\!$ & $\!\!\!\!$ 0.03 $\!\!\!\!$ & $\!\!\!\!$ 0.0002 $\!\!\!\!$ & $\!\!\!\!$ 0.0002 $\!\!\!\!$ & $\!\!\!\!$ 0.0003 $\!\!\!\!$ & $\!\!\!\!$ 0.0005 $\!\!\!\!$ & $\!\!\!\!$ 0.0000 \\ 
{} $\!\!\!\!$ & $\!\!\!\!$ -5 $\!\!\!\!$ & $\!\!\!\!$ 8.623 $\!\!\!\!$ & $\!\!\!\!$ 8700 $\!\!\!\!$ & $\!\!\!\!$ 10534.9 $\!\!\!\!$ & $\!\!\!\!$ 0.09 $\!\!\!\!$ & $\!\!\!\!$ 0.12 $\!\!\!\!$ & $\!\!\!\!$ 0.0043 $\!\!\!\!$ & $\!\!\!\!$ 0.07 $\!\!\!\!$ & $\!\!\!\!$ 0.0035 $\!\!\!\!$ & $\!\!\!\!$ 0.55 $\!\!\!\!$ & $\!\!\!\!$ 0.10 $\!\!\!\!$ & $\!\!\!\!$ 0.0002 $\!\!\!\!$ & $\!\!\!\!$ 0.0225 $\!\!\!\!$ & $\!\!\!\!$ 0.0225 $\!\!\!\!$ & $\!\!\!\!$ 0.0200 $\!\!\!\!$ & $\!\!\!\!$ 0.0000 \\ 
{} $\!\!\!\!$ & $\!\!\!\!$ -3 $\!\!\!\!$ & $\!\!\!\!$ 8.760 $\!\!\!\!$ & $\!\!\!\!$ 8100 $\!\!\!\!$ & $\!\!\!\!$ 11060.7 $\!\!\!\!$ & $\!\!\!\!$ 0.00 $\!\!\!\!$ & $\!\!\!\!$ 0.01 $\!\!\!\!$ & $\!\!\!\!$ 0.0000 $\!\!\!\!$ & $\!\!\!\!$ 0.01 $\!\!\!\!$ & $\!\!\!\!$ 0.0015 $\!\!\!\!$ & $\!\!\!\!$ 0.68 $\!\!\!\!$ & $\!\!\!\!$ 0.14 $\!\!\!\!$ & $\!\!\!\!$ 0.0002 $\!\!\!\!$ & $\!\!\!\!$ 0.0500 $\!\!\!\!$ & $\!\!\!\!$ 0.0500 $\!\!\!\!$ & $\!\!\!\!$ 0.0750 $\!\!\!\!$ & $\!\!\!\!$ 0.0000 \\ 
{} $\!\!\!\!$ & $\!\!\!\!$ 4.8 $\!\!\!\!$ & $\!\!\!\!$ 8.868 $\!\!\!\!$ & $\!\!\!\!$ 7000 $\!\!\!\!$ & $\!\!\!\!$ 9313.8 $\!\!\!\!$ & $\!\!\!\!$ 0.00 $\!\!\!\!$ & $\!\!\!\!$ 0.01 $\!\!\!\!$ & $\!\!\!\!$ 0.0000 $\!\!\!\!$ & $\!\!\!\!$ 0.00 $\!\!\!\!$ & $\!\!\!\!$ 0.0000 $\!\!\!\!$ & $\!\!\!\!$ 0.72 $\!\!\!\!$ & $\!\!\!\!$ 0.07 $\!\!\!\!$ & $\!\!\!\!$ 0.0002 $\!\!\!\!$ & $\!\!\!\!$ 0.0650 $\!\!\!\!$ & $\!\!\!\!$ 0.0650 $\!\!\!\!$ & $\!\!\!\!$ 0.0675 $\!\!\!\!$ & $\!\!\!\!$ 0.0067 \\ 
{} $\!\!\!\!$ & $\!\!\!\!$ 12.9 $\!\!\!\!$ & $\!\!\!\!$ 8.607 $\!\!\!\!$ & $\!\!\!\!$ 3450 $\!\!\!\!$ & $\!\!\!\!$ 10342.2 $\!\!\!\!$ & $\!\!\!\!$ 0.00 $\!\!\!\!$ & $\!\!\!\!$ 0.00 $\!\!\!\!$ & $\!\!\!\!$ 0.0000 $\!\!\!\!$ & $\!\!\!\!$ 0.00 $\!\!\!\!$ & $\!\!\!\!$ 0.0000 $\!\!\!\!$ & $\!\!\!\!$ 0.84 $\!\!\!\!$ & $\!\!\!\!$ 0.00 $\!\!\!\!$ & $\!\!\!\!$ 0.0002 $\!\!\!\!$ & $\!\!\!\!$ 0.0350 $\!\!\!\!$ & $\!\!\!\!$ 0.0350 $\!\!\!\!$ & $\!\!\!\!$ 0.0250 $\!\!\!\!$ & $\!\!\!\!$ 0.0600 \\ 
\hline
\end{tabular} 
\begin{flushleft}
${}^{\textrm{a)}}$ The abundances of Fe, Co and Ni in our models are assumed to be the sum of \Nifs\ and its decay chain products (\Cofs\ and \Fefs) on the one hand, and directly synthesised / progenitor Fe on the other hand. Thus, they are conveniently given in terms of the \Nifs\ mass fraction at $t=0$ [$X($\Nifs$)_0$], the Fe abundance at $t=0$ [$X(\textrm{Fe})_0$], and the time from explosion onset $t$.
\end{flushleft}
\end{table*}

\end{document}